\RequirePackage{silence}
\documentclass[fleqn,usenatbib,useAMS]{mnras} 
\usepackage{graphicx}
\usepackage{amsmath}

\DeclareMathOperator{\sech}{sech}
\usepackage{amsfonts}
\usepackage{float}
\usepackage{bm}
\usepackage{cancel}
\usepackage{ae,aecompl}
\usepackage{array}
\usepackage{soul}
\usepackage{mathtools}
\usepackage{multirow}
\usepackage[utf8]{inputenc}
\usepackage{booktabs}
\usepackage{graphicx}
\usepackage{float}
\usepackage{makecell}

\DeclareRobustCommand{\appropto}{\mathrel{\vcenter{
		\offinterlineskip\halign{\hfil$##$\cr 
			\propto\cr\noalign{\kern2pt}\sim\cr\noalign{\kern-2pt}}}}}

\title [Star forming main sequence galaxies in MOND]{Simulations of star forming main sequence galaxies in Milgromian gravity} 

\author[S. T. Nagesh et al.]{Srikanth T. Nagesh,$^{1,4}$\thanks{Email: \href{mailto:tnsrikanth1998@gmail.com}{tnsrikanth1998@gmail.com}}
Pavel Kroupa,$^{1,2}$\thanks{Email: \href{mailto:pkroupa@uni-bonn.de}{pkroupa@uni-bonn.de}}
Indranil Banik,$^{3}$\thanks{Email: \href{mailto:indranilbanik1992@gmail.com}{indranilbanik1992@gmail.com}}
Benoit Famaey,$^{4}$
Neda Ghafourian,$^{5}$
\newauthor
Mahmood Roshan,$^{5,6}$ 
Ingo Thies,$^{1}$
Hongsheng Zhao,$^{3}$ and
Nils Wittenburg,$^{1}$
\\
$^{1}$Helmholtz-Institut f\"ur Strahlen- und Kernphysik, Universit\"at Bonn,
Nussallee 14-16, 53115 Bonn, Germany \\
$^{2}$Astronomical Institute, Faculty of Mathematics and Physics, Charles University, V Hole\v{s}ovi\v{c}k\'ach 2, CZ-180 00 Praha 8, Czech Republic \\
$^{3}$Scottish Universities Physics Alliance, University of Saint Andrews, North Haugh, Saint Andrews, Fife, KY16 9SS, UK \\
$^{4}$Universit\'{e} de Strasbourg, CNRS UMR 7550, Observatoire astronomique de Strasbourg, 11 rue de l'Universit\'{e}, 67000 Strasbourg, France\\
$^{5}$Department of Physics, Faculty of Science, Ferdowsi University of Mashhad, P.O. Box 1436, Mashhad, Iran\\
$^{6}$Institute for Research in Fundamental Sciences (IPM), P. O. Box 19395-5531, Tehran, Iran}

\pubyear{2022}
\begin{document}
\label{firstpage}
\pagerange{\pageref{firstpage}--\pageref{lastpage}}
\maketitle

\begin{abstract} 
We conduct hydrodynamical MOND simulations of isolated disc galaxies over the stellar mass range $M_{\star}/M_\odot = 10^7 - 10^{11}$ using the adaptive mesh refinement code \textsc{phantom of ramses} (\textsc{por}), an adaptation of the \textsc{ramses} code with a Milgromian gravity solver. The scale lengths and gas fractions are based on observed galaxies, and the simulations are run for 5~Gyr. The main aim is to see whether existing sub-grid physics prescriptions for star formation and stellar feedback reproduce the observed main sequence and reasonably match the Kennicutt-Schmidt relation that captures how the local and global star formation rates relate to other properties. Star formation in the models starts soon after initialisation and continues as the models evolve. The initialized galaxies indeed evolve to a state which is on the observed main sequence, and reasonably matches the Kennicutt-Schmidt relation. The available formulation of sub-grid physics is therefore adequate and leads to galaxies that largely behave like observed galaxies, grow in radius, and have flat rotation curves $-$ provided we use Milgromian gravitation. Furthermore, the strength of the bars tends to be inversely correlated with the stellar mass of the galaxy, whereas the bar length strongly correlates with the stellar mass. Irrespective of the mass, the bar pattern speed stays constant with time, indicating that dynamical friction does not affect the bar dynamics. The models demonstrate Renzo's rule and form structures at large radii, much as in real galaxies. In this framework, baryonic physics is thus sufficiently understood to not pose major uncertainties in our modelling of global galaxy properties.

\end{abstract}


\section{Introduction}
\label{Introduction}

Galaxies are rather simple systems \citep{Disney_2008} that obey well-defined scaling relations \citep{Sanders_1990}. These include the mass discrepancy-acceleration relation (MDAR)/radial acceleration relation \citep[RAR;][]{McGaugh_Lelli_2016, Lelli_2017}, the Faber-Jackson relation \citep{Faber_1976, McGaugh_2004, Lelli_2017}, the baryonic Tully-Fisher Relation \citep[BTFR;][]{Tully_Fisher_1977, McGaugh_2000, McGaugh_2012}, the main sequence (MS) of galaxies \citep{Speagle_2014}, and the Kennicutt-Schmidt (KS) relation \citep{Schmidt_1959, Kennicutt_1998}. The formation of stars in the galaxy and their feedback on the gas need to be treated numerically with sub-grid physics algorithms given that these processes act on scales $\la 1$~pc, much below the typical numerical resolution of around $10 - 100$~pc. However, it is gravitation which governs how the star-forming gas settles to form a galaxy. This creates a problem because gravitation is the least understood of the fundamental forces. Based on Solar System data and the work of \citet{Newton_1687}, gravity was interpreted to be a geometric effect caused by the distortion of spacetime itself \citep{Einstein_1916}, causing deviations from Newtonian gravity in the strong field regime that explain, e.g., the orbit of Mercury. More recently, \citet{Verlinde_2011} suggested gravitation to be an entropic force which arises from differences in the information content of space, while \citet{Stadtler_2020} suggest gravitation to be a consequence of the wave nature of matter. Given the lack of a deep physical understanding of gravitation and its relation to the quantum mechanical properties of spacetime and matter, it is perhaps not surprising that significant deviations from the non-relativistic Newtonian formulation arise in the rotation curves (RCs) of galaxies \citep[][and references therein]{Faber_1979}. After all, this flat RC problem arises on much larger scales (and within a much weaker gravitational acceleration field) than the Solar System scales that yielded the empirical constraints used in the formulation of General Relativity more than a century ago.

\subsection{Milgromian dynamics (MOND)}

A modern non-relativistic theory of gravitation was proposed by \citet{Bekenstein_Milgrom_1984} following \citet{Milgrom_1983}, who noted that deviations from Newtonian dynamics appear only when the Newtonian gravitational acceleration $g_{_N} \la a_{_0} \approx 3.8$~pc/Myr$^2$ \citep{Begeman_1991, Gentile_2011}. Galaxies very often fall in this regime. Their RCs can be matched rather well if the gravitational acceleration $g = \sqrt{a_{_0} g_{_N}}$, which leads to flat RCs because $g$ then declines only inversely with distance. Another immediate prediction is that all galaxies must be on the BTFR \citep{Milgrom_1983b}, i.e. that the RC of an isolated galaxy with baryonic mass $M_b$ must become asymptotically flat at the level
\begin{eqnarray}
    v_{_f} = \left( {G M_b a_{_0}} \right)^{1/4} \, \, ,
    \label{eq:asymptoticvelocity}
\end{eqnarray}
where $M_b$ is the total baryonic mass of the galaxy. MOND uses only the observed distribution of baryons to compute the gravitational potential. This procedure works quite well \citep{Kroupa_2018, Sanders_2019}. Indeed, MOND successfully predicted tight scaling relations that were subsequently observed \citep{Famaey_McGaugh_2012, Lelli_2017}.

Due to the acceleration-dependent gravity law in MOND, the central surface density $\Sigma_0$ is an important quantity. The vertical Newtonian gravity at the centre of a thin disc galaxy is $g_{_{N,z}} = 2 \mathrm{\pi} G \Sigma_0$. Low surface brightness galaxies (LSBs) have $\Sigma_0 \leq \Sigma_\dagger$, where $\Sigma_\dagger$ is the critical surface density in MOND.
\begin{eqnarray}
    \Sigma_\dagger ~\equiv~ \dfrac{a_{_0}}{2 \mathrm{\pi} G} ~=~ 137 \, M_\odot/\text{pc}^2 \, .
    \label{eq:crit_surface_density}
\end{eqnarray}
Galaxies with a lower central surface density show a non-Newtonian behaviour ($g > g_{_N}$). MOND has been successful in explaining the dynamics of galaxies in this regime \citep{McGaugh_2020}. This is also true for the case of AGC~114905, which was claimed to be problematic \citep{Mancera_2022} $-$ but it can be reconciled with MOND if the inclination has been overestimated, a rather plausible scenario \citep{Banik_fake_inclination_2022}.

Although MOND was originally phrased as a unique relation between $g$ and $g_{_N}$, it was obvious from the start that this can only hold in spherical symmetry. That is why \citet{Bekenstein_Milgrom_1984} formulated a generalized classical Lagrangian and derived a Milgromian version of the Poisson equation that supersedes the usual linear Poisson equation. This approach uses an aquadratic Lagrangian \citep[AQUAL;][]{Bekenstein_Milgrom_1984}. The other available approach for such a generalization of the classical Lagrangian and Poisson equation uses an auxiliary field sourced by the baryonic matter. This is called the quasi-linear formulation of MOND \citep[QUMOND;][]{QUMOND}.

Both formalisms have been implemented numerically: \citet{Tiret_2008} developed an \textit{N}-body solver for the AQUAL formulation to study the evolution of spiral galaxies using pure stellar discs. Gas dynamics was later included using a sticky particle scheme \citep{Tiret_2008_gas}. \textsc{raymond} \citep{Candlish_2015} is another \textit{N}-body and hydrodynamics solver that can solve both the AQUAL and QUMOND formulations of MOND. There are multiple \textit{N}-body solvers that have been used to investigate diverse scenarios in MOND \citep{Brada_1999, Brada_2000, Londrillo_2009, Angus_2011, Wu_2013}. Among these, a highly efficient publicly available algorithm has been developed to handle only the less computationally intensive QUMOND approach. This is the \textsc{phantom of ramses} (\textsc{por}) solver developed in Bonn \citep{Lughausen_2015, Nagesh_2021}, which we use in this study. It is a modification of the adaptive mesh refinement (AMR) code \textsc{ramses} \citep{Teyssier_2002} that is widely used to simulate astrophysical problems assuming Newtonian gravity. Importantly for our work, only the gravity solver is modified in \textsc{por} $-$ the non-gravitational baryonic physics is not modified. \textsc{por} has been applied to model interacting systems like the Antennae \citep{Renaud_2016}, the Sagittarius tidal stream around the Milky Way \citep[MW;][]{Thomas_2017}, tidal tails of open star clusters in the Solar neighbourhood \citep{Kroupa_2022}, the shell galaxy NGC~474 \citep{Bilek_2022b}, and the Local Group satellite planes, which in MOND condensed out of tidal debris expelled by a past MW-M31 encounter \citep{Bilek_2018, Banik_2022_satellite_plane}. \textsc{por} has also been used to simulate the formation of galaxies, with the result that rotating gas clouds naturally collapse into exponential disc galaxies \citep{Wittenburg_2020}. Non-rotating clouds on the other hand form elliptical galaxies on the observed short time scales \citep{Eappen_2022}.

For in-depth reviews of MOND, we refer the reader to \citet{Famaey_McGaugh_2012} and \citet{Banik_Zhao_2022}, while \citet{Merritt_2020} discusses the philosophical aspects of the missing gravity problem.

\subsection{Star formation}

Stars form in molecular clouds (MCs) containing mostly molecular hydrogen ($H_2$). The overall star formation rate (SFR) of a galaxy is mainly dependent on its ability to form MCs, which have a wide range of mass, size, and density. MCs are sufficiently dense regions of the interstellar medium (ISM) of a galaxy to contain sub-regions that may collapse under their own gravity, which is possible if the cloud mass $M_\textrm{cloud}$ exceeds the thermal Jeans mass $M_J$ \citep{Jeans_1902}. Once $M_\textrm{cloud} > M_J$, the gas cloud collapses and condenses to form stars. Processes like magnetic fields and turbulence regulate the star forming efficiency by acting against this collapse. But the most important mechanism that regulates collapse is self-regulation \citep{Yan_2023}: when stars form in MCs, protostellar winds and massive OB stars destroy the clouds via ionization, heating by ultraviolet photons, stellar winds, and supernova blast waves. Although these processes can quench star formation locally, the shock waves can compress gas in another cloud, inducing star formation globally. These processes play an important role that regulates galaxy formation and evolution. The interplay of only a few observable parameters like luminosity, stellar mass, and gas surface density $\Sigma_g$ leads to the emergence of some tight empirical relations such as the MS \citep[which relates the SFR to the stellar mass $M_{\star}$;][]{Speagle_2014} and the KS relation \citep[which relates the gas surface density $\Sigma_g$ to the SFR surface density $\Sigma_{\textrm{SFR}}$;][]{Kennicutt_1998}. The semi-analytic study of \citet{Zoonozi_2021} provides a theoretical comparison between the KS law and the galaxy MS in Newtonian and Milgromian gravity based on the assumption that clouds convert a fixed proportion of their mass to stars per free-fall time.

\subsection{Aim}

The main aim of this article is to investigate how well the existing numerical implementation of star formation and feedback available in \textsc{por} \citep[as incorporated in the 2015 version of \textsc{ramses};][]{Teyssier_2002} allows the reproduction of real galaxies by numerical models. The intent is to compute model galaxies using the available sub-grid physics and then test how well they resemble real galaxies with different levels of sophistication of the sub-grid physics. Additionally, a study of the bars formed in these simulations will also be performed to check if MOND might alleviate significant tensions with the Lambda-Cold Dark Matter \citep[$\Lambda$CDM;][]{Efstathiou_1990, Ostriker_Steinhardt_1995} standard model of cosmology \citep{Roshan2021_2}.

The article is structured as follows: Section~\ref{sec:Numerical-methods} gives a brief introduction to the numerical methods. The setup of the models is described in Section~\ref{sec:Models_setup}. Their results are presented and discussed in Section \ref{sec:Results}. We conclude in Section~\ref{sec:Discussion}.

\section{Numerical Methods}
\label{sec:Numerical-methods}

\textsc{por} is a patch to the publicly available code \textsc{ramses} \citep{Teyssier_2002}, a grid-based code that uses the AMR technique. \textsc{ramses} also has a hydrodynamical solver, enabling simulations with gas and star formation. \textsc{por} is a patch to the 2015 version of \textsc{ramses} that was added by \citet{Lughausen_2015} in order to solve a numerical implementation of QUMOND \citep{QUMOND}, whose field equation for the potential $\Phi$ is
\begin{eqnarray}
    \nabla^2 \Phi ~\equiv~ - \nabla \cdot \bm{g}  ~=~ - \nabla \cdot \left( \nu \bm{g}_{_N} \right) \, ,
    \label{eq:poisson}
\end{eqnarray}
where $\bm{g}$ is the true gravity, $\bm{g}_{_N}$ is the Newtonian gravity, and $\nu$ is an interpolation function between the Newtonian and MOND regimes such that $\bm{g} = \nu \bm{g}_{_N}$ in spherical symmetry. We use the simple interpolating function \citep{Famaey_Binney_2005} as this seems to work well with recent observations \citep{Gentile_2011,Iocco_Bertone_2015, Banik_2018_Centauri, Chae_2018}.
\begin{eqnarray}
    \nu ~=~ \frac{1}{2} + \sqrt{\frac{1}{4} + \frac{a_{_0}}{g_{_N}}} \, ,
    \label{eq:Interpolationfunc}
\end{eqnarray}
with $\bm{g}_{_N}$ being calculated from the baryon density $\rho_\textrm{b}$ using the standard Poisson equation
\begin{eqnarray}
    \nabla \cdot \bm{g}_{_N} ~=~ -4 \mathrm{\pi} G \rho_b \, .
\end{eqnarray}
Throughout this work, we use the notation that $v \equiv \lvert \bm{v} \rvert$ for any vector $\bm{v}$, with $N$ subscripts denoting Newtonian quantities.

The \textsc{por} package is publicly available \footnote{\label{PoRbit}\url{https://bitbucket.org/SrikanthTN/bonnPoR/src/master/}}, along with a published user manual to set up isolated and interacting disc galaxy simulations and a brief review of all hitherto performed research with \textsc{por} \citep{Nagesh_2021}.

\subsection{Star formation recipe} 
\label{star_formation_ramses}

Activating the \textsc{por} patch changes the Poisson solver from Newtonian to MONDian, but the hydrodynamical solver used remains the default one that is available in the 2015 version of \textsc{ramses}. It uses a second-order Godunov scheme with a Riemann solver for the conservative Euler equations \citep{Teyssier_2002, Teyssier_2006}. As explained in section~3.1 of \citet{Dubois_2008} and section~2.2.1 of \citet{Wittenburg_2020}, gas in each cell is converted to stellar particles such that the star formation rate density
\begin{eqnarray}
    \rho_{\mathrm{SFR}} ~=~ \frac{\rho}{t_{\star}} ~ \textrm{if}~ \rho > \rho_0 \, ,
    \label{eq:Schmidt-law}
\end{eqnarray}
where $\rho$ is the mean gas volume density in the cell, $\rho_0$ is the density threshold for star formation, and $t_{\star}$ is the star formation timescale, which is proportional to the local free-fall time $t_{\mathrm{ff}}$.
\begin{eqnarray}
    t_{\star} ~=~ t_{\mathrm{ff}} \left(\dfrac{\rho}{\rho_0}\right)^{-1/2} \, , \quad t_{\mathrm{ff}} ~=~ \sqrt{\frac{3 \mathrm{\pi}}{32 G \rho}} \, .
    \label{eq:FFTime} 
\end{eqnarray}
The justification to use the Newtonian free-fall time here is that, in the relevant regime, it is essentially identical to the MONDian one \citep[see fig.~8 of][]{Zoonozi_2021}. Hence the computation of this quantity was not modified in the \textsc{por} version of the \textsc{ramses} code. We refer the reader to their study for a detailed analytic discussion of free-fall times in MOND and what this implies for the SFR.

In \textsc{ramses}, a cell is ready to form stars if $\rho > \rho_0$. The number of formed stellar particles is $N$, which is drawn from a Poisson distribution with mean $\overline{N}$. This corresponds to eq.~12 of \citet{Dubois_2008}, but we reproduce this here with slightly different notation as their version contains a typo (missing power of $N$). The Poisson probability distribution function
\begin{eqnarray}
\label{eq:PPDF}
    P \left( N \right) ~=~ \dfrac{\overline{N}^N}{N!} \exp \left( -\overline{N} \right) \, , \quad \overline{N} ~=~ \frac{\rho_{\textsc{sfr}} d^3\bm{x} \, dt}{m_{\star}} \, .
\end{eqnarray}
The mean value $\overline{N}$ depends on the local star formation rate density $\rho_{\textsc{sfr}}$ (Eq.~\ref{eq:Schmidt-law}), the timestep $dt$, the volume $d^3\bm{x}$ of the gas cell, and the mass $m_{\star} = \rho_0 d^3\bm{x}$ of the newly formed stellar particles, using an instantaneous recycling scheme to account for supernovae (SNe) and stellar winds (Sec. \ref{Feedback}). At each timestep, \textsc{ramses} checks if $\rho > \rho_0$ and if so, it uses the Poisson distribution (Eq.~\ref{eq:PPDF}) to determine how many stellar particles should be formed. As a result, the possibility remains that $N = 0$ randomly even if $\rho > \rho_0$ and thus $\overline{N} > 0$.

Since the mass of a newly formed particle depends on the volume of the corresponding gas cell and this decreases as the refinement level increases, star formation in the simulation depends on the resolution. If the cell size is large, it takes a significant amount of time for the density to exceed the threshold $\rho_0$, which causes star formation to be modelled by a few rarely formed but individually massive star particles. In the case of higher resolution, the cells are smaller and the star particles that form in these cells have a lower mass. A higher resolution simulation is also better able to resolve density contrasts, making it more likely to have cells where $\rho > \rho_0$. This causes the density of gas to exceed $\rho_0$ in less time, thus forming more particles and increasing the star formation rate. Therefore, the mass of the formed stellar particles and how many there are depend on the spatial resolution of the code. The amount of gas consumed from a gas cell to make a stellar particle is also dependent on the star forming efficiency. In addition, \textsc{ramses} has a safety catch such that at most 90\% of the gas in a cell is consumed by star formation \citep{Dubois_2008, Wittenburg_2020}.

\subsection{Feedback prescription}
\label{Feedback}

Feedback from SNe is one of the key phenomena that impacts star formation in a galaxy. The thermal and kinetic energy from a supernova affects the ISM through small scale effects like turbulence, thermal instability, and metal enrichment. These processes affect subsequent star formation. \textsc{ramses} handles the supernova mass removal by assuming that each time a stellar particle of mass $m_{\star}$ is created, the mass removed from the gas cell is $m_{\star} \left( 1 + \eta_{\textrm{sn}} \right)$, with $\eta_{\textrm{sn}}$ accounting for the extra mass of the stellar particle that goes into SNe. This mass of value $m_{\star} \left( 1 + \eta_{\textrm{sn}} \right)$  is removed from the gas cell immediately and $m_{\star} \eta_{\textrm{sn}}$ is released back into the ISM some time $t_\textrm{sne}$ after the formation of the particle \citep{Dubois_2008}. In our models, $\eta_{\textrm{sn}} = 0.1$ and $t_\textrm{sne} = 10$~Myr, which are the default values. This corresponds to an invariant initial mass function (IMF) which represents the long-lived stellar particle of mass $m_{\star}$, with a fraction $\eta_{\textrm{sn}}$ being lost in mass from the birth stellar population through SNe that detonate over a characteristic timescale $t_\textrm{sne}$. \citet{Wittenburg_2020} provides a brief comparison of different star formation prescriptions available in \textsc{por} and their effects on disc galaxy formation from monolithic collapse of a rotating gas cloud evolved in MOND \citep[an application to the formation of ellipticals can be found in][]{Eappen_2022}. \textsc{ramses} offers three different types of feedback prescriptions which can be classified as simple, intermediate, and complex. These are briefly described below.

\subsubsection{Simple feedback}
\label{simple_feedback}

In this case, all the energy from SNe is deposited as thermal energy into the ISM. When the gas density $\rho > \rho_\textrm{0}$, a certain fraction of the gas is converted into stellar particles (Section~\ref{star_formation_ramses}), while the remaining gas is handled by a polytropic equation of state which forces the gas temperature $T$ to satisfy
\begin{eqnarray}
    T ~\geq ~ T_\textrm{0} \left( \frac{\rho}{\rho_\textrm{0}} \right)^{\gamma_0 - 1} \, ,
    \label{EOS}
\end{eqnarray}
with equality arising if $\rho < \rho_0$. Here, $T_\textrm{0}$ is the average ISM temperature and $\gamma_\textrm{0}$ is the polytropic index, which we take to be $5/3$ as appropriate for a monoatomic gas \citep[see section 3.2 of][]{Dubois_2008}. Since the characteristic timescale for radiative losses in star-forming regions is shorter than the numerical timestep, the thermal energy just radiates away before having a significant effect on the local ISM \citep{Dubois_2008}.  

\subsubsection{Intermediate feedback}
\label{inter_feedback}

The intermediate (complexity) feedback prescription allows one to specify the fraction of SNe energy to be released into the ISM as kinetic energy, with a radial energy injection scale $r_\textrm{sn}$. This injection scale is called the supernova bubble radius. For every stellar particle created, there is a corresponding blast wave that carries supernova ejecta and gas out to the blast radius, with a velocity computed using the local Sedov spherical blast wave solution. SNe then release energy, mass, momentum, and metallicity into the respective cells. The remaining thermal energy is accounted for in the polytropic equation of state (Eq.~\ref{EOS}). By default, the 2015 version of \textsc{ramses} uses a 100\% kinetic energy feedback fraction and calculates the maximum radius of supernova ejecta using the length of the cell under consideration. Pure thermal feedback can be activated by setting the kinetic feedback fraction to zero.

\subsubsection{Complex feedback}
\label{complex_feedback}

The complex feedback prescription involving radiative transfer is \textsc{ramses-rt} \citep{Rosdahl_2013}, which is available in \textsc{ramses}. In this case, \textsc{ramses} computes radiative cooling and heating processes separately, without the need to change the complete hydrodynamical solver. This is because the thermochemistry mainly depends on the gas density, temperature, metallicity, and ionization state $-$ but by default, collisional ionization equilibrium (CIE) is assumed, enabling ionization states to be calculated using temperature and density alone. Thus, the code does not track the ionization states.

In the case of our models, the cooling function is computed using look-up tables from the \citet{Sutherland_1993} cooling model in the temperature and metallicity plane, while tables from \citet{Courty_2004} are used for different cooling/heating processes. The heating term mainly includes photo-ionization, while the cooling term includes recombination, collisional excitation and ionization, Compton scattering, and Brehmsstrahlung. All these processes are not computed in detail as the code assumes CIE and calculates cooling, heating, and ionization rates using temperature and density alone. The temperature and energy density are updated at the end along with the Euler equations \citep{Wittenburg_2020}. 

As mentioned earlier in Section~\ref{complex_feedback}, \textsc{ramses-rt} (unlike \textsc{ramses}) uses a first-order Godunov scheme to solve the Euler equations. \textsc{ramses-rt} handles radiative transfer differently to \textsc{ramses}. The code keeps track of ionization states that are computed carefully keeping track of photons, collisions, and most importantly a non-equilibrium thermochemistry \citep{Rosdahl_2013, Wittenburg_2020}. 

We do not run models with the complex feedback prescription as it is computationally expensive. Not running this prescription also helps to emphasize that Milgromian disc galaxy models do not need complex feedback prescriptions to behave like real galaxies.

\section{Models and setup}
\label{sec:Models_setup}

Our aim here is not to perform galaxy formation simulations but to set up already formed rotating disc galaxies with realistic mass distributions and gas fractions. The aim is to check if the models develop star formation activity comparable to that of observed galaxies.\footnote{\label{Movies}{Movies showing the evolution of the models are available here:} \url{https://www.youtube.com/playlist?list=PL2mtDSIH4RQhLvF2cxuOI72XLQFqgCsb0}} If this occurs, then the existing sub-grid algorithms described above would be adequate in the context of MOND.

\subsection{Initial conditions}

We present 5 models of disc galaxies with $M_{\star}/M_\odot$ in the range $10^7 - 10^{11}$ and gas fractions as dictated by observations \citep{SPARC}. The models have two main components: an inner stellar disc with a radial scale length $R_d$ and an outer gas disc with radial scale length $R_{\rm g}$, which we set to $2.5 \, R_d$ for reasons discussed below.

\subsubsection{Calculation of the stellar and gas disc scale lengths}
\label{disc_scale_length}




The independent parameter of our models is the stellar mass $M_{\star}$. This is used to calculate the luminosity $L_{3.6}$ at $3.6 \, \mu$m with an assumed mass-to-light ratio of 0.5 Solar units at this mid-infrared wavelength \citep{SPARC, Schombert_2022}. $L_{3.6}$ is then used to calculate the gas mass $M_g = 1.33 \, M_{HI}$ (to account for primordial helium) using eq.~4 of \citet{SPARC}:
\begin{eqnarray}
    \log_{10} M_{HI} ~=~ 0.54 \log_{10} L_{3.6} + 3.90 \, ,
    \label{Sparc4}
\end{eqnarray}
where masses and luminosities are in Solar units and disc scale lengths are in kpc, a convention used throughout this work. The so-obtained $M_{HI}$ is used to calculate the radius $R_1$ at which the gas surface density is $1 \, M_\odot$/pc\textsuperscript{2} by inverting eq.~3 of \citet{SPARC}, which gives
\begin{eqnarray}
    \log_{10} R_1 ~=~ \frac{\log_{10} \left( M_{HI} \right) -  7.20}{1.87} \, .
    \label{Sparc3}
\end{eqnarray}
Note that the sign in front of the 7.20 was incorrect in \citet{SPARC}, a mistake which has been rectified above. The stellar $R_d$ is also found from the gas mass by combining eqs.~3 and 6 of \citet{SPARC}, which yields
\begin{eqnarray}
    \log_{10} R_d ~=~ 0.62 \log_{10} M_{HI} - 5.40 \, .
    \label{stellar_rd}
\end{eqnarray}
We now have the stellar $R_d$, but we still do not know the exponential scale length $R_g$ of the gas component. We therefore guess this and substitute the guess into the gas surface density profile
\begin{eqnarray}
    \Sigma_g ~=~ \frac{M_g}{2 \mathrm{\pi} {R_g}^2} \exp \left( -R/R_g \right) \, .
    \label{gas_surface_density}
\end{eqnarray}
Since we already know that $\Sigma_g$ should be $1 \, M_\odot$/pc\textsuperscript{2} at the radius $R=R_1$, we can confirm whether our guess for $R_g$ is correct. We use a Newton-Raphson procedure to vary $R_g$ in order to ensure the surface density reaches the desired level at the radius $R_1$. In this way, we found that a very good approximation is
\begin{eqnarray}
    R_g ~=~ 2.5 \, R_d \, .
\end{eqnarray}
The initial conditions obtained in this manner are listed in Table~\ref{tab:SPARC}, which shows the fraction of the total mass in stars ($f_s$) and in gas ($f_g$). We use these to define the effective radius $\widetilde{R}_{\textrm{eff}}$ of a simulated galaxy by taking a weighted mean of its initial stellar and gas disc scale lengths.
\begin{eqnarray}
    \widetilde{R}_{\textrm{eff}} ~\equiv~ R_d f_s + R_g f_g \, .
    \label{Effective_radius}
\end{eqnarray}

\begin{table*}
    \begin{tabular}{ccccccccc}
    \hline
     Model name & $M_\textrm{tot}$ ($M_\odot$) & $M_\star$ ($M_\odot$) & $f_g$ & $R_d$ (kpc) & $R_1$ (kpc) & $R_\textrm{g}$ (kpc) & $\widetilde{R}_{eff} $ &$\Sigma_0/\Sigma_\dagger$\\ \hline
    1e7 & $1.02 \times 10^{8}$  & $1.0 \times 10^7$ & 90.2\% & 0.28 & 2.20 &  0.72 & 0.68 & 0.34\\
    1e8 & $3.44 \times 10^8$ & $1.0 \times 10^8$ & 71.0\% & 0.62 & 4.28 & 1.56 & 1.29 & 0.50\\
    1e9 & $2.11 \times 10^9$ & $1.0 \times 10^9$ & 52.6\% & 1.35 & 8.33  & 3.38 & 2.42 & 1.28\\
    1e10 & $1.38 \times 10^{10}$ & $1.0 \times 10^{10}$ & 27.8\%  & 2.92 & 16.21 & 7.36 & 4.13 & 3.74 \\
    1e11 & $1.13 \times 10^{11}$ & $1.0 \times 10^{11}$ & 11.8\% & 6.33 & 31.52 & 15.82 & 7.45  & 9.21 \\ \hline
    \end{tabular}
    \caption{Parameters calculated using the SPARC scaling relations and the Newton-Raphson method (Section~\ref{disc_scale_length}) for the galaxy models, named according to their initial stellar mass in Solar units. $M_\textrm{tot} = M_\star + M_g$ is the total baryonic mass of the model with stellar mass $M_\star$ and gas fraction $f_\textrm{g}$. The stellar disc exponential scale length is $R_\textrm{d}$ (Eq.~\ref{stellar_rd}), $R_1$ is the radius at which $\Sigma_g = 1 \, M_\odot$/pc$^{2}$, while $R_\textrm{g}$ is the exponential gas disc scale length. The last column is the ratio between the initial central surface density and the MOND critical surface density (Eq.~\ref{eq:crit_surface_density}).}
    \label{tab:SPARC}
\end{table*}

\subsubsection{Setting up a disc in Milgromian gravity}

We set up the galaxy simulations using a version of \textsc{disk initial conditions environment} \citep[\textsc{dice};][]{Perret_2014} adapted to MOND gravity \citep[as discussed in more detail in][]{Banik_2020_M33}. The modified version is publicly available.\textsuperscript{\ref{PoRbit}} We adapted the template for the MW as that is already structurally most similar to the models we wish to consider. In particular, our models have an inner stellar disc and an outer gas disc, as explained above. To ensure a stable disc, it must not be completely dynamically cold. The minimum velocity dispersion is set by the Toomre condition \citep{Toomre_1964}, whose generalization to MOND was given in eq.~7 of \citet{Banik_2020_M33} based on earlier analytic results \citep{Banik_2018_Toomre}. In \textsc{dice}, we set a floor of 1.25 on the MOND Toomre parameter. The implementation of the aforementioned modifications has been discussed extensively in section~2 of \citet{Banik_2020_M33}.

In the \textsc{dice} hydrodynamical template for the MW, it is possible to specify the gas fraction in the disc and its temperature. The template is structured such that the required parameters for the inner and outer components can be set independently. We obtain the required parameters from Table~\ref{tab:SPARC} and adopt a uniform initial temperature of $T = 25000$~K (25~kK) for all the models. This is not the absolute temperature $-$ it is actually a measure of the 1D gas velocity dispersion $\sigma_g$, which \textsc{dice} calculates \citep{Banik_2020_M33} using
\begin{eqnarray}
    \sigma_g ~=~ \sqrt{\frac{kT}{\mu m_p}} \, ,
    \label{eq:velocitydispersion}
\end{eqnarray}
where $k$ is the Boltzmann constant, $T$ is the temperature of the gas, $\mu = 7/4$ is the mean molecular weight of the gas, and $m_p$ is the mass of a proton. Thus, our models all have $\sigma_g = 10.9$~km/s initially.

\subsection{Simulation setup}

The \textsc{dice} outputs were provided as inputs for \textsc{por}, which adds the gas using the \texttt{condinit} routine \citep*{Teyssier_2010} based on the parameters specified in the namelist file. Initially, all models have $2.0 \times 10^6$ particles. All the models were set up using a radial double exponential profile (one for the stars and one for the gas) in which both stellar and gas discs are modelled as exponential. The thickness of the disc is modelled using a $\sech^2$ profile, which is detailed in section~2.3.1 of \citet{Banik_2020_M33} for the gas component where the thickness changes with radius. The initial gas temperature \textit{T2\_ISM} = $T_\textrm{gas} = 25$~kK, though the gas can subsequently cool down to a minimum temperature of \textit{T2\_star} = 10~kK as the calculation proceeds. We adopt a star formation efficiency of 2\% \citep{Dubois_2008}. The number density threshold $n_0 = 0.1/$cm\textsuperscript{3} translates to a density threshold of $\rho_0 \approx 0.1 \, H/$cm\textsuperscript{3}, which is adapted from \citet{Dubois_2008}. Every time the gas density or particle number density exceeds this threshold, the grid is refined, i.e., the cells in this grid split into 2\textsuperscript{3} child cells in 3D.

All of the presented models use the intermediate feedback prescription in which supernovae provide feedback with a kinetic energy fraction of 50\% at the supernova bubble radius $r_{\rm{sn}} = 150$~pc \citep{Dubois_2008}. The smallest allowed cell is smaller than the diameter of the supernova bubble, while the Jeans length is resolved by at least four cells. Without resolving the Jeans length, the gas in a collapsing region might fragment artificially \citep{Truelove_1998}. We run our highest mass model with one extra level of refinement to demonstrate numerical convergence (Appendix~\ref{Effect_of_resolution}).

\begin{figure}
    \includegraphics[width=8.5cm]{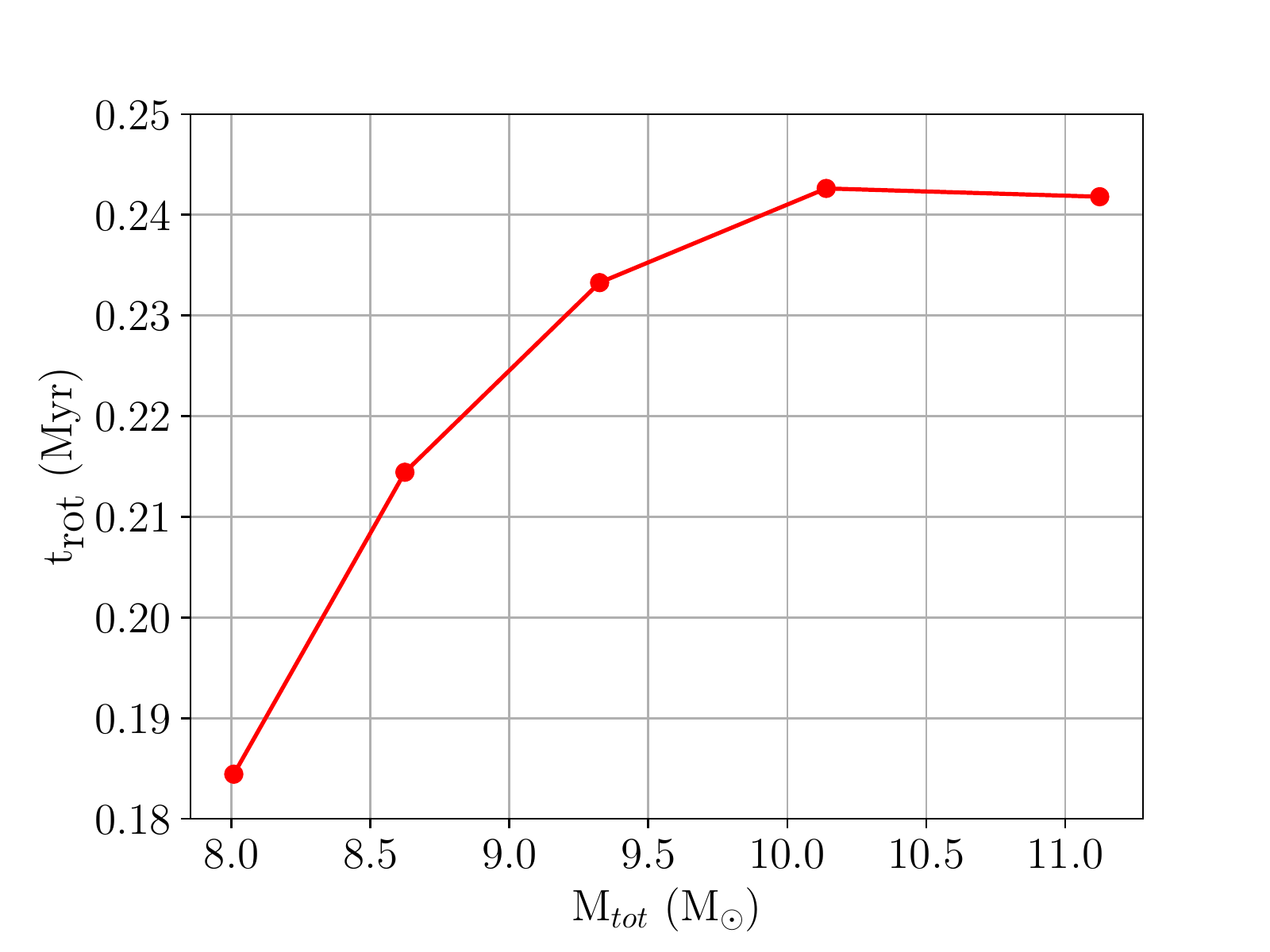}
    \caption{The rotational period $t_\textrm{rot}$ (Eq.~\ref{eq:trot}) at the effective radius $\widetilde{R}_{\rm{eff}}$} (Eq.~\ref{Effective_radius}), shown as a function of the initial total baryonic mass $M_\textrm{tot}$ of each model.
    \label{fig:trotvmass}
\end{figure}

The galaxy models are advanced for 5~Gyr with an output frequency interval of 100 Myr.\footnote{The outputs are sometimes not exactly 100~Myr apart}, but the temporal deviations are small. The rotational period at any radius $r$ is calculated using
\begin{eqnarray}
    t_{\textrm{rot}} \left( r \right) ~=~ \frac{2\mathrm{\pi}r}{v_c \left( r \right)} \, ,
    \label{eq:trot}
\end{eqnarray}
where $v_c \left( r \right)$ is the circular velocity as returned by \textsc{dice}. The rotational period of each model at its effective radius differs little between models (Fig.~\ref{fig:trotvmass}). Assuming that $t_{\textrm{rot}} \approx 225$~Myr, the model galaxies complete $\approx 22$ revolutions during the 5~Gyr simulations.

The minimum number of refinement levels is $levelmin = 7$, while the maximum is $levelmax = 10$ for models 1e7 and 1e8 and $levelmax = 11$ for models 1e9, 1e10, and 1e11. The cell size in the best-resolved regions is determined by $levelmax$ (Table~\ref{tab:Resolution}), while the cell size in the least resolved regions is set by $levelmin$. Since our main interest is the sub-grid physics of star formation in \textsc{por}, all parameters are kept constant between our models except the total mass, stellar and gas disc scale length, box length, and gas fraction. The larger $levelmax$ in the higher mass models is required because their larger size means a higher resolution is needed to reach the same highest spatial resolution in pc.

\begin{table}
    \begin{tabular}{ccccc}
    \hline
    Model & Box size & Highest spatial \\
    name & (kpc) & resolution (pc) \\ \hline
    1e7 & 100 & 97.65 \\ 
    1e8 & 160 & 156.25 \\ 
    1e9 & 300 & 146.64\\ 
    1e10 & 350 & 170.89 \\ 
    1e11 & 400 & 195.31 \\ \hline
    \end{tabular}
    \centering
    \caption{The box size and highest spatial resolution of each model. Since results are converged for model 1e11, we assume that we have sufficient resolution for the other models (Appendix~\ref{Effect_of_resolution}).}
    \label{tab:Resolution}
\end{table}

\subsection{Data extraction and barycentre adjustment}

The mass, position, velocity, acceleration, and birth time of each particle is extracted using the \textsc{extract\_por} software. The gas data is extracted using \textsc{rdramses}, which treats gas cells as particles at their cell centres and prints out the results. \textsc{extract\_por} and \textsc{rdramses} are publicly available.\textsuperscript{\ref{PoRbit}}

After extracting the particle and gas data for all the models, the barycentre position $\bm{R}$ and velocity $\bm{V}$ are calculated at every snapshot.
\begin{eqnarray}
    \bm{R} &=& \frac{1}{M_p + M_g} \left( \sum_{i=1}^{N_p} m_{p, i}  \bm{r}_{p, i} + \sum_{i=1}^{N_g}  m_{g, i} \bm{r}_{g, i} \right) \, , \\
    \bm{V} &=& \dfrac{1}{M_p + M_g} \left( \sum_{i=1}^{N_p} m_{p, i} \bm{v}_{p, i} + \sum_{i=1}^{N_g}  m_{g, i} \bm{v}_{g, i} \right) \, ,
    \label{eq:Barycenter-v}
\end{eqnarray}
where $M_p$ ($M_g$) is the total mass of stellar (gas) particles, $N_p$ ($N_g)$ is the number of stellar (gas) particles, $\bm{r}_{p, i}$ ($\bm{r}_{g, i}$) is the position of a stellar (gas) particle labelled by the index $i$, and $\bm{v}_{p, i}$ ($\bm{v}_{g, i}$) is the velocity of a stellar (gas) particle. In the analyses presented next, the barycentre position and velocity are subtracted for the snapshot under consideration. This corrects for barycentre drift due to numerical effects, though we note that the drift is small for the isolated models considered here \citep[see footnote 14 to][]{Banik_2020_M33}.

\section{Results}
\label{sec:Results}

\subsection{The main sequence of galaxies}

To analyse the SFR, one of the most important variables is the birth time of each particle. All particles present initially have a negative timestamp (birth time) set by \textsc{extract\_por} for safety reasons. Particles formed during the simulation have their birth time written out in Myr. The SFR and star formation history (SFH) are calculated by looping over all particles and binning them in time according to their timestamp whilst skipping the initial set of particles. The masses of all particles formed within any temporal bin are summed up and divided by its duration to get the SFR, which is then plotted at the centre of the corresponding interval. The SFH obtained in this way for model 1e11 is shown in Fig.~\ref{fig:SFH_1e11}, while the SFHs of the other models are shown in Appendix~\ref{SFH_Four_models}. We only consider data after $\approx 4$ revolutions, which corresponds to $\approx 900$~Myr. This allows the models to reach dynamical equilibrium, reducing numerical effects. For analyses related to star formation, we consider the data up to 4.5 Gyr as the SFR reaches zero for some models after this.

\begin{figure}
    \includegraphics[width=8.5cm]{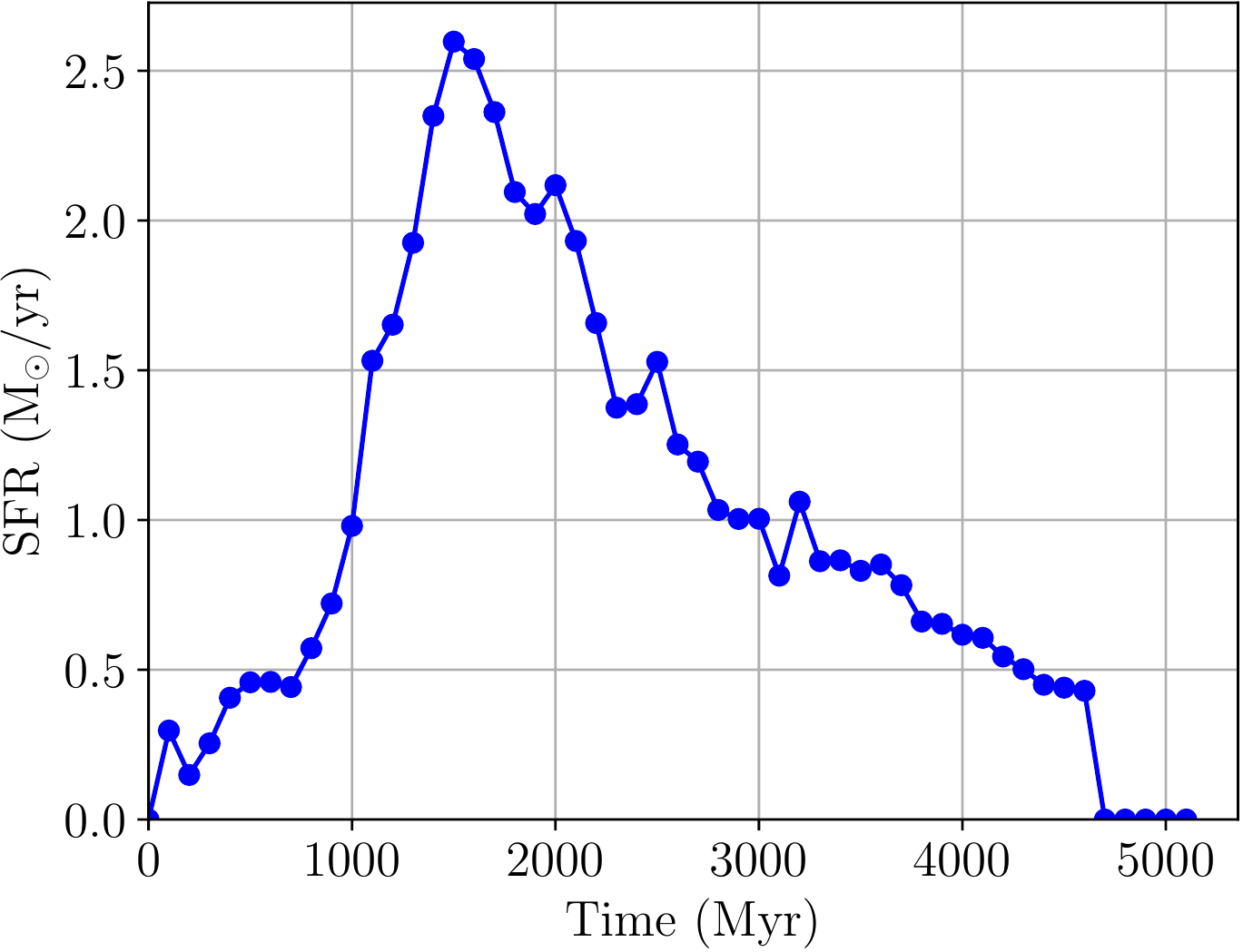}
    \caption{Star formation rate as a function of time in model 1e11.}
    \label{fig:SFH_1e11}
\end{figure}

\begin{figure}
    \includegraphics[width=8.5cm]{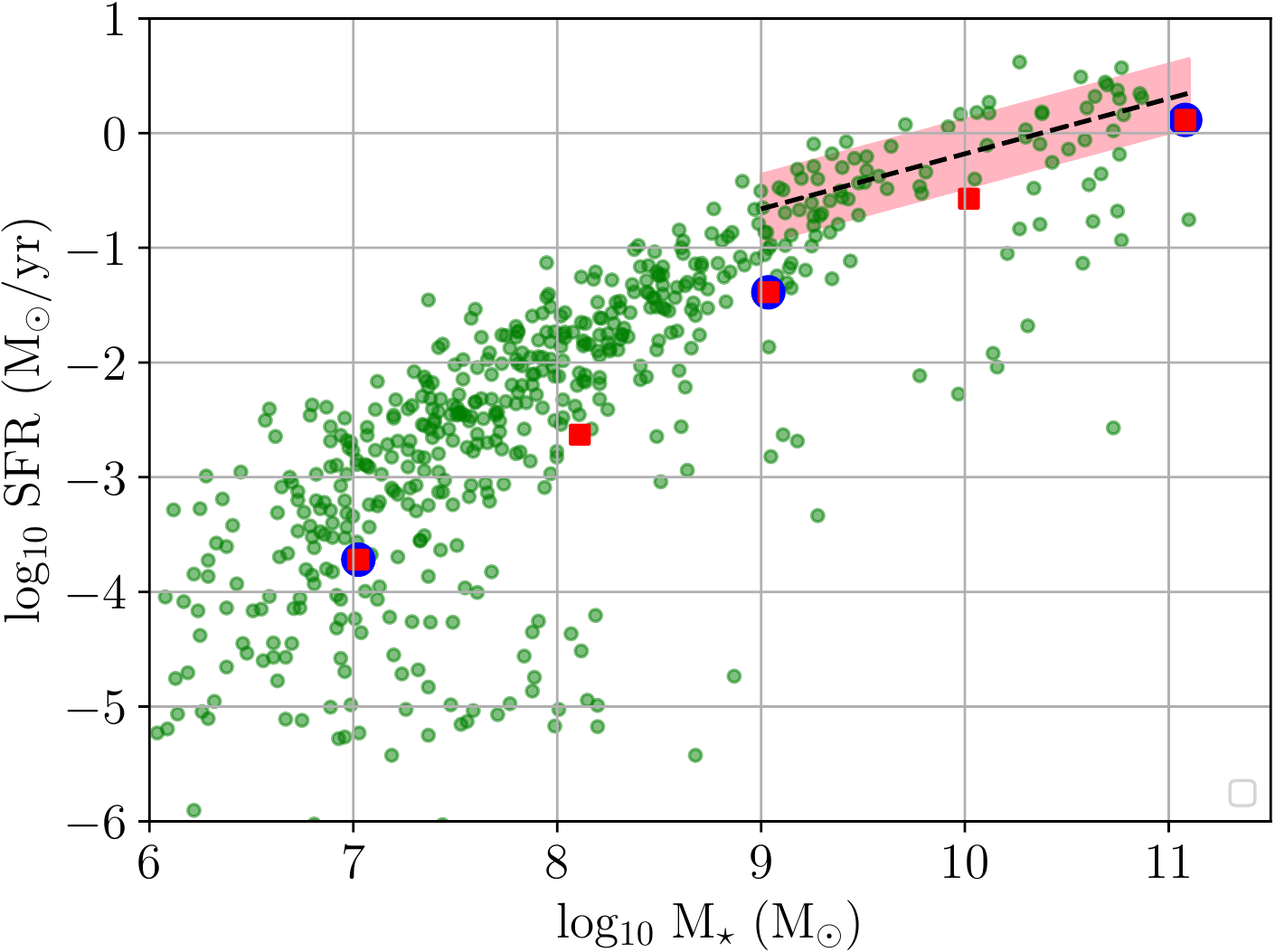}
    \caption{The galaxy MS, relating $M_{\star}$ and the global SFR. The green circles are observed galaxies in the Local Cosmological Volume \citep{Karachentsev_2013}, while the red (blue) points show our models with intermediate (simple) feedback. The SFR is the average over the period $1 - 4.5$~Gyr. The dashed line (Eq.~\ref{eq:MS-fit}) is the MS from \citet{Speagle_2014}, with the shaded magenta band showing a scatter of $\pm 0.3$~dex.}
    \label{fig:Main-sequence}
\end{figure}

Fig.~\ref{fig:Main-sequence} shows the average SFR over the period $1 - 4.5$~Gyr as a function of the the average $M_{\star}$.  We also show observed values for galaxies within 11~Mpc (the Local Cosmological Volume) based on \citet{Karachentsev_2013}. The best fit to the MS in a larger sample is given by eq.~28 of \citet{Speagle_2014} which, neglecting uncertainties, is
\begin{eqnarray}
    \log_{10} \rm{SFR} &=& \left( 0.84 - 0.026 \, t \right) \log_{10} \left( \frac{M_\star}{M_\odot} \right) \nonumber \\
    &-& \left( 6.51 - 0.11 \, t \right) \, ,
    \label{eq:MS-fit}
\end{eqnarray}
where the SFR is in $M_\odot$/yr and $t$ is the age of the universe in Gyr. We show this as the dashed black line in Fig.~\ref{fig:Main-sequence}, with the shaded magenta region showing a scatter of $\pm 0.3$~dex based on the uncertainties quoted in the above coefficients in \citet{Speagle_2014}.

All 5 models are run with the intermediate feedback prescription, while three models (1e7, 1e9, and 1e11) are also run with simple feedback in which there is no kinetic feedback from SNe (Section~\ref{simple_feedback}). The results of these eight simulations are shown in Fig.~\ref{fig:Main-sequence}, where the models with intermediate (simple) feedback are plotted as red squares (blue circles). It is immediately apparent that the type of feedback has little effect on the SFR, suggesting that the overall evolution of the galaxy models is not very sensitive to the sub-grid physics.

It has been shown that galaxies in the Local Cosmological Volume have a nearly constant SFH \citep{Kroupa_2020a}. Our models are isolated and so do not accrete gas from their surroundings. This prevents our model galaxies from remaining on the MS throughout their evolution. To check whether galaxies would be on the MS in MOND after a Hubble time, it is necessary to conduct a cosmological MOND simulation that includes gas hydrodynamics, work which is currently in progress (N. Wittenburg et al., in preparation).

\subsection{Gas depletion timescale}
\label{sec:Gas_depletion_times}

The gas depletion timescale $\tau_g$ measures the time taken by a galaxy to exhaust its gas content $M_g$ given the current SFR \citep{Pflamm-Altenburg_2009}. We employ two methods to determine $\tau_g$. In the first method, we find
\begin{eqnarray}
    \tau_{g, 1} ~=~ \frac{M_g}{\dot{M}_\star} \, ,
    \label{eq:Gas-depletion}
\end{eqnarray}
where $M_g$ is the neutral gas mass at the desired time and $\dot{M}_\star$ is the SFR then. Since the gas supply gets exhausted in a finite time, we calculate the quantities entering $\tau_{g,1}$ using the average of the snapshots in the period $1 - 4.5$~Gyr. The idea is to estimate what the SFR would be if a galaxy had the observed mass distribution and gas fraction, not to check whether that is feasible in MOND in the first place $-$ addressing the latter would require a cosmological simulation.

\begin{figure}
    \includegraphics[width=8.5cm]{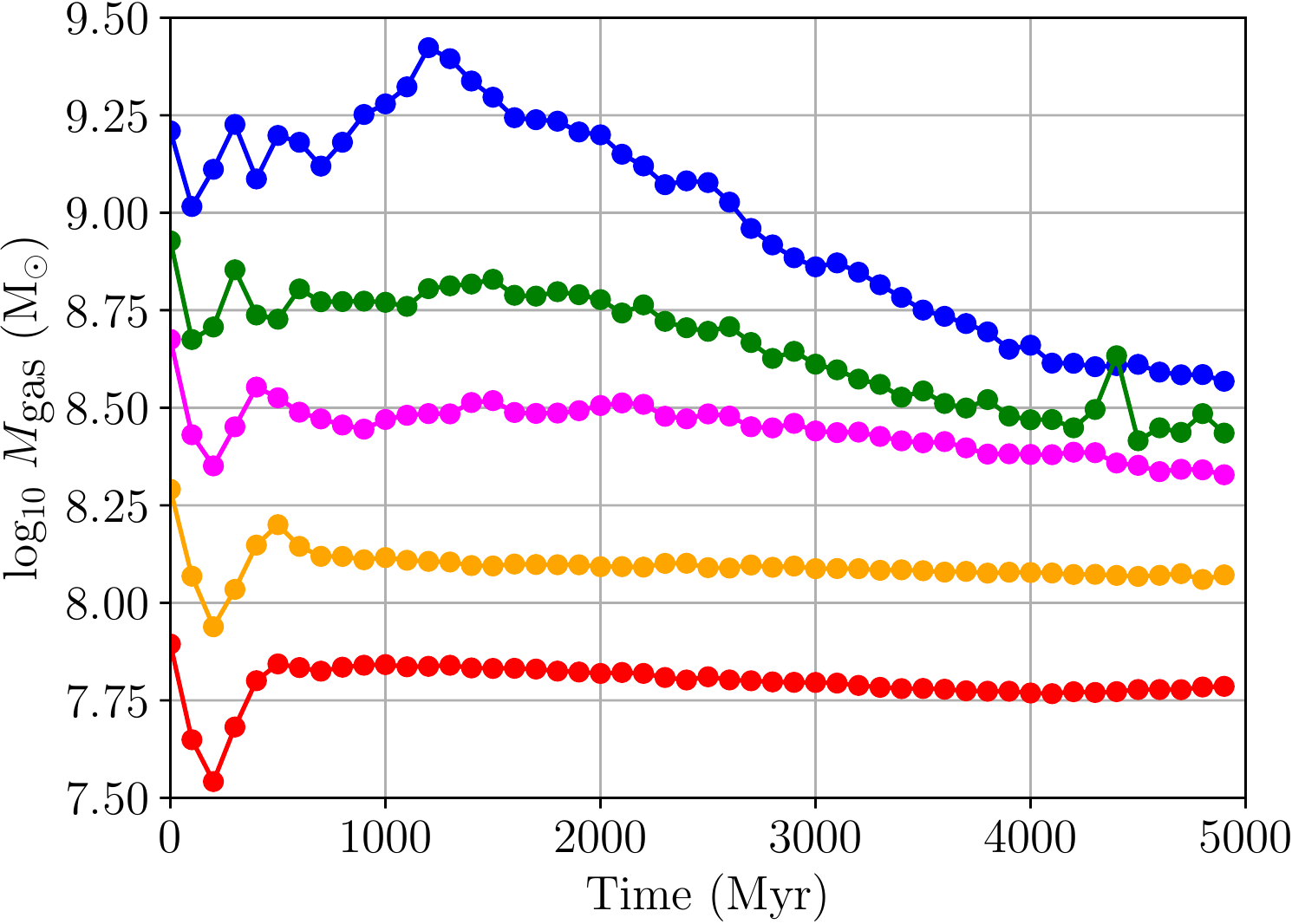}
    \caption{Gas mass in the disc region (out to $10 ~ \widetilde{R}_\textrm{eff}$) as a function of time. The blue, green, magenta, orange, and red curves correspond to models 1e11, 1e10, 1e9, 1e8, and 1e7, respectively.}
    \label{fig:Gas-mass}
\end{figure}

The above method is quite sensitive to fluctuations in the SFR, an issue that we attempt to address with our second method to find $\tau_g$. The gas mass within a cylindrical radius of $10 \, \widetilde{R}_\textrm{eff}$ is plotted as a function of time (Fig.~\ref{fig:Gas-mass}). After an initial `settling down' phase that lasts $\approx 1$~Gyr, the gas mass starts to decrease roughly exponentially. We perform a linear regression between time and the logarithm of the gas mass over the period $1-4.5$~Gyr and take $\tau_{g,2}$ to be the inverse of the slope.

\begin{figure}
    \includegraphics[width=8.5cm]{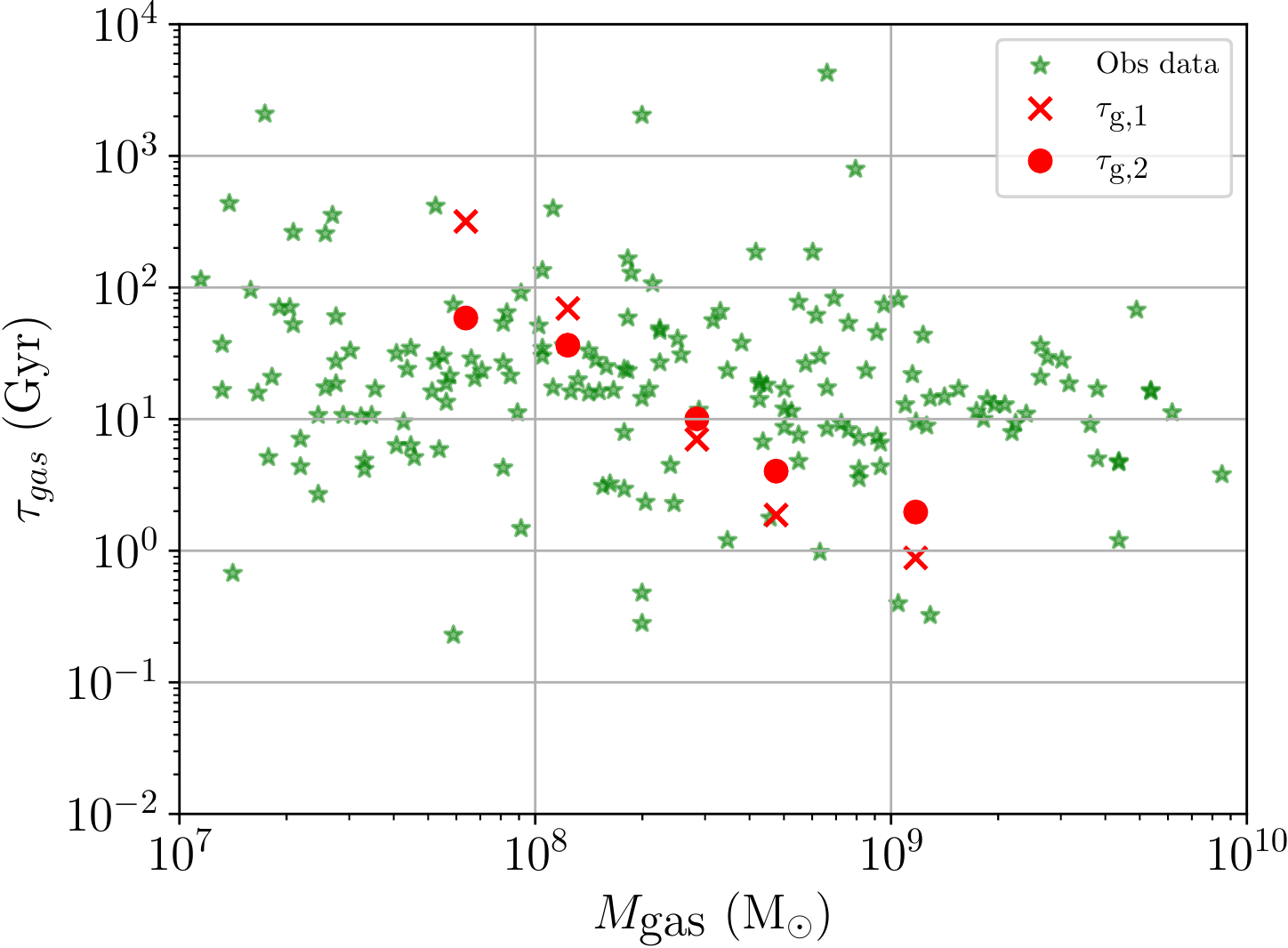}
    \caption{Gas depletion time $\tau_g$ as a function of gas mass. The green dots are observed data from \citet{Pflamm-Altenburg_2009} assuming an invariant IMF, while the red crosses (dots) show simulated $\tau_{g,1}$ ($\tau_{g, 2}$) values (see the text).}
    \label{fig:Gas-depletion-times}
\end{figure}

Fig.~\ref{fig:Gas-depletion-times} shows the gas depletion times obtained using these methods, with $\tau_{g, 1}$ ($\tau_{g, 2}$) values shown using red crosses (dots). The smaller green stars show observational results \citep{Pflamm-Altenburg_2009}. It is clear that both $\tau_{g ,1}$ and $\tau_{g, 2}$ are comparable to those of observed galaxies. It should be noted that these values of gas depletion timescales inherently assume an invariant galaxy-wide IMF. This is at present not a tangible proposition \citep{Kroupa_2013, Jerabkova_2018, Kroupa_Jerabkova_2021, Yan_2021_IMF}. The calculation by \citet{Pflamm-Altenburg_2009} using the integrated galactic IMF \citep[IGIMF;][]{Kroupa_2003, Weidner_2006} theory predicts $\tau_{g} \approx 3$~Gyr for all late-type galaxies (Fig. 6 of \citet{Pflamm-Altenburg_2009}). Note that the green points in Fig. \ref{fig:Gas-depletion-times} are not corrected for the IGIMF-effect.

\subsection{The Kennicutt-Schmidt relation}
\label{sec:KS_law}

\begin{figure*}
    \hspace{-1cm}
    \includegraphics[width=17.5cm]{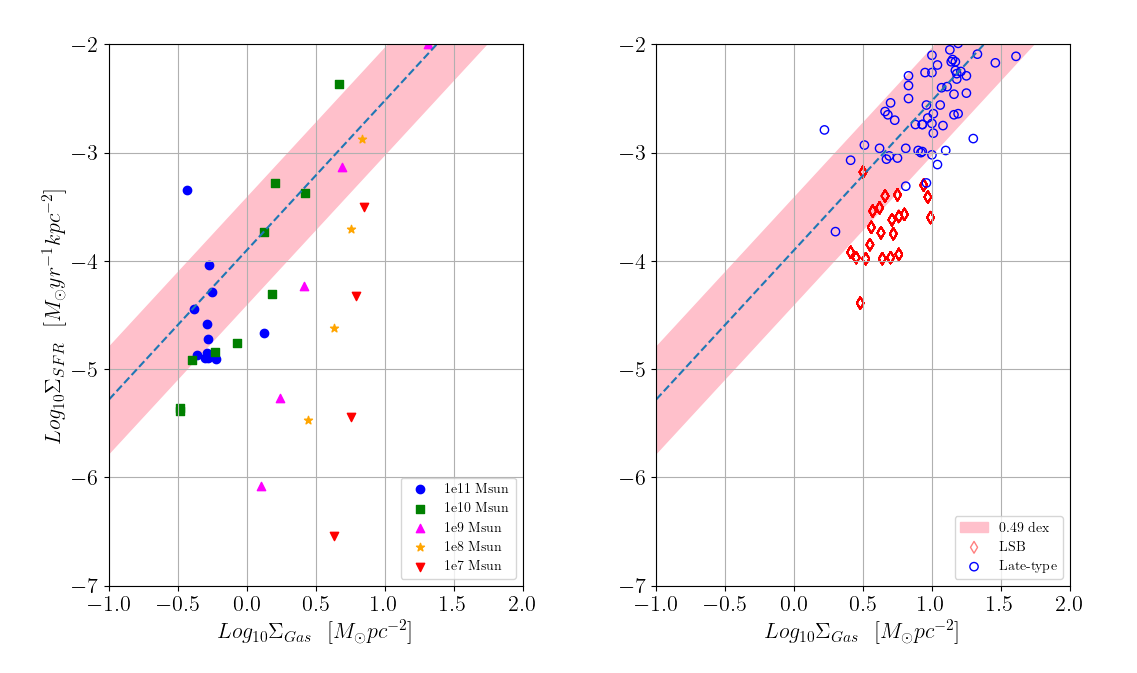}
    \caption{The Kennicutt-Schmidt diagram relating the surface densities of gas and of star formation. \emph{Left}: Different annuli in the models. Blue, green, magenta, orange, and red correspond to models 1e11, 1e10, 1e9, 1e8, and 1e7, respectively. \emph{Right}: Observed LTs (red) and LSBs (blue), taken from \citet{Shi_2011}. The dashed blue line (Eq.~\ref{eq:KS-law}) is taken from \citet{Kennicutt_1998}.}
    \label{fig:KS_Obs}
\end{figure*}

The KS law is an empirical relation between the star formation rate surface density $\Sigma_{\textrm{SFR}}$ and the gas surface density $\Sigma_g$ for disc galaxies \citep{Schmidt_1958, Kennicutt_1998}. The KS diagram is used to investigate disc-averaged SFRs and gas densities. It has also been shown that the KS relation can be applied to sub-kpc scale star forming regions within galaxies \citep{Bigiel_2008}. Here we investigate whether different regions of the models agree with the KS relation. For the models, $\Sigma_{\textrm{SFR}}$ and $\Sigma_g$ are calculated  by binning in cylindrical polar radius. These annular bins have a constant width $\Delta R$, which we set to $\approx 10 \times$ the highest spatial resolution. The bins go out to a maximum radius of $R_{\rm{max}} = 5 \, \widetilde{R}_{\rm{eff}}$ for the corresponding model. The number of radial bins for each model is then $R_{\rm{max}}/\Delta R$. We find the SFR in each annulus over the time interval $1 - 4.5$~Gyr. This SFR is divided by the area of the bin, which gives $\Sigma_{\textrm{SFR}}$. Similarly, by calculating the mass of gas particles in each bin and dividing this by the area of the bin, we can obtain $\Sigma_g$. An example of this analysis is shown in Appendix \ref{KS_analysis}.

$\Sigma_{\textrm{SFR}}$ and $\Sigma_g$ are plotted in Fig.~\ref{fig:KS_Obs}, where models are shown in the left panel. The right panel shows observed late-type galaxies (LTs) and LSBs taken from \citet{Shi_2011}. To represent the observed galaxies, we use eq.~28 in \citet{Kennicutt_1998}, with a scatter in the original data of 0.49 dex.
\begin{equation}
    \Sigma_{\textrm{SFR}} ~=~ 2.5 \times 10^{-10} {\Sigma_g}^{1.4} \, ,
    \label{eq:KS-law}
\end{equation}
where $\Sigma_g$ is in $M_\odot$/pc\textsuperscript{2} and $\Sigma_{\textrm{SFR}}$ is in $M_\odot$/pc\textsuperscript{2}/yr, with uncertainties omitted for clarity. Fig.~\ref{fig:KS_Obs} shows different star-forming regions within each model galaxy, giving a better idea of their contribution to the global SFR. Some simulated regions fall on the KS relation, some are within the range of the data, and some are below it. Notice how most regions are compatible with the KS relation for models 1e10 and 1e11. Star-forming regions in the lower mass models are typically below the KS relation. This follows the trend in observed LSBs, which are also slightly below the MS relation (Fig.~\ref{fig:Main-sequence}).

\subsection{Renzo's rule}

Since Milgromian galaxies are purely baryonic, the distribution of baryons dictates the gravitational field both locally and globally. A major implication is that features in the baryonic surface density $\Sigma$ should reflect on the dynamics of the galaxy. Renzo's rule is the observation that any feature in the luminosity profile of a galaxy has an imprint on the RC, and vice versa \citep{Renzo_2004}. While this makes sense in Newtonian gravity at high surface brightness where the baryons dominate the gravitational potential, Renzo's rule has been observed in galaxies independently of their surface brightness, including in LSBs where conventionally the RC is mostly not due to the baryons. This is discussed in \citet{Famaey_McGaugh_2012} and references therein. 

\begin{figure*}
    \includegraphics[width=0.495\textwidth]{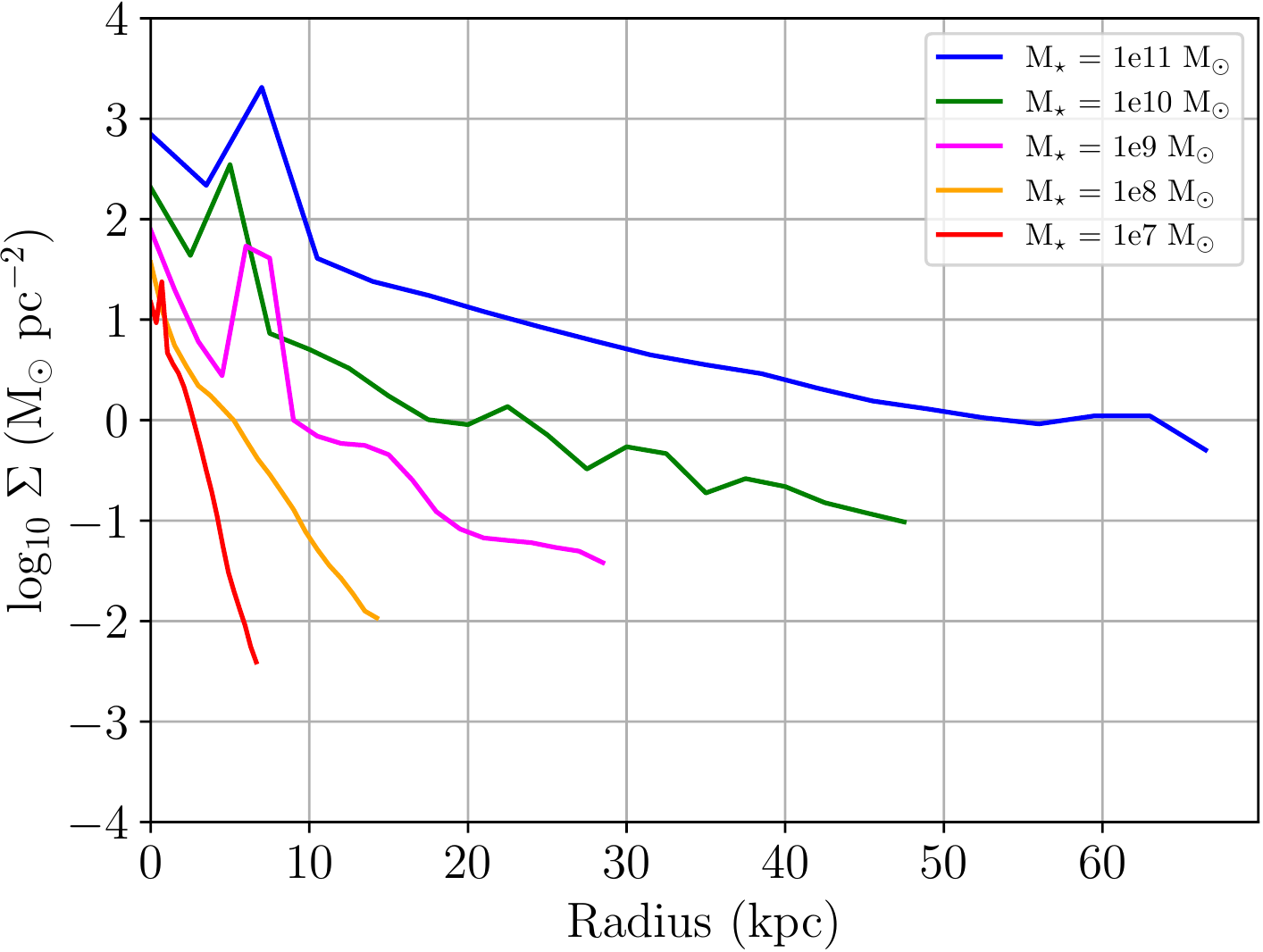}
    \hfill
    \includegraphics[width=0.495\textwidth]{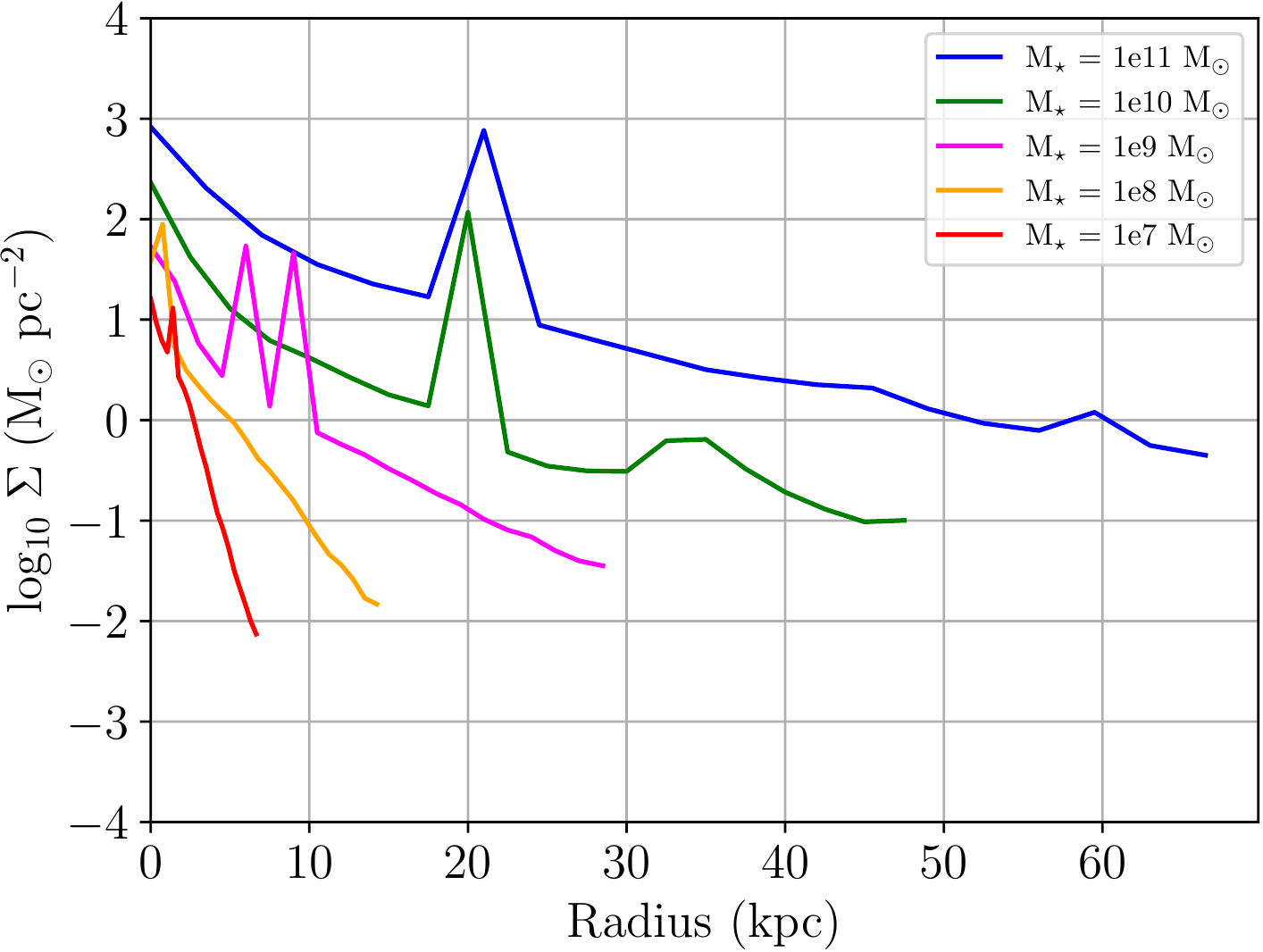}
    \caption{The total surface density of all our models after 3~Gyr (left) and 5~Gyr (right), as indicated in the legend.}
    \label{fig:surface_density}
\end{figure*}

\begin{figure*}
    \includegraphics[width=0.495\textwidth]{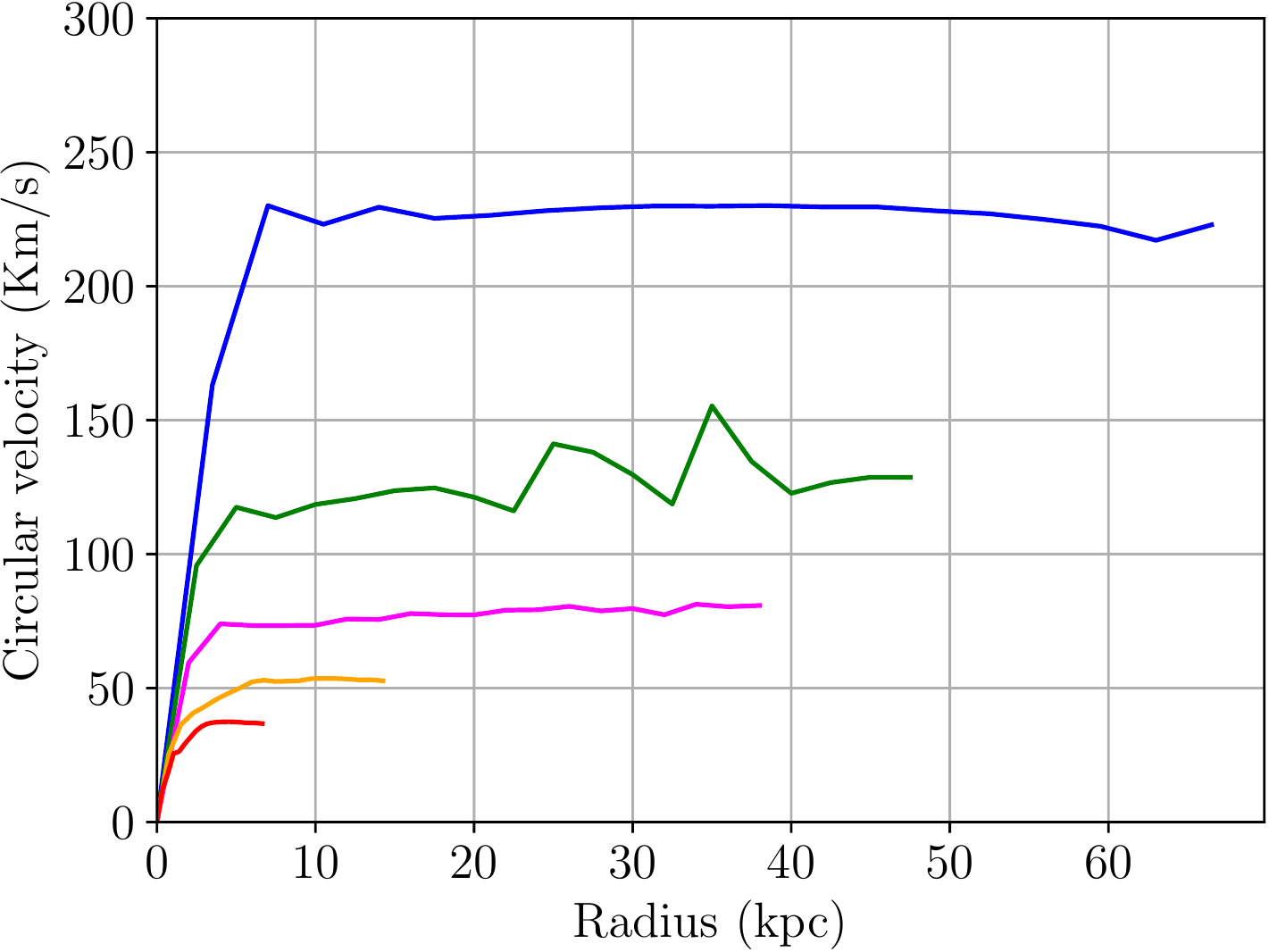}
    \hfill
    \includegraphics[width=0.495\textwidth]{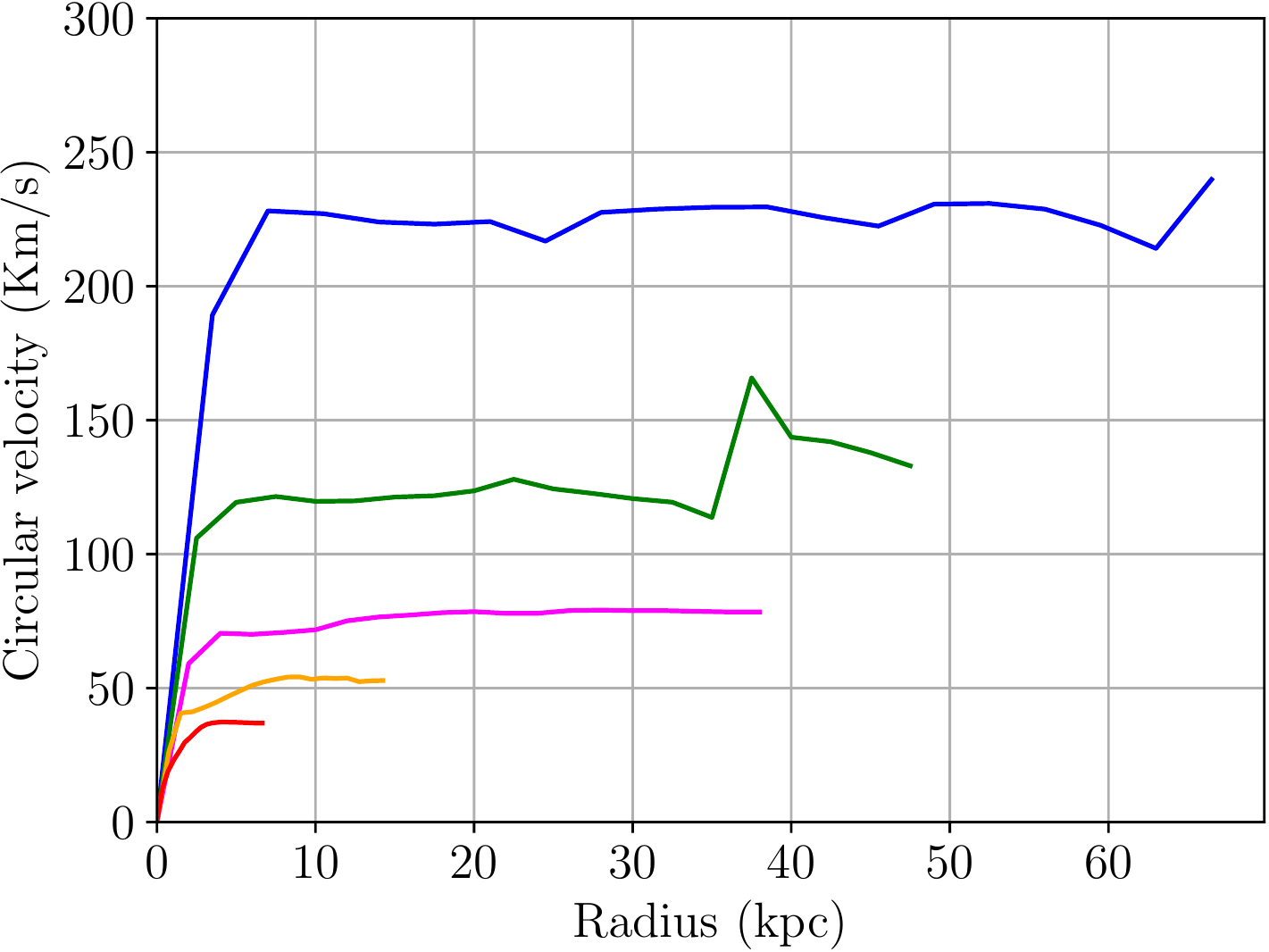}
    \caption{The rotation curve of every model at 3~Gyr (left) and 5~Gyr (right). The same colour scheme is used as in Fig.~\ref{fig:surface_density} to show the different models.}
    \label{fig:Rot_curves}
\end{figure*}

We computed RCs for our models at two different times in their evolution. The circular velocity $v_c = \sqrt{ - \bm{r} \cdot \bm{g}}$, where $\bm{r}$ is the position of a particle and $\bm{g}$ is its acceleration. The $v_c$ estimates from different particles were averaged in each annulus. Fig.~\ref{fig:Rot_curves} shows the RCs of all the models out to 10 effective radii at 3~Gyr and 5~Gyr. The surface density profiles at these times are shown in Fig.~\ref{fig:surface_density}. These profiles are not completely smooth. Local star-forming gas clumps in the disc gravitationally perturb the stellar particles around them, so the localised gravitational field superimposes itself on the global galactic gravitational field. The effect is seen as kinks and bumps in the RCs (Fig.~\ref{fig:Rot_curves}), much as in real low-mass galaxies $-$ although a bit more pronounced here than in most observed massive discs.

\subsection{Vertical velocity dispersion}
\begin{figure*}
    \includegraphics[width=0.495\textwidth]{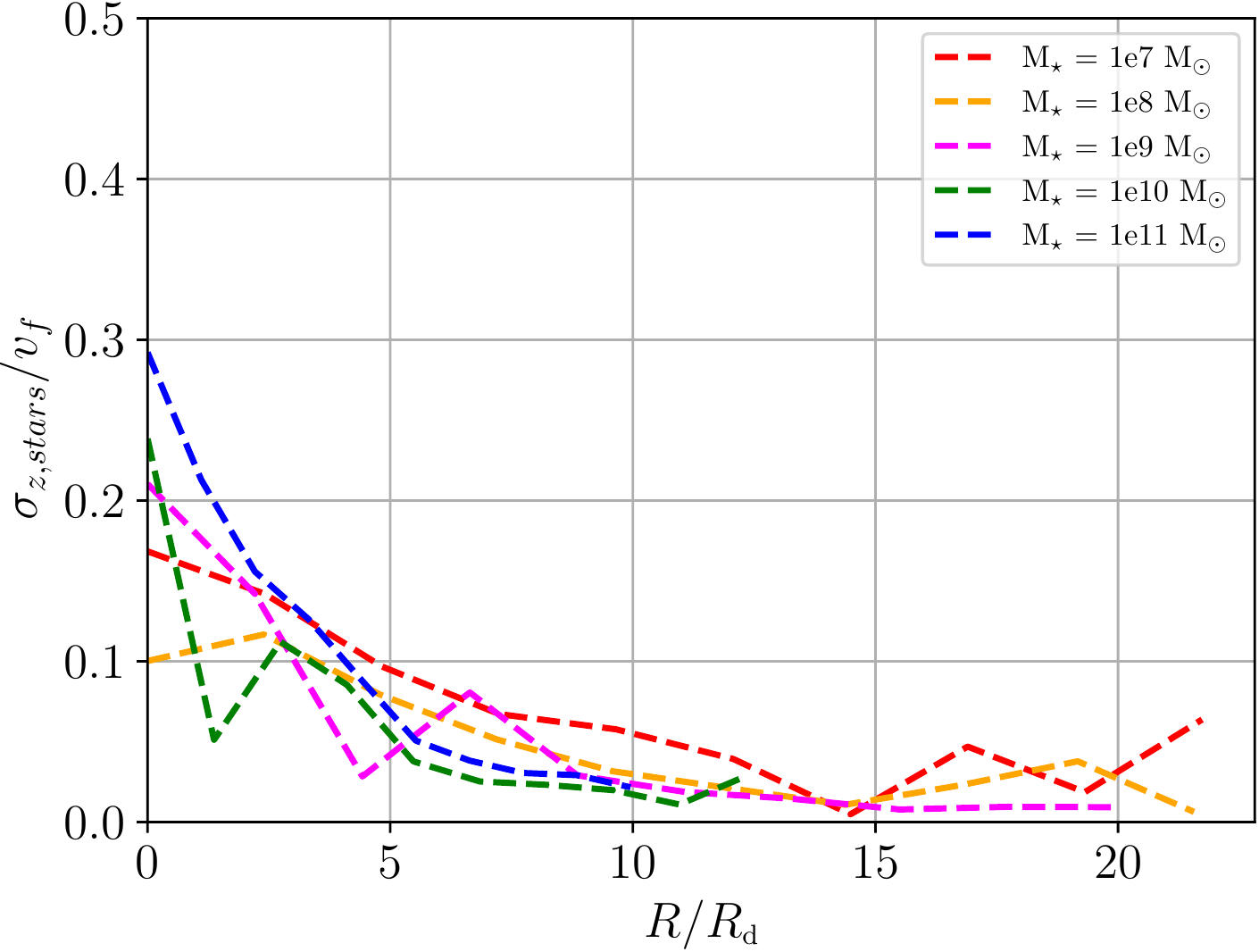}
    \hfill
    \includegraphics[width=0.495\textwidth]{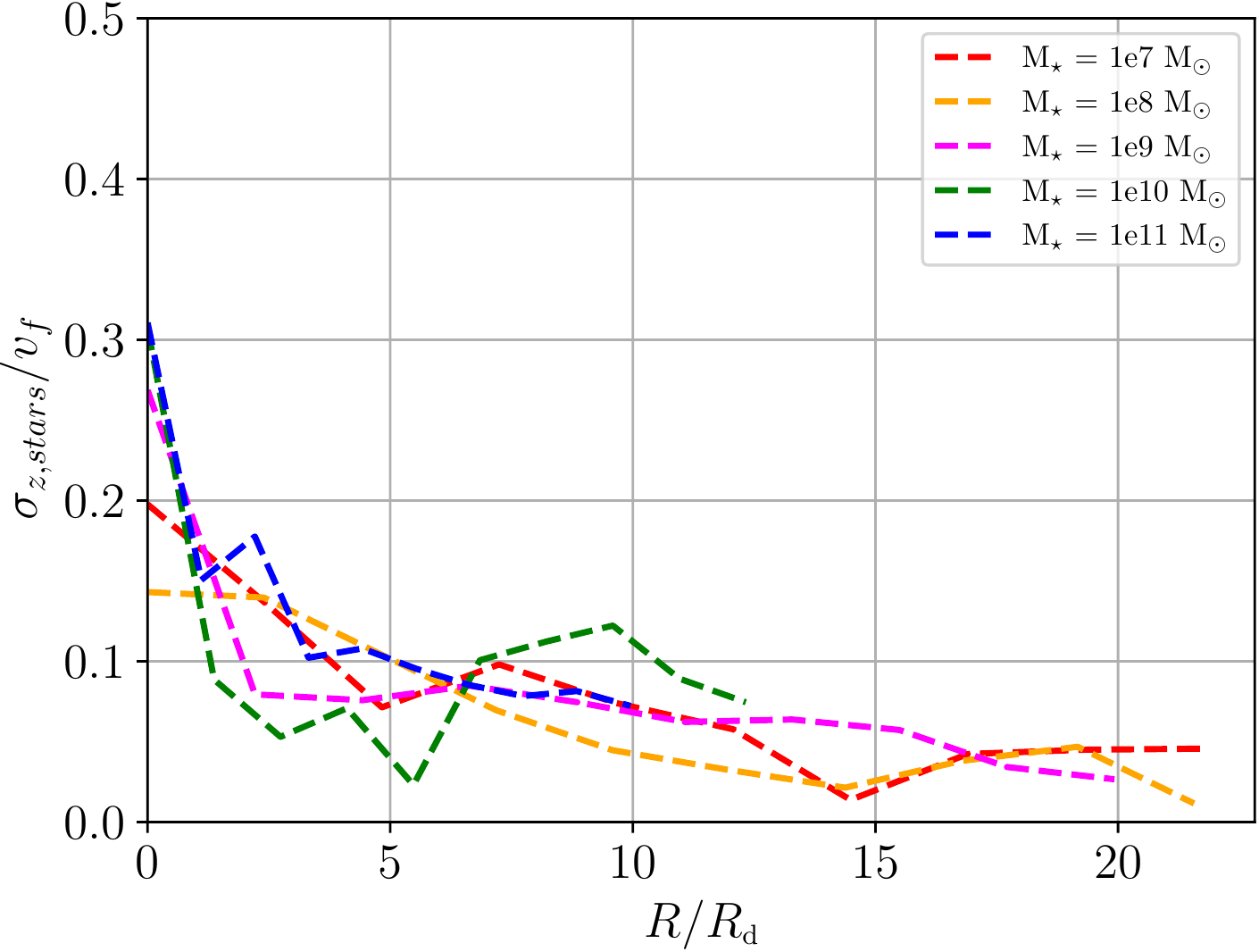}
    \caption{$\sigma_z/v_{_f}$ of all stellar particles as a function of galactocentric distance after 1 Gyr (left) and 5 Gyr (right).}
    \label{fig:Vert_Vel_r_disp}
\end{figure*}

The vertical velocity dispersion $\sigma_z$ can be used as a measure of how dynamically hot the disc is. We find the mass-weighted $\sigma_z$ of all the stellar particles using eq.~22 of \citet{Banik_2020_M33}.\footnote{We also tried not mass-weighting the particles. Both techniques give very similar results.}

\begin{eqnarray}
    \sigma_z^2 ~=~ \frac{{\sum_i m_{\star, i} \sum_i m_{\star, i} v_{z, i}^2 - \left( \sum_i m_{\star, i} v_{z, i} \right)^2}}{\left( \sum_i m_{\star, i} \right)^2 - \sum_i m_{\star, i}^2 } \, ,
    \label{eq:velocity-dispersion}
\end{eqnarray}
where $i$ is the particle index, $m_{\star, i}$ is the mass of particle $i$, and its vertical velocity relative to that of the barycentre is $v_{z, i}$. Once we have found $\sigma_z$ in this way, we divide it by the asymptotic rotational velocity $v_{_f}$ (Eq.~\ref{eq:asymptoticvelocity}), of the corresponding model, in order to obtain a dimensionless measure of how dynamically hot the disc is.\footnote{Note that the calculation needs to include particles that were initially present and those that form during the simulation.} This calculation was repeated for different snapshots and different annuli. The results are shown in Fig.~\ref{fig:Vert_Vel_r_disp} at 1~Gyr (shortly after the disc settles down) and at 5~Gyr, when the simulation ends. The majority of the discs have a ratio $\sigma_z/v_{_f} \leq 0.2$, implying that the stellar discs remain dynamically cold throughout their evolution.

\begin{figure*}
    \includegraphics[width=0.495\textwidth]{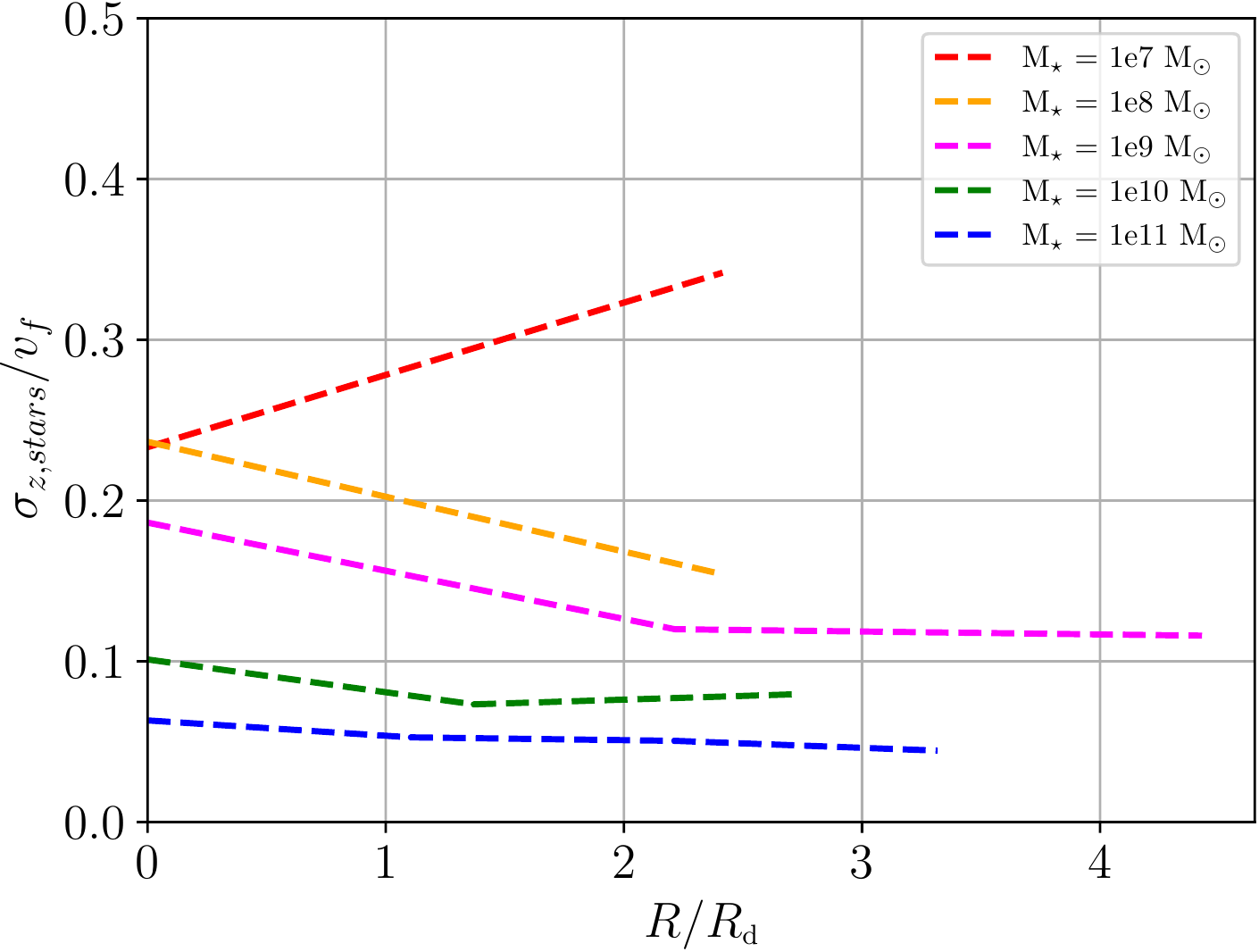}
    \hfill
    \includegraphics[width=0.495\textwidth]{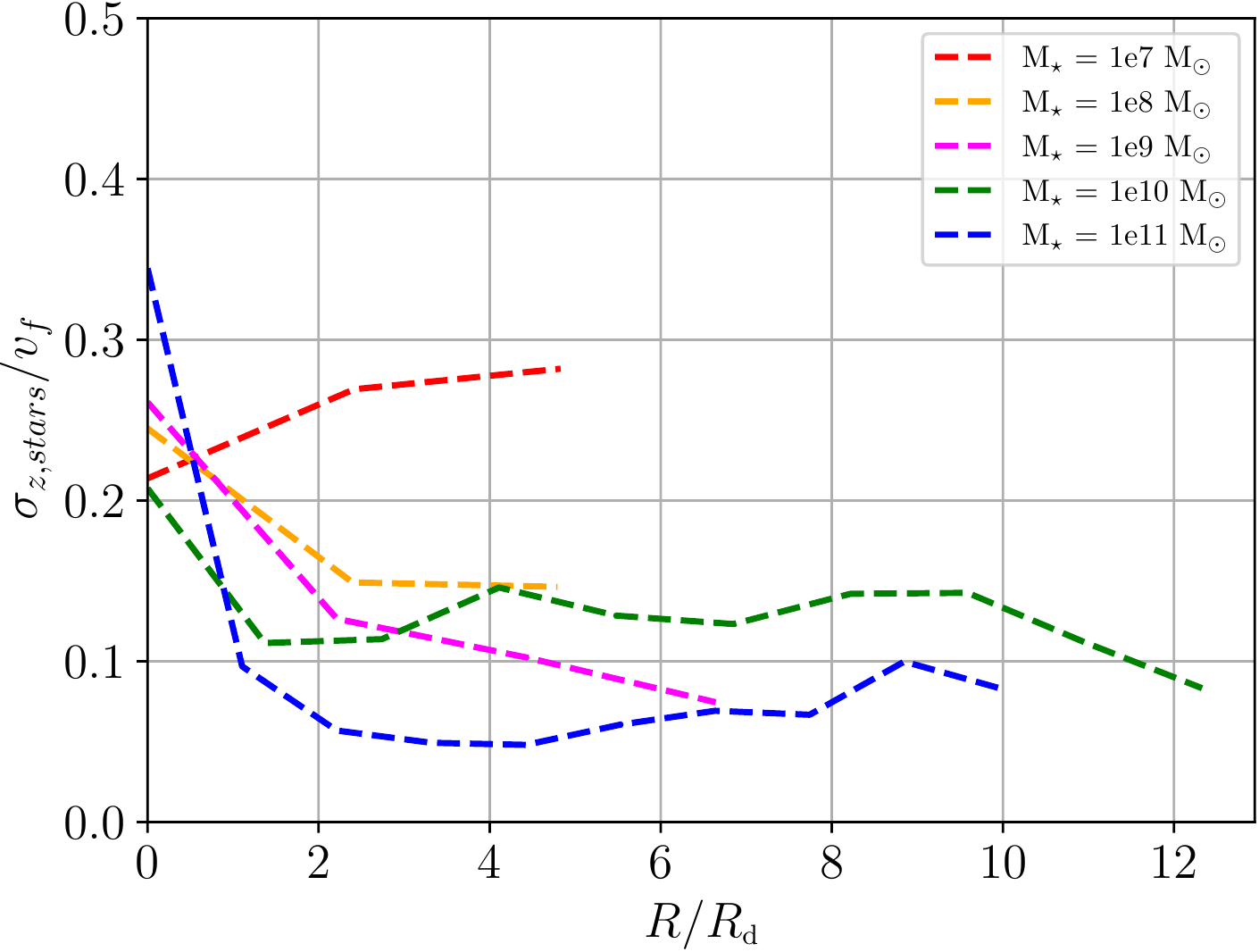}
    \caption{$\sigma_z/v_{_f}$ of newly formed particles as a function of galactocentric distance after 1 Gyr (left) and 5 Gyr (right).}
    \label{fig:Vert_Vel_r_disp_new}
\end{figure*}

The gas is a diffuse component that gets redistributed by feedback from SNe. To get a sensible estimate of the gas disc thickness, we had to restrict attention to gas cells within some maximum distance in the vertical direction. We tried a cutoff of some rational multiple of $\widetilde{R}_\textrm{eff}$, but the choice of cutoff seems to bias the result. We therefore show the edge-on view of the gas to show its thickness as well as that of the stellar component. It is evident from Table~\ref{tab:Final_parameters} and Appendix~\ref{Face_on_view} that the gas mostly lies in a thin disc and that stars form within it.

\begin{table}
    \begin{tabular}{cccccc}
    \hline
    Model & $M_\star$ & $M_{tot}$ & $R_d$ & $R_g$ & $z_{\textrm{rms}}$ \\ 
    name & ($M_\odot$) & ($M_\odot$) & (kpc) & (kpc) & (kpc) \\ \hline
    1e7 & $1.10 \times 10^7$ & $8.37 \times 10^7$ & 0.37 & 1.08 & 0.20 \\
    1e8 & $1.10 \times 10^8$ & $2.51 \times 10^8$ & 0.64 & 2.64 & 0.20 \\
    1e9 & $1.16 \times 10^9$ & $1.47 \times 10^9$ & 1.43 & 5.24 & 0.55 \\
    1e10 & $1.10 \times 10^{10}$ & $1.14 \times 10^{10}$ & 5.31 & 10.73 & 0.68 \\
    1e11 & $1.05 \times 10^{11}$ & $1.05 \times 10^{11}$ & 9.71 & 22.57 & 1.95 \\ \hline
    \end{tabular}
    \caption{Parameters of all models at the end of their evolution (after 5~Gyr). The columns show the model name, total stellar mass, total galaxy mass within $10 ~ \widetilde{R}_\textrm{eff}$, the exponential scale lengths for the stellar and gas discs, and the root mean square thickness of the stellar component.}
    \label{tab:Final_parameters}
\end{table}

Newly formed stars are expected to be dynamically colder than the general stellar population due to their recent formation out of the dissipative gas component. Fig.~\ref{fig:Vert_Vel_r_disp_new} shows $\sigma_z$ of the particles that formed during the simulation. In the first Gyr, most particles form within the plane of the disc. By the end of the simulation, the newly formed particles have a higher $\sigma_z$. This is due to secular heating of the stellar component by fluctuations in the gravitational potential. Part of the reason is also that the gas gets heated by SNe, which thickens the gas disc and slightly affects the in-plane star formation process. Note also that for this reason, in the lower mass models, the newly formed stars form in a thicker and dynamically hotter state than in models 1e10 and 1e11. In the lowest mass model, $\sigma_z$ of the newly formed stars actually increases with radius due to the shallow gravitational potential of the outer discs in such models and their fragility to SNe heating. Face-on and edge-on views of all the models are shown in Appendix~\ref{Face_on_view}, where it can nevertheless be seen that star formation largely occurs in the disc plane.

\subsection{Outer disc streams}
\label{Stellar_streams}

\begin{figure}
    \includegraphics[width=0.495\textwidth]{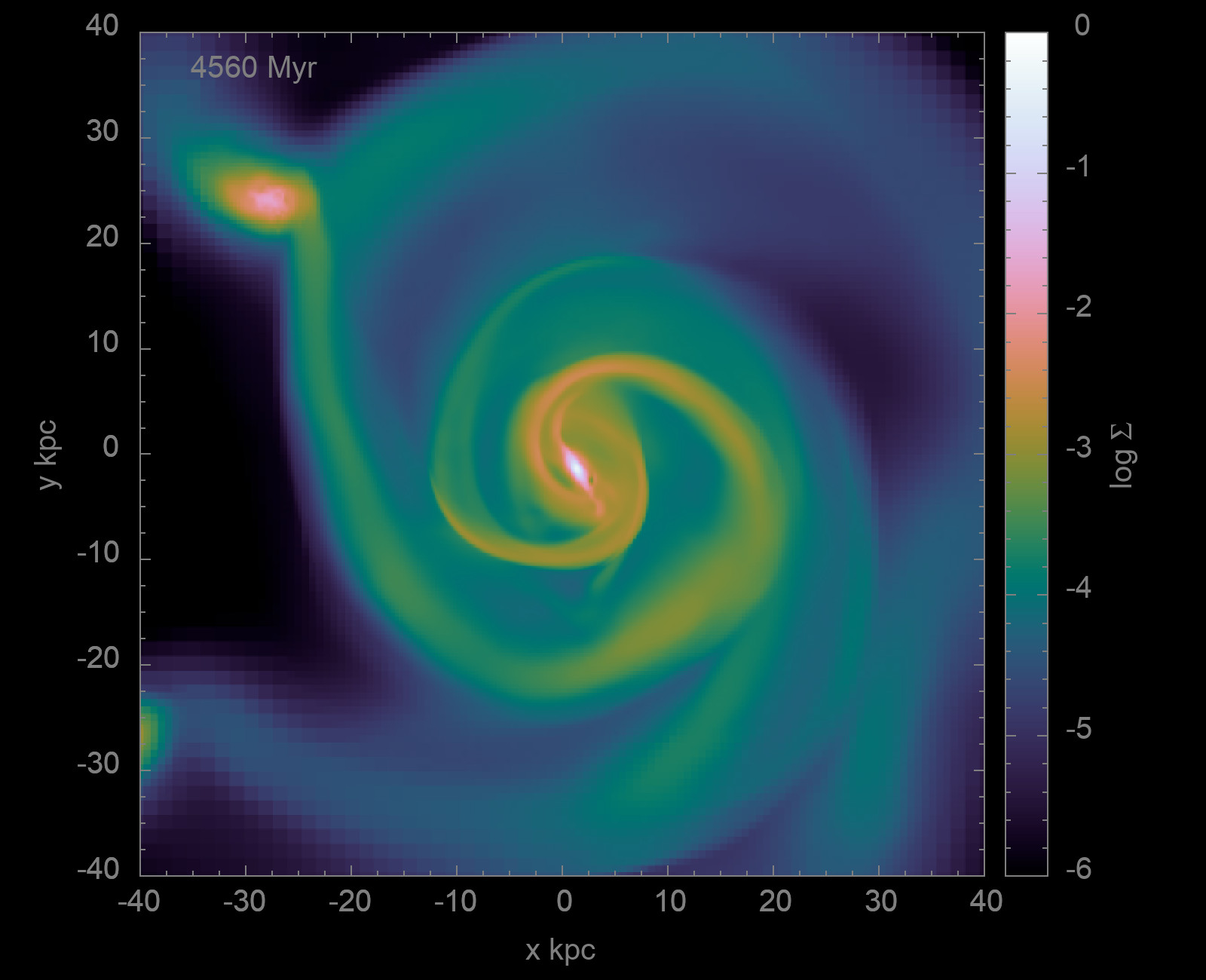}
    \caption{Gas distribution of model 1e10 at 4560 Myr. Notice the clump towards the top left which appears like a satellite, even though the simulation is of an isolated disc galaxy.}
    \label{fig:Streams_1e10}
\end{figure}

Massive models like 1e10 and 1e11 develop structures in their discs at large radii (Fig.~\ref{fig:Streams_1e10}). These structures have self-sustaining star formation activity. Fig.~\ref{fig:Face_edge_1e11} shows that such overdense regions have younger stars than the typical disc, thus leading one to interpret them as distinct non-disc entities.

Observationally, some structures in the MW disc are candidates to have formed due to past mergers of dwarf satellite galaxies with the MW, though many recent studies tend to show that these are mostly made of stellar populations characteristic of the outer disc itself \citep[e.g.,][and references therein]{Laporte_2020}. In the case of our models, these structures are initially part of the outer disc, from where they orbit outwards to large radii and later turn around and merge with the disc. Our models thus demonstrate that such structures naturally arise in MOND without any external perturbation, thus raising the possibility that observed similar structures around disc galaxies are parts of their outer disc rather than external objects, which can be seen in the movies \textsuperscript{\ref{Movies}}.

\subsection{Bar analysis}
\label{s.BA}

In this section, we study the properties of the central galactic bar in models 1e11, 1e10, and 1e9. The lower-mass models 1e8 and 1e7 are not considered here due to their more turbulent behavior, which reduces the reliability of the results. We quantify the bar in terms of its length, strength, and pattern speed at different times. We then use this information to report the $\mathcal{R}$ parameter (Eq.~\ref{R_definition}) and compare its distribution to observations and to cosmological $\Lambda$CDM simulations. The procedure described in this section is very similar to that applied in \citet{Roshan2021_2}.

\subsubsection{Bar strength}
\label{s.BS}

To quantify the strength of the bar, we consider the azimuthal Fourier expansion of the stellar surface density, which is generally what observers use to analyse bars. The disc is divided into annuli with a fixed width of $\Delta r = 0.5$~kpc. To confine the calculations to the bar region, we consider the disc out to $R = 10$~kpc. In each annulus, the Fourier coefficients are calculated as
\begin{eqnarray}
	a_{\rm{m}} \left( R \right) &\equiv& \frac{1}{M \left( R \right)} \sum_{k=1}^{N} m_k \cos \left( \rm{m} \phi_k \right), ~ \rm{m} = 1, 2, .. \, , \\
	b_{\rm{m}} \left( R \right) &\equiv& \frac{1}{M \left( R \right)} \sum_{k=1}^{N} m_k \sin \left( \rm{m} \phi_k \right), ~ \rm{m} = 1, 2, .. \, , 
\end{eqnarray}
where the annulus contains $N$ particles. The mass of particle $k$ is $m_k$ and the corresponding azimuthal angle is $\phi_k$. The mean radius of the annulus is $R$ and the stellar mass within it is $M \left( R \right)$.

\begin{figure}
	\centering
	\includegraphics[width=0.45\textwidth]{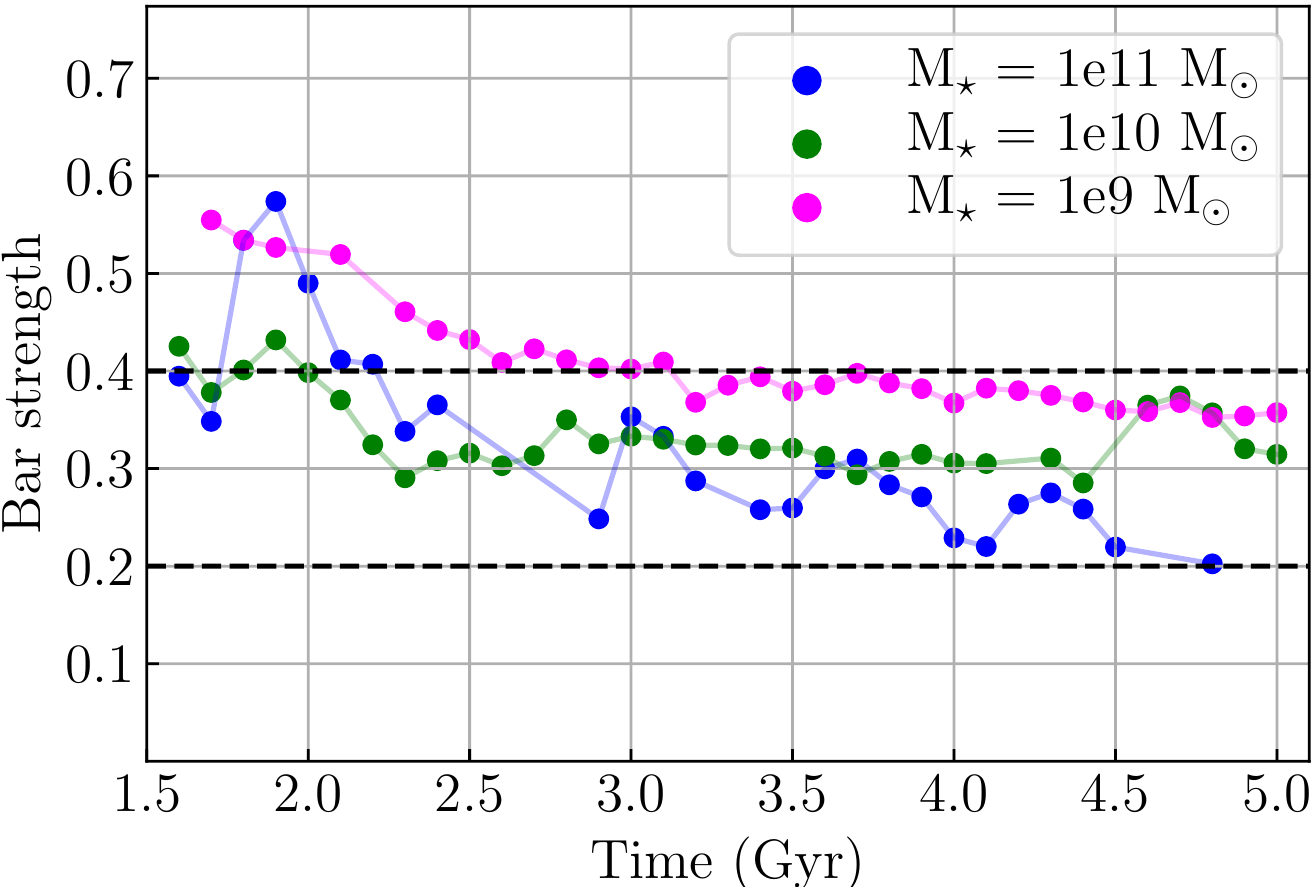}
	\caption{Bar strength in model 1e11 (blue), 1e10 (green), and 1e9 (magenta). All models show a similar time evolution. Bar strengths in the range $0.2 - 0.4$ are weak while values $\geq 0.4$ constitute strong bars.}
	\label{fig:BS}
\end{figure}

We then focus on the case $\rm{m} = 2$ and find
\begin{eqnarray}
    A_{\rm{2}} \left( R \right) ~\equiv~ \sqrt{a_{\rm{2}} \left( R \right)^2 + b_{\rm{2}} \left( R \right)^2} \, .
    \label{A_2_def}
\end{eqnarray}
The maximum value over different annuli, $A^{\text{max}}_2$, is defined as the bar amplitude. This definition is common in the literature \citep[e.g.,][]{Guo2019, Rosas_2020}. According to this criterion, the galactic bars can be classified as weak bars with $0.2 \leq A_2^{\text{max}} < 0.4$ and strong bars with $A_2^{\text{max}} \geq 0.4$. Fig.~\ref{fig:BS} illustrates the evolution of the bar strength in models 1e11 (blue), 1e10 (green), and 1e9 (magenta). From this figure, it is apparent that in all three models, the strength of the bar shows a rather strong decrease in the beginning stages of the evolution before shifting to a more steady trend after the system reaches a more stable state. The initial decrease is expected since the strength of the $\rm{m} = 2$ mode is affected by the existence of spiral arms in the system, which are strong initially but disappear after some evolution of the disc. Furthermore, after the disc reaches the more stable state, the bar strength shows a dependence on the mass of the system. Although all three models are mainly in the weak regime, model 1e9 shows a higher strength and evolves near the edge of the strong bar regime, while model 1e11 has the weakest bar. This is broadly consistent with the fact that the bar fraction tends to be smaller for galaxies with a higher stellar mass \citep[see the S$^4$G observations in][]{Erwin2018}.

An important aspect of our results is that bars form naturally in MOND even when it predicts a significant enhancement to Newtonian gravity at all radii \citep{McGaugh_1998b}. This is because all Milgromian galaxies are self-gravitating. However, this is not the case in $\Lambda$CDM for a galaxy dominated by a stabilising dark halo \citep{McGaugh_1998a}. As a result, these `sub-maximal discs' are expected to only very rarely have a strong bar, as shown recently with TNG50 \citep*{Kashfi2023}. Those authors showed that this result is in contradiction with the observed fact that bars in sub-maximal discs are quite common in the SPARC sample, even though this selects against barred galaxies. While some bars might be triggered externally, environmental effects should already be included in a cosmological simulation.

\subsubsection{Bar length}
\label{s.BL}

A common method to estimate the bar length also uses the azimuthal Fourier decomposition of the surface density in different annuli \citep{Ohta1990, Aguerri2000, Guo2019}. This is known to give an appropriate estimate of the bar length in simulations \citep{Lia2002}. In this method, the bar length is defined as the outer radius at which the ratio of the surface densities in the bar ($A_b$) and inter-bar ($A_{ib}$) regions satisfies the relation
\begin{eqnarray}
    \frac{A_b}{A_{ib}} ~>~ 0.5 \left[ \left(\frac{A_b}{A_{ib}}\right)_{\rm{max}} + \left(\frac{A_b}{A_{ib}}\right)_{\rm{min}}\right]  \, ,
	\label{Bar_interbar_ratio}
\end{eqnarray}
where $A_b = A_0 + A_2 + A_4 + A_6$, $A_{ib} = A_0 - A_2 + A_4 - A_6 $, and $A_{\rm{m}}$ is the Fourier strength of azimuthal mode $\rm{m}$, which we find using eq.~\ref{A_2_def} (in this system, $A_0 \equiv 1$).

\begin{figure}
	\centering
	\includegraphics[width=0.45\textwidth]{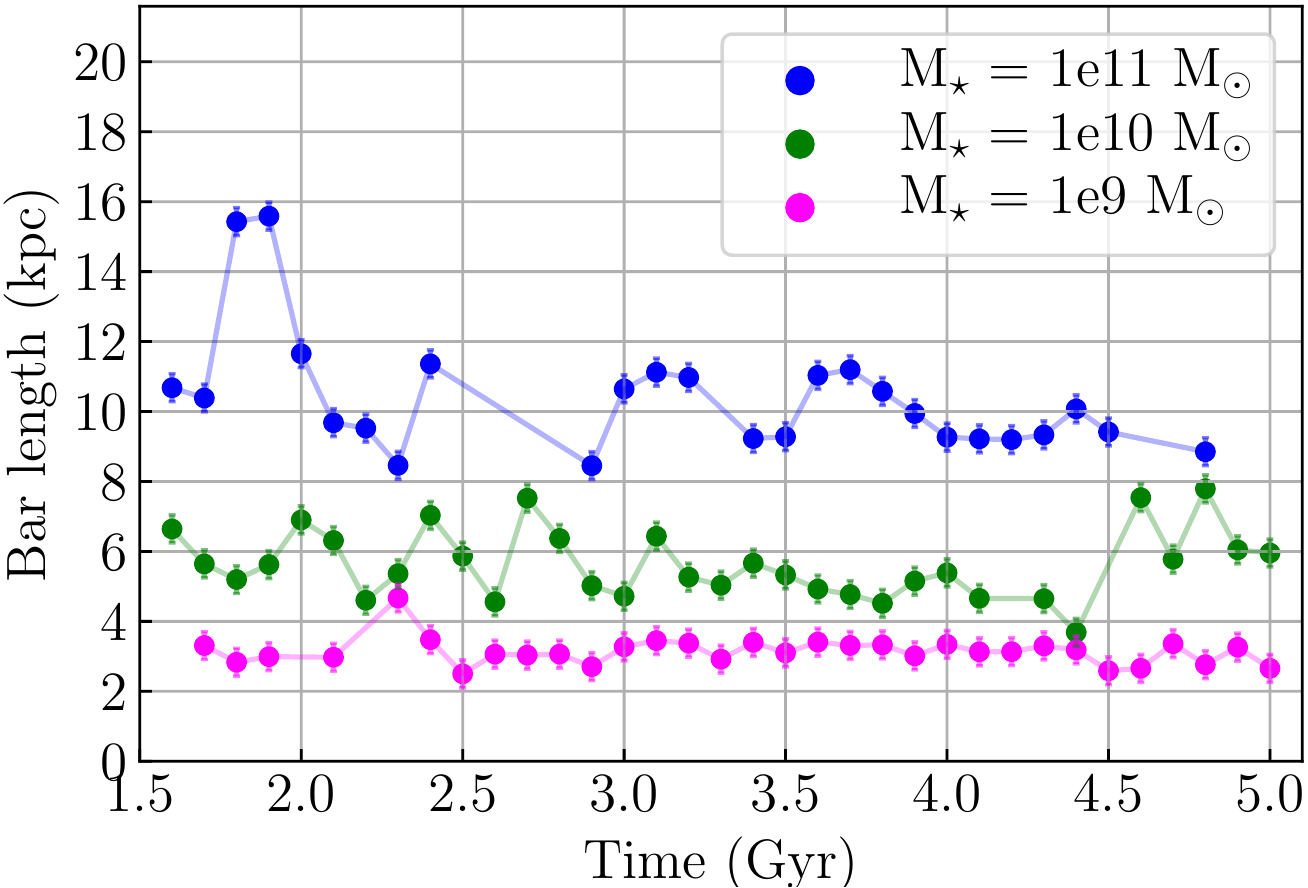}
	\caption{Bar length using Fourier amplitude decomposition (Eq.~\ref{Bar_interbar_ratio}) for models 1e11 (blue), 1e10 (green), and 1e9 (magenta).}
	\label{fig:BLF}
\end{figure}

Applying this procedure, Fig.~\ref{fig:BLF} shows the bar length for model 1e11 (blue), 1e10 (green), and 1e9 (magenta). This figure shows that the bar length is positively correlated with the mass of the system. All three models follow a similar and almost steady evolution over time.

It should be mentioned that \citet{Kim2021} study 379 observed galaxies with redshift $z$ in the range $0.2 < z \leq 0.835$ and masses $10 < \log_{10} \left( M_{\star}/M_{\odot} \right) < 11.4$. They report that the bar length strongly depends on the galaxy mass, (they rise together). Furthermore, they outline that the bar length shows no significant change over this redshift range. They conclude that little or no evolution in the bar length is present for the last $\approx 7$~Gyr. Moreover, \citet{Perez2012} and \citet{Lee2022} report a similar result when studying the bar length and its relation with redshift. Our results are thus in good compliance with observations regarding the dependence of the bar length on galaxy mass and time.

\subsubsection{Pattern speed}
\label{s.OP}

Galaxy bars are expected to rotate almost as a solid body, leading to a well-defined angular rotation rate or pattern speed $\Omega_p$. It is not straightforward to calculate this from observations as only one snapshot in the galaxy's evolution is accessible. Different techniques have been used in the literature to get around this difficulty. The only model-independent technique is known as the Tremaine-Weinberg (TW) method \citep{Tremaine1984}. This has been successfully used in both observational studies \citep[e.g.,][]{Aguerri_2015, Cuomo2019b, Williams_2021} and theoretical studies \citep[e.g.,][]{gerssen2007, Zou2019}. In this method, the pattern speed is obtained from the ratio $\Omega_p \sin i = \langle V \rangle/\langle X \rangle$, where $i$ is the disc inclination with respect to the sky plane, $\langle V \rangle$ is the luminosity-weighted average line of sight (LOS) velocity $V_{\text{LOS}}$, and $\langle X \rangle$ is the luminosity-weighted average position $X$ parallel to the major axis of the galactic disc. $\langle V \rangle$ and $\langle X \rangle$ are usually referred to as the kinematic and photometric integrals, respectively.

To have results that can be compared with observations, we employ the TW method to calculate the pattern speed in our simulations. We follow the same procedure as that described in \citet{Roshan2021_2}. The density and the velocity of the simulated stellar particles are used as the tracers in the pseudo-slits that are implemented for calculating $\langle V \rangle$ and $\langle X \rangle$. Since the TW method works more precisely with intermediate values of the disc inclination angle ($i$) and bar Position Angle (PA), we choose $i=45^{\circ}$ and PA$=60^\circ$ \citep{Debattista2003, Zou2019, Cuomo2019b}. Then the number of evenly spaced slits $N_s$, their width $\Delta_s$, height $h_s \geq N_s \Delta_s$, and length $l_s$ are varied until we find a convergent value for the pattern speed. The mean of the obtained values is then calculated, with the largest deviation from the mean used as the reported error of the calculation.

It should be noted that for some snapshots, the pattern speed seems to converge to two or more separate values. The appearance of multiple pattern speeds for a single galaxy has previously been noted for the cases where in addition to the bar, the galactic disc hosts extra features such as rings and/or spiral arms \citep{Debattista2002, Meidt2008}. When such cases arise in our simulation, the pattern speed is reported with a rather large uncertainty. This is mostly seen in model 1e11.

\begin{figure}
	\centering
	\includegraphics[width=0.45\textwidth]{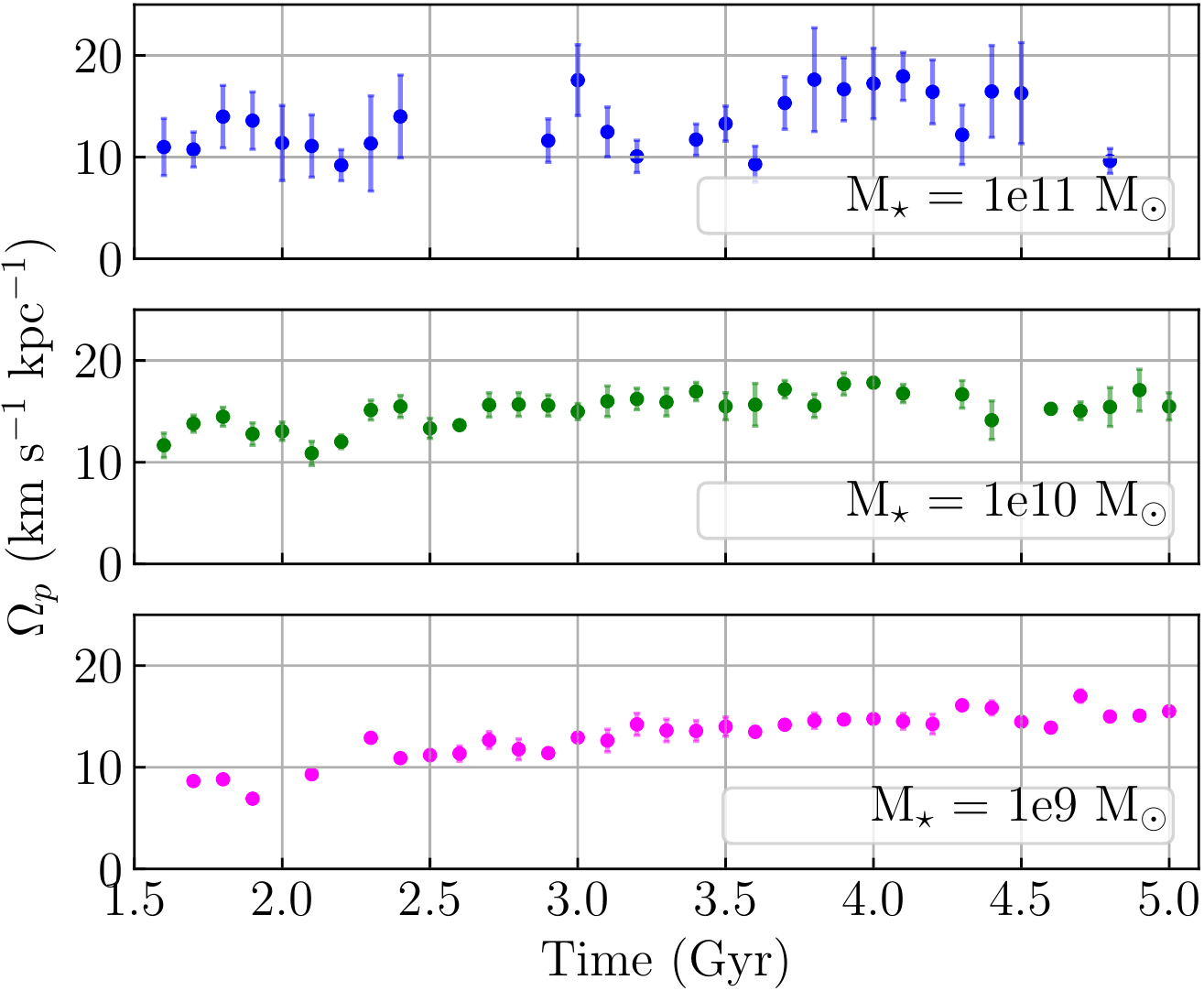}
	\caption{Bar pattern speed using the TW method for model 1e11 (blue), 1e10 (green), and 1e9 (magenta). The models follow a similar trend with time in that $\Omega_p$ remains almost constant.}
	\label{fig:OP}
\end{figure}

Fig.~\ref{fig:OP} demonstrates the pattern speed results for models 1e11, 1e10, and 1e9 as blue, green, and magenta dots, respectively. The pattern speed in all three models traces a similar trend and remains almost constant during the system's evolution. This is in contrast to the evolution of the pattern speed in isolated simulations conducted within the context of CDM, where due to the dynamical friction caused by the dark matter particles, the bar loses its angular velocity and slows down \citep{Ghafourian_2020, Roshan2021_1}.

\subsubsection{\texorpdfstring{$\mathcal{R}$}{R} parameter}
\label{s.R}

A well-known way to quantify the bar rotational speed is the dimensionless $\mathcal{R}$ parameter,
\begin{eqnarray}
    \mathcal{R} ~\equiv~ \frac{R_{\text{CR}}}{R_{\text{bar}}} \, ,
    \label{R_definition}
\end{eqnarray}
where $R_{\text{bar}}$ is the bar length and $R_{\text{CR}}$ is the corotation radius, the radius at which the disc particles rotate with the same angular speed as the bar. Therefore, this is the radius where $v_c \left( R_{\text{CR}} \right) = R_{\text{CR}} \Omega_p$, which must be solved iteratively for $R_{\text{CR}}$. Bars with $\mathcal{R} = 1 - 1.4$ are fast bars that almost extend to their corotation radius. If $\mathcal{R} > 1.4$, the bar would be short and in the slow regime. For the cases with $\mathcal{R} < 1$, the bar surpasses the corotation radius. Such cases are considered to be unphysical and are generally thought to indicate that the bar length is overestimated and/or the corotation radius is underestimated \citep{Cuomo2021}.

Observational studies indicate that bars in real galaxies are mostly fast \citep{Corsini2011, Aguerri_2015, Cuomo2019b, Guo2019}. However, in the CDM scenario, simulations of isolated galaxies \citep{Debattista2000, Athanassoula_2003} as well as cosmological simulations \citep{Algorry_2017, Roshan2021_2} lead to bars rotating mainly in the slow regime. Investigations regarding this inconsistency are still ongoing, As already mentioned, according to studies of isolated disc galaxies, bars slow down during the secular evolution of a galaxy mainly due to Chandrasekhar dynamical friction caused by the CDM particles. This is highlighted by considering extended gravity theories where galaxies lack a DM halo or by considering CDM models in which dynamical friction with the halo is unphysically suppressed by using a rigid halo. In this case, there is no massive halo absorbing angular momentum from the baryonic disc, eliminating the increasing trend of the $\mathcal{R}$ parameter with time \citep{Ghafourian_2020, Roshan2021_1}.

In the CDM scenario, dynamical friction causes the pattern speed to slow down, moving the corotation resonance outwards. Since $\mathcal{R}$ cannot have been much smaller than its typical present value of $\approx 1$, it should not have increased substantially over time. To avoid such an increase, the bar length would have to rise with time in order to keep up with the corotation radius as it migrates outward. However, as already mentioned in Section~\ref{s.BL}, studies show that the observed sizes of bars at higher redshift do not differ substantially from those of bars in the local Universe. Therefore, the increasing trend of the $\mathcal{R}$ parameter with time in $\Lambda$CDM simulations is at odds with the near constancy of bar lengths, which is more easily understood as due to bar lengths and corotation radii having changed little over the past several Gyr.

It is worth mentioning that earlier studies \citep[e.g.,][]{Debattista2000, Athanassoula_2003} state that the decrease in the bar rotational velocity could be avoided if the CDM halo density is low in the centre of the disc, which is possible with a maximal disc. A similar result was reported by \citet{Fragkoudi_2021} based on 16 galaxies from the Auriga zoom-in galaxy simulations \citep{Grand_2017}. These galaxies have fast bars. However, such a selection does not follow abundance matching $-$ the galaxies have too little DM, leading to selection bias. The only reason to consider such a small sample size given other studies with hundreds of galaxies \citep[e.g.,][]{Roshan2021_1} is if the resolution is much higher, but \citet{Fragkoudi_2021} indicated that resolution is not the reason for their somewhat perplexing results. It is also not the case that high-resolution simulations in the $\Lambda$CDM paradigm inevitably lead to fast bars \citep[e.g.,][]{Zana_2018, Zana_2019}.

Some studies claim that the decreasing pattern speed is not necessarily due to dynamical friction from the CDM halo. For example, \citet{Bi2021} consider a limited number of galaxies in very high-resolution zoom-in simulations with the \textsc{gizmo} code \citep{Hopkins2017} at higher redshifts and claim that the bar properties are significantly affected by the environment, especially mergers and close flybys. The large values of $\mathcal{R}$ are attributed to the above-mentioned mechanisms rather than dynamical friction. Although this subject is out of the scope of our paper, it should be stressed that irrespective of the underlying physics, their result is still inconsistent with pattern speed observations. On the other hand, using isolated simulations it is claimed in \cite{Beane2022} that a gas fraction of about 5\% can prevent the bar from slowing down in Milky Way-like galaxies. However, the gas component is already implemented in state-of-the-art cosmological simulations $-$ it does not operate to stabilize the bar pattern speed. One may speculate that the way in which gas should be modelled in dark matter cosmological simulations still needs modifications. However, it is important to note that no work has shown a CDM galaxy to develop a fast bar if simulated at high resolution but to develop a slow bar if simulated at the resolution of large cosmological hydrodynamical $\Lambda$CDM simulations like TNG50 \citep{Pillepich_2018, Nelson_2019}. Therefore, its results should be trusted as representative of the $\Lambda$CDM cosmology.

\begin{figure}
	\centering
	\includegraphics[width=0.45\textwidth]{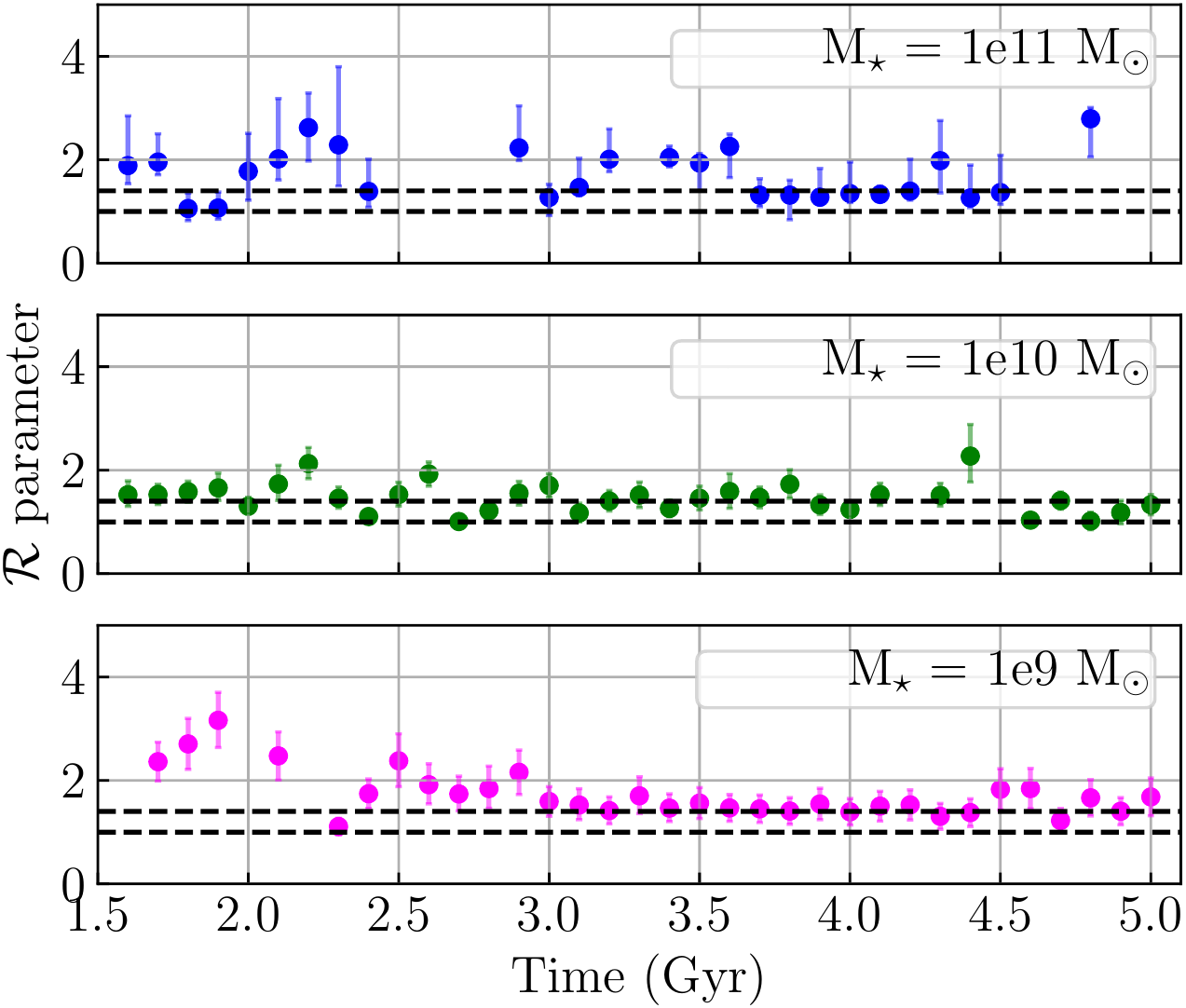}
	\caption{$\mathcal{R}$ parameter for the models 1e11 (blue), 1e10 (green), and 1e9 (magenta). All the models behave similarly in the sense that $\mathcal{R}$ remains almost constant during their evolution. The dashed lines show the fast bar regime ($1 \leq  \mathcal{R} \leq 1.4 $)}.
	\label{fig:R}
\end{figure}

Fig.~\ref{fig:R} shows the evolution of the $\mathcal{R}$ parameter in our models 1e11 (blue), 1e10 (green), and 1e9 (magenta). As expected, the $\mathcal{R}$ parameter remains almost constant with time. The mean value of $\mathcal{R}$ is $1.7^{+0.6}_{-0.4}$ (model 1e11), $1.5^{+0.2}_{-0.2}$ (model 1e10), and $1.7^{+0.3}_{-0.3}$ (model 1e9), which is rather close to the fast bar regime.

Given the lack of secular evolution, it is reasonable to consider our models viewed at different times as representative of a population of galaxies viewed at the same time. This is helpful because observationally, the $\mathcal{R}$ parameter is only known at low $z$. To do a comparison of this sort, we used the reported values for the bar length and corotation radius of the 104 galaxies in \citet{Cuomo2020} to calculate the $\mathcal{R}$ parameter and its error for each observed galaxy in this sample. Then the range for the above-mentioned mean $\mathcal{R}$ and its error for each simulated model is considered and the number of observed galaxies having an overlap with this range is counted. The number of observed galaxies that match this criterion is $54$ (51.9\%), $64$ (61.5\%), and $52$ (50.0\%) for models 1e11, 1e10, and 1e9, respectively. Our galaxy models therefore agree reasonably well with observations.

\begin{figure}
	\centering
	\includegraphics[width=0.45\textwidth]{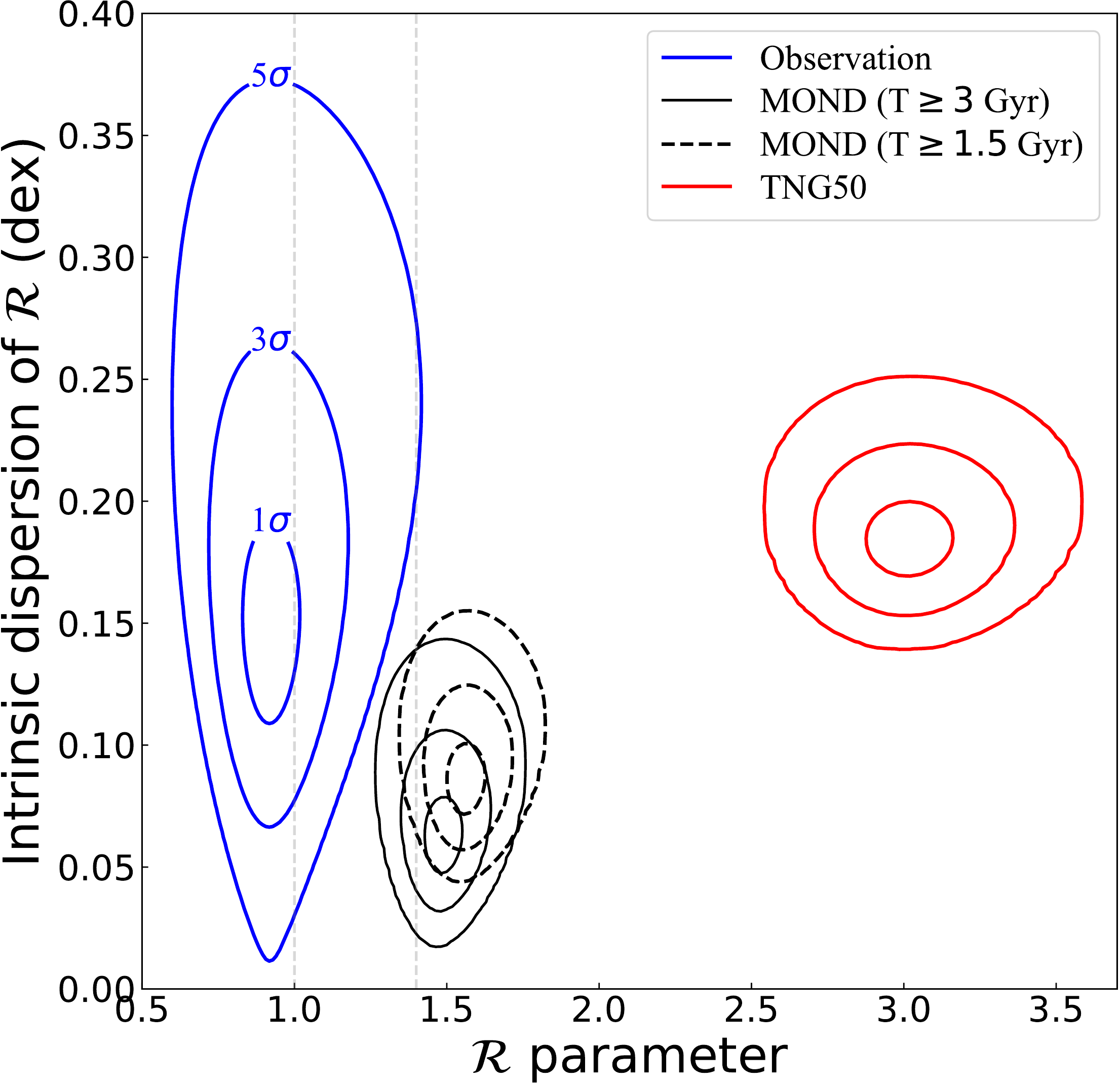}
	\caption{The posterior inference on the population logarithmic mean $\mathcal{R}$ and its intrinsic dispersion for the MOND simulation data at $t \geq 1.5$~Gyr (dashed black) and $t \geq 3$~Gyr (solid black) in comparison to the observations (blue) and TNG50 simulation results (red). The vertical dashed gray lines indicate the fast bar region.}
	\label{fig:contours}
\end{figure}

For a better comparison, we study the statistical distribution of the $\mathcal{R}$ parameter for our models using the exact same method applied in \citet{Roshan2021_2}. For more information regarding the employed method, we refer the reader to their section 5. In Fig.~\ref{fig:contours}, we plot the posterior inference on the mean value of $\mathcal{R}$  and its related intrinsic dispersion in log-space. In this figure, the black contours are calculated for the MOND simulation for all the data points resulting from the three mass models 1e11, 1e10, and 1e9 M$_\odot$ at $t \geq 1.5$ Gyr (dashed) and $t \geq 3$ Gyr (solid). The galaxies in these contours are contrasted to the red contours yielded by galaxies in the TNG50 simulation \citep{Roshan2021_2} and the galaxies in the blue contours of the observational results \citep{Cuomo2020}. Note that the MOND simulations studied here are isolated, so their comparison to $\Lambda$CDM cosmological simulations and observational data can only provide a general perspective on the fast bar issue. One obvious shortcoming of our results is that we only consider a small number of galaxies, so a more complete MOND simulation should lead to a larger intrinsic dispersion in $\mathcal{R}$, which would be more in line with the observations. Improving the resolution should also slightly reduce the typical value of $\mathcal{R}$ \citep[see fig.~21 of][]{Roshan2021_1}. Despite the isolated nature of our simulations and other shortcomings, it is clear that MOND holds great promise in explaining the observed fast bars, which however falsify $\Lambda$CDM cosmology at $13\sigma$ significance \citep{Roshan2021_2} based on TNG50 or TNG100.

The tension with $\Lambda$CDM is mostly due to the simulated bars being too short \citep{Frankel2022}. This tension is greatly reduced in our MOND simulations, but not completely eliminated. The somewhat high $\mathcal{R}$ value in our simulations might be influenced by underestimation of the bar length according to the employed method for measuring the bar size, as suggested by \citet{Frankel2022}. As illustrated in Fig.~\ref{fig:contours}, by giving the disc enough time to pass the phase of the initially strong spirals and settle down completely, even better compatibility to the fast bar regime is obtained in the simulation data. Furthermore, the overall constancy of the $\mathcal{R}$ parameter over time complies with observational findings \citep{Perez2012, Kim2021, Lee2022}.

\begin{figure}
	\centering
	\includegraphics[width=0.5\textwidth]{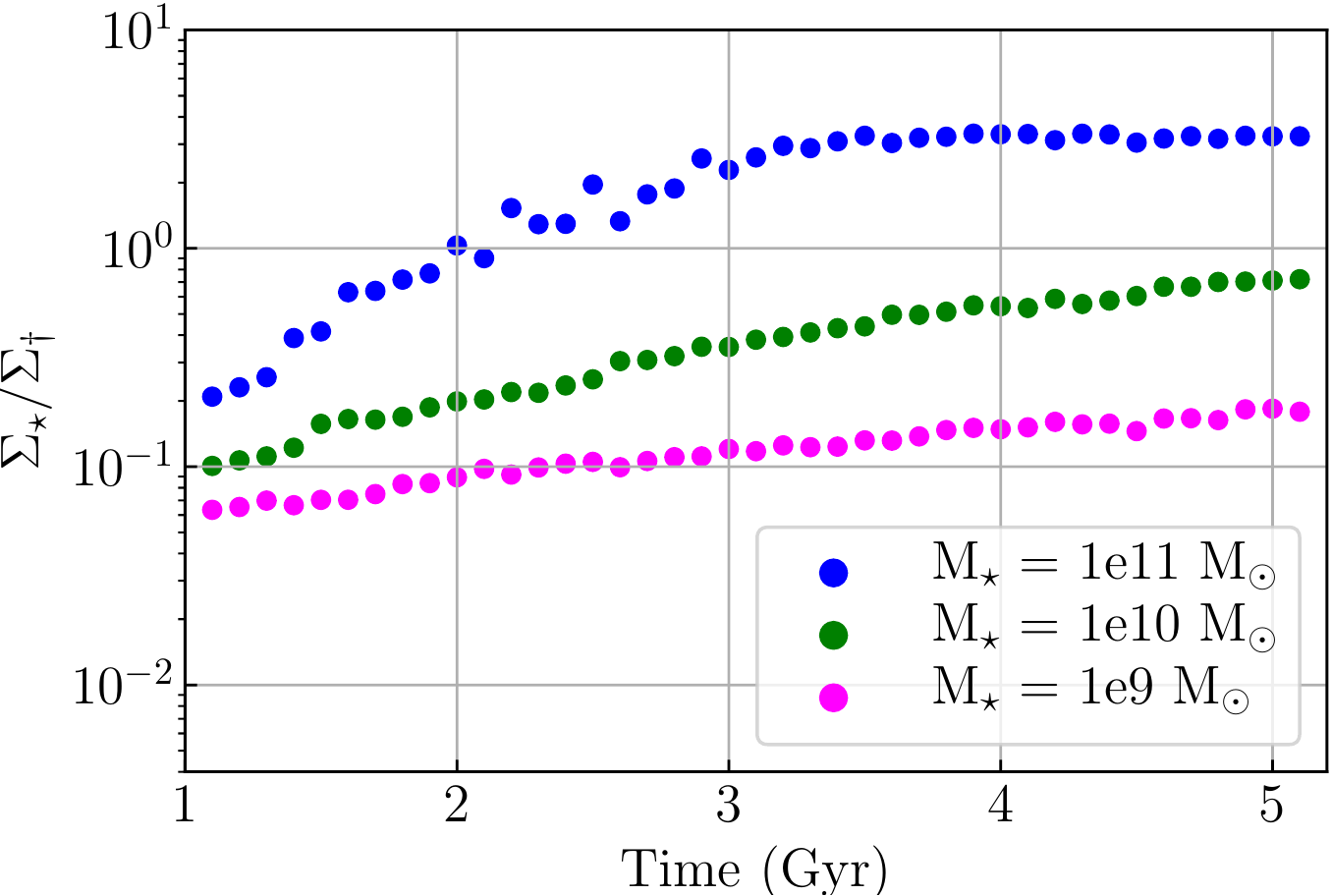}
	\caption{Evolution of the central stellar surface density in the models shown (see the legend). Values are shown in units of the critical surface density in MOND (Eq.~\ref{eq:crit_surface_density}). Some central concentration of mass is evident in all cases, though no bulge forms (see} Appendix~\ref{Face_on_view}).
	\label{fig:Sigma_ratio_evolution}
\end{figure}

Our results regarding the $\mathcal{R}$ parameter are similar to the MOND simulation in \citet{Roshan2021_1}, where $\mathcal{R}$ stayed roughly constant with time (see their fig.~22). However, the galaxy bar in their work was mostly in the fast regime, whereas our galaxy bars are somewhat slower. One possible reason for this difference is that they considered a galaxy where the central surface density is $10 \, \Sigma_\dagger$ initially, whereas our models start with lower values (Table~\ref{tab:SPARC}). The central density of the stellar component rises only modestly and by an almost similar factor in all cases (Fig.~\ref{fig:Sigma_ratio_evolution}), so the initial values in Table~\ref{tab:SPARC} give a good indication of how deep the galaxies are into the MOND regime. However, this is likely not the issue because model 1e11 is not too distinct from the model in \citet{Roshan2021_1} in this respect. Another difference is that they considered a purely stellar model with $5 \times 10^6$ particles. Our models have a lower resolution in the gas component because it is computationally very expensive to allow enough refinement levels to have that many gas cells in the disc region. Moreover, our results could also differ due to the hydrodynamic nature of our simulations and the fact that they include star formation and supernova feedback. The existence of gas in the system affects the bar's evolution because of its effect on the distribution of angular momentum \citep[][and references therein]{Frankel2022}. Further work will be required to clarify if bars are indeed expected to be slow in the MOND framework, which could pose a problem similar to that faced by $\Lambda$CDM \citep{Roshan2021_2}. In general, unlike the cosmological simulations considered in their work, the simulations presented here are not yet at a stage where they can be directly compared to observational surveys. However, we do expect that a lack of dynamical friction from a massive halo should improve the agreement with observations of fast galaxy bars. This is evident in the much lower values of $\mathcal{R}$ in our MOND simulations compared to TNG50.

\section{Discussion and conclusions}
\label{sec:Discussion}

We presented hydrodynamical simulations of Milgromian disc galaxies that include star formation and stellar feedback. Our models cover the stellar mass range $M_{\star}/M_\odot = 10^7 - 10^{11}$, with the disc scale lengths and gas fractions chosen based on the SPARC dataset \citep{SPARC}. Our aim was to check if combining MOND with the sub-grid physics encoded in \textsc{ramses} leads to star formation activity that adequately resembles the observed SFRs of galaxies. The disc galaxies were set up using \textsc{dice} \citep{Banik_2020_M33} and then advanced for 5~Gyr using \textsc{por} \citep{Lughausen_2015, Nagesh_2021}. To avoid fine-tuning the model parameters in an attempt to reproduce observations, all our models have the same star formation parameters (efficiency, ISM temperature, floor temperature, and feedback prescriptions). Our main results are as follows:
\begin{enumerate}
    \item The models agree reasonably well with the observed MS (Fig.~\ref{fig:Main-sequence}) $-$ their global SFR is reasonable given their $M_{\star}$.

    \item The simple and intermediate complexity feedback prescriptions (Sections~\ref{simple_feedback} and \ref{inter_feedback}, respectively) give rather similar results, with the SFR being close to the MS value with either prescription. Thus, Milgromian galaxies are not very sensitive to the choice of feedback prescription and do not expel a substantial portion of their gas, in line with observational studies \citep{Recchi_2015_chemistry, Marasco_2023}.
    
    \item The gas depletion timescales calculated using two methods agree fairly well with observations (Fig.~\ref{fig:Gas-depletion-times}). Less massive galaxies have a longer gas depletion timescale for an invariant canonical IMF, a trend also seen in our models.

    \item Our galaxies deplete their gas in a small fraction of a Hubble time, whereas observed galaxies have a roughly constant SFH \citep{Kroupa_2020a}. This deficiency of our models is caused by the fact that to maintain a constant SFR, a galaxy needs to accrete gas, a process not considered here. Including gas accretion from the surroundings would require a cosmological MOND simulation.

    \item The gas mass in the disc decreases roughly exponentially with time in models 1e11, 1e10, and 1e9. The gas mass seems to hit a floor in models 1e8 and 1e7, indicating inefficiency in the star formation process. This could be due to low-mass star-forming regions not being resolved. The minimum mass of stellar particles is $\approx 10^4 \, M_\odot$ for models 1e11, 1e10, and 1e9, but this drops to $10^3 \, M_\odot$ for models 1e8 and 1e7.

    \item The KS relation \citep{Kennicutt_1998} is a tight empirical relation between the local surface densities of gas and of star formation in disc galaxies. By dividing our galaxies into annuli, we found that most regions of model 1e11 lie on the KS relation (Fig.~\ref{fig:KS_Obs}), which is expected because it lies on the MS. However, models 1e10 and 1e9 fall below the MS. As a result, many regions of these models fall below the KS relation. This may not be problematic for our models because the KS relation was only tested with high surface brightness galaxies. Observed LSBs deviate below the KS relation \citep{Kennicutt_1998, Bigiel_2008}. It is thus good that models 1e8 and 1e7 (typical LSBs) also deviate below the KS relation. However, there are regions from models 1e11, 1e10, and 1e9 that deviate more significantly from the KS relation. These regions typically contain gas-rich clumps with a low SFR.
    
    \item Since Milgromian galaxies are purely baryonic, our models demonstrate a correspondence between features in the baryonic surface density profile and in the RC (Fig.~\ref{fig:Rot_curves}). This correspondence has been observed in real galaxies \citep{Famaey_McGaugh_2012} and is called Renzo's Rule \citep{Renzo_2004}. Its inevitability in MOND is related to its important prediction that baryons dominate the gravitational potential both locally and globally.

    \item The importance of self-gravity at all radii explains why our models naturally form structures at large radii (Fig.~\ref{fig:Streams_1e10}). This gives new insight into such structures, which are often thought of as non-disc entities like an infalling satellite. These structures might need to be reinterpreted as part of the outer disc, though we note that they should still be moving within the disc plane.

    \item Fig.~\ref{fig:Vert_Vel_r_disp} shows that our models remain as thin discs throughout their evolution. The final set of parameters are shown in Table~\ref{tab:Final_parameters}, which reveals that $R_d$ and $R_g$ have increased over the 5~Gyr duration of our simulations. Expansion of the disc was also noted in the MOND simulations presented in \citet{Roshan2021_1}. This implies that Milgromian galaxies naturally expand during their evolution, so sophisticated feedback prescriptions are not required to prevent the baryons from collapsing to the centre \textsuperscript{\ref{Movies}}.

    \item The evolution of the bar properties over time is illustrated in Figs.~$\ref{fig:BS} - \ref{fig:R}$ for the three models with a stellar mass of $10^9 \, M_\odot$, $10^{10} \, M_\odot$, and $10^{11} \, M_\odot$. According to these plots, all three models produce weak bars with a length that is almost unchanged during the disc's evolution. The bar length is larger for the more massive models, which is consistent with observational results. Moreover, employing the TW method, it has been shown that after some initial fluctuations, the bar pattern speed remains constant for all three models until the end of the simulation. A related finding is that due to the absence of DM particles in our models, the $\mathcal{R}$ parameter shows no increasing trend. The overall constancy of the bar length, pattern speed, and $\mathcal{R}$ parameter over time shows consistency with observational findings, although the bars are slower than in previous stellar-only MOND models \citep{Roshan2021_1}. This is probably related to the inclusion of hydrodynamics. A more detailed study of the influence of hydrodynamics and gas fraction on bar properties in MOND is left to further work. The MOND models remain simplified since they are not in a cosmological setting, but indicate better agreement with the observed galaxies. Our results show that MOND holds great promise in explaining why galaxies typically have fast bars, a result which is in severe tension with $\Lambda$CDM (see Fig.~\ref{fig:contours}).
\end{enumerate}

The main sub-grid parameters include the star formation efficiency, the supernova bubble radius, the fraction of energy from SNe injected as kinetic energy, and various other feedback parameters. These were previously tested in \textsc{ramses} \citep{Teyssier_2002, Teyssier_2006, Dubois_2008, Wittenburg_2020}. Our work demonstrates that in the context of MOND, the adopted values for these parameters in 2015 already allow us to adequately reproduce the properties of disc galaxies. Therefore, our main conclusion is that significant fine-tuning of the sub-grid parameters is not required to achieve fairly realistic model galaxies in terms of their SFR. Despite some deviations, our models agree with observations within uncertainties, especially at the low-mass end where relatively little prior work has been done. Moreover, changing the feedback prescription has little effect on the results, backing up the hypothesis that galaxies are simple systems \citep{Disney_2008}. Though they are undoubtedly subject to complex feedback processes, these play a relatively minor role \citep{Kroupa_2015}. This can be understood through an analogy to how Newtonian gravity works in planetary systems despite our still incomplete understanding of the complex processes underpinning the formation of stars and planets out of gas clouds. The relatively minor role of feedback is of importance for cosmological MOND simulations because it means that, in principle, these can be run safely with the simple or intermediate feedback prescriptions. It remains to be seen if such prescriptions will remain appropriate in order to produce realistic galaxy populations (rather than individual galaxies) in a cosmological context, which will be the obvious next step in our endeavours (Wittenburg et al., in prep).

\section*{Acknowledgements}

This work was performed as an MSc thesis project at the University of Bonn. IB is supported by Science and Technology Facilities Council grant ST/V000861/1, which also partially supports HZ. IB acknowledges support from a `Pathways to Research' fellowship from the University of Bonn. IT acknowledges support through the Stellar Populations and Dynamics research (SPODYR) group at the University of Bonn. PK thanks the DAAD-Eastern Europe exchange program for support. NG thanks Tahere Kashfi for providing the codes calculating the posterior inferences on population bar properties. The authors thank Moritz Haslbauer, and Stacy McGaugh for useful discussions. They are also grateful for comments from the anonymous referee which helped to substantially improve this manuscript.

\section*{Data availability}

The algorithms used to prepare and run \textsc{por} simulations of disc galaxies and to extract their results into human-readable form are publicly available.\textsuperscript{\ref{PoRbit}} A user manual is available describing the operation of these codes \citep{Nagesh_2021}. Movies showing the evolution of the models are publicly available \textsuperscript{\ref{Movies}}.

\bibliographystyle{mnras}
\bibliography{Star_formation_article}

\begin{appendix}

\section{Star formation histories}
\label{SFH_Four_models}

\begin{figure*}
    \includegraphics[width=0.495\textwidth]{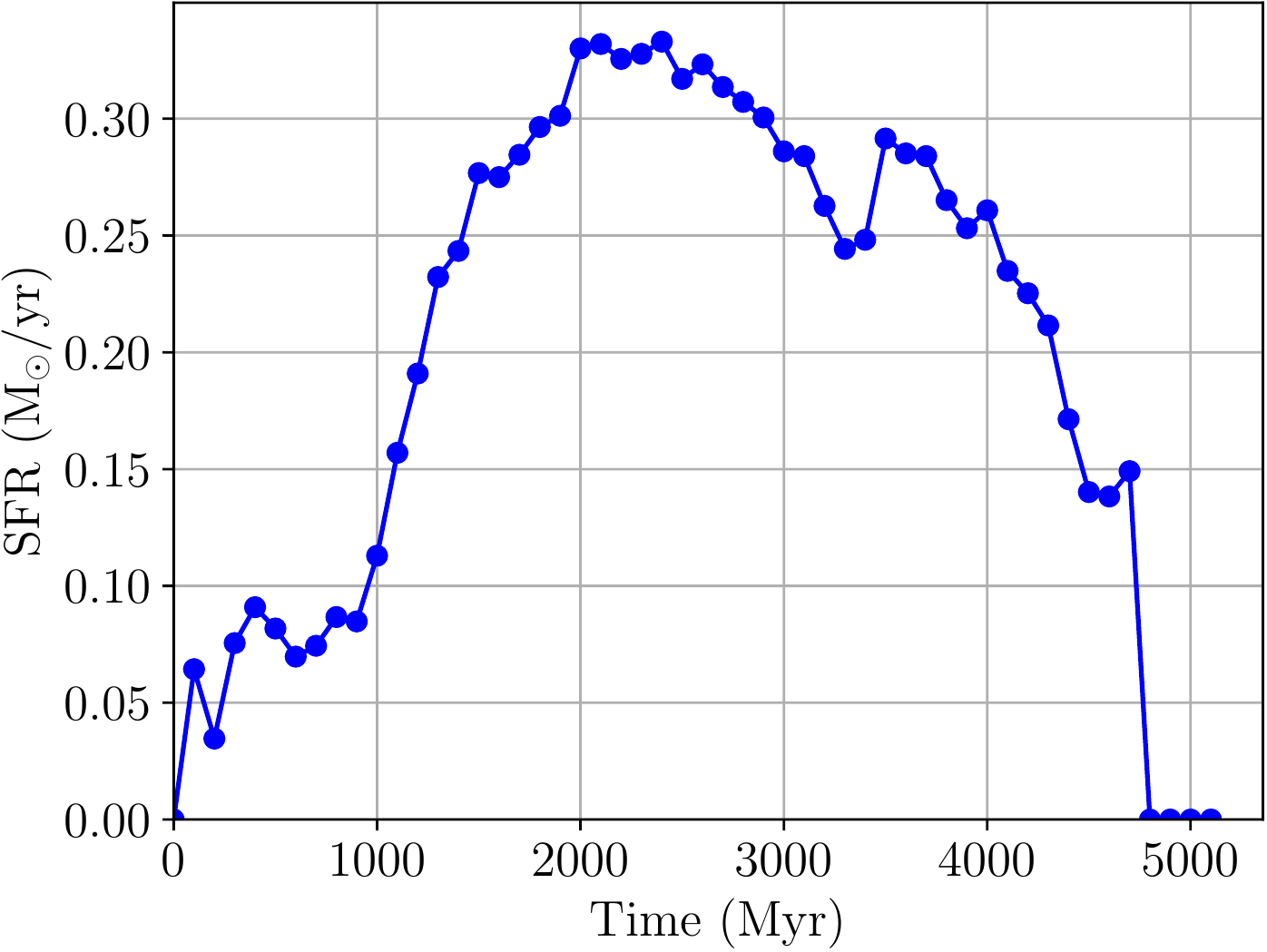}
    \hfill
    \includegraphics[width=0.495\textwidth]{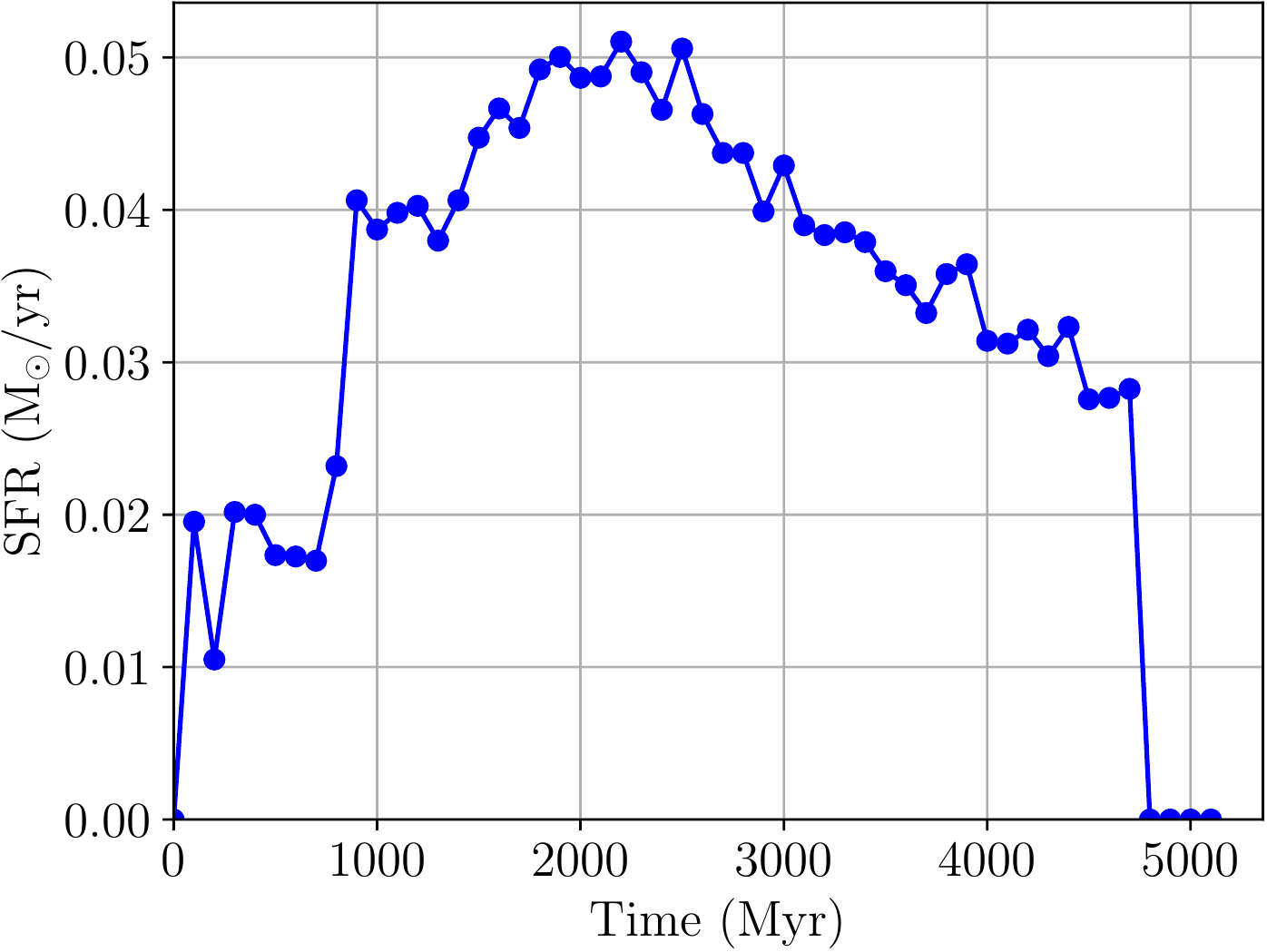}
    \includegraphics[width=0.495\textwidth]{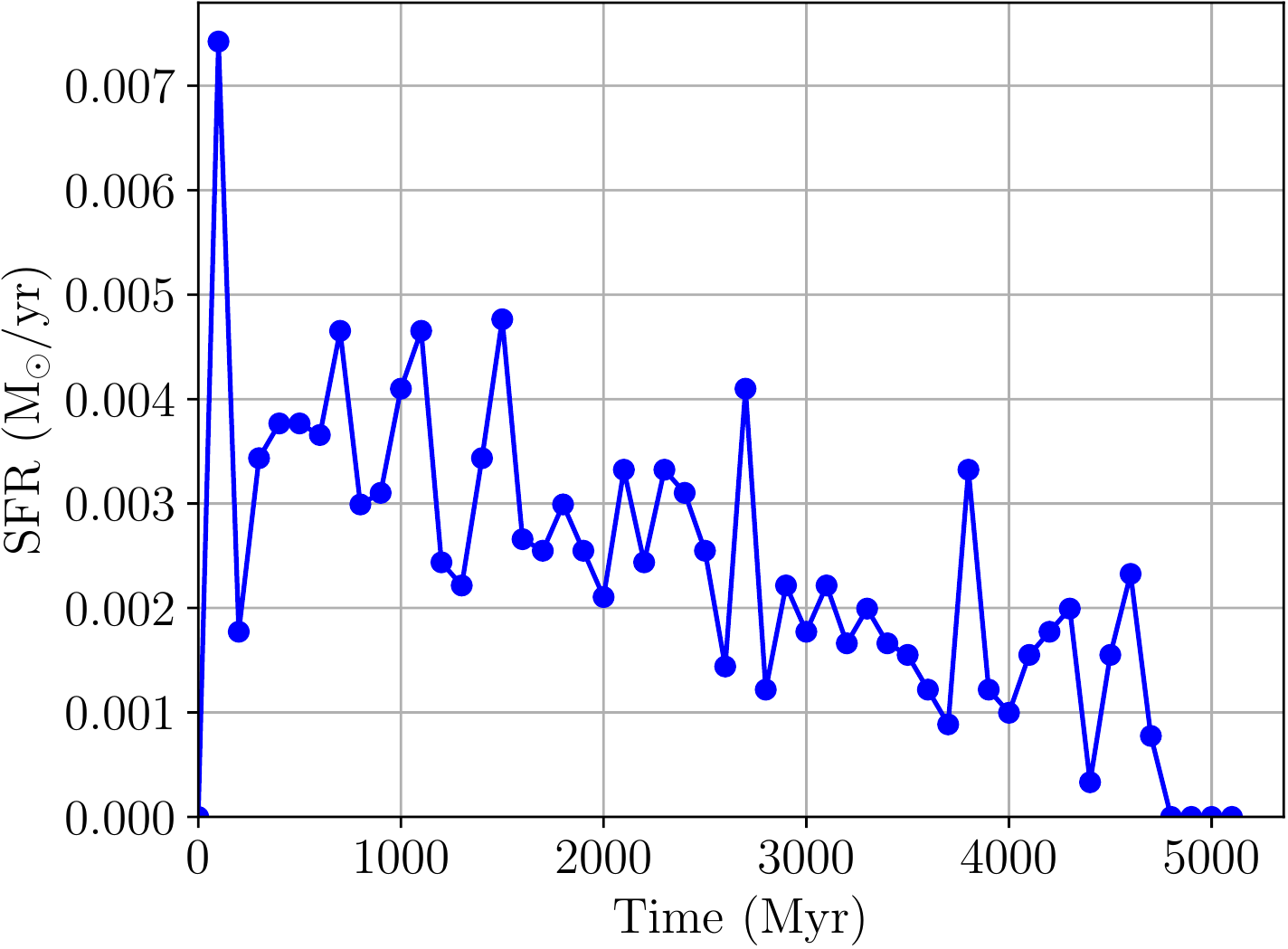}
    \hfill
    \includegraphics[width=0.495\textwidth]{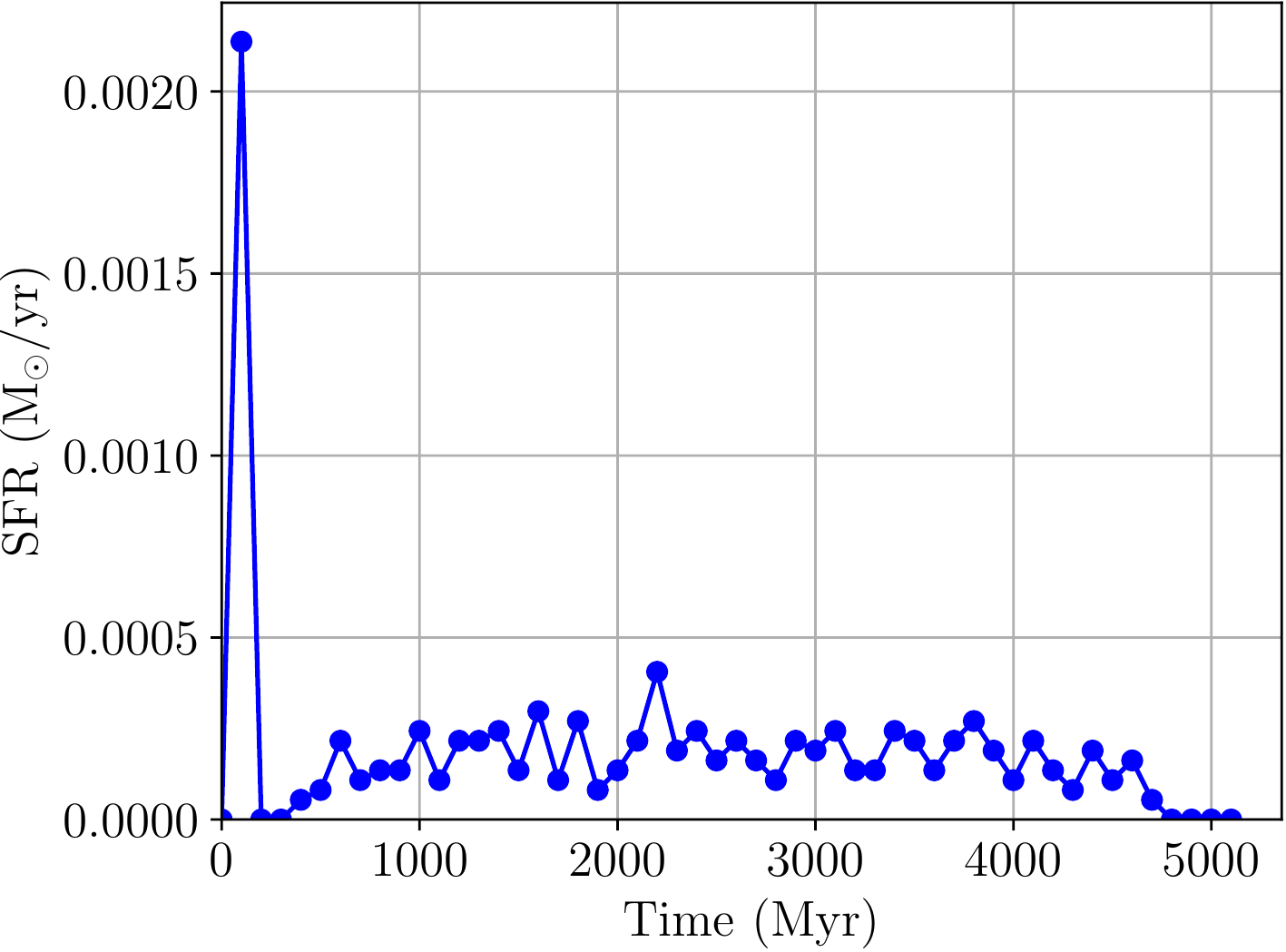}
    \caption{Star formation history of model 1e10 (top left), 1e9 (top right), 1e8 (bottom left), and 1e7 (bottom right). Model 1e11 is shown in Fig.~\ref{fig:SFH_1e11}.}
    \label{fig:SFH_1e10_1e7}
\end{figure*}

For model 1e11 (Fig.~\ref{fig:SFH_1e11}), the SFR rises smoothly until it peaks at $\approx 1.5$~Gyr, after which it declines as the gas is used up. For models 1e10, and 1e9 (top panels of Fig.~\ref{fig:SFH_1e10_1e7}), the SFR rises and declines gradually, in contrast to the other models. For models 1e8 and 1e7 (bottom panels of Fig.~\ref{fig:SFH_1e10_1e7}), there is an instant burst of star formation within the first 100 Myr, i.e., the SFR peaks at the first time step, declining or staying constant thereafter. At the end of their evolution, the SFR becomes zero. This is not due to exhaustion of gas mass, but rather occurs because the density of gas becomes smaller than the threshold required for star formation.

\section{Radial velocity dispersion}
\label{Radial_disperison}

\begin{figure*}
    \includegraphics[width=0.495\textwidth]{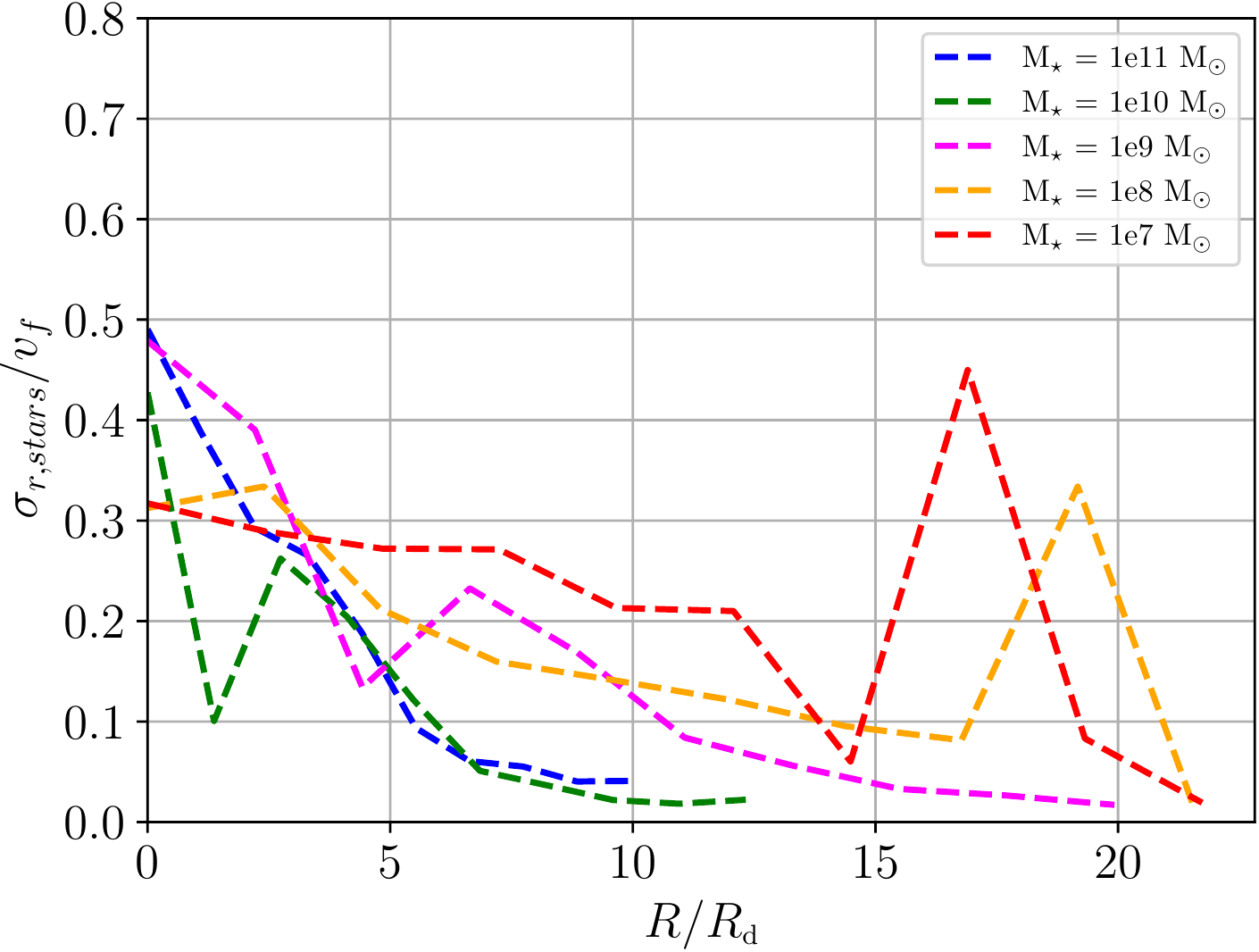}
    \hfill
    \includegraphics[width=0.495\textwidth]{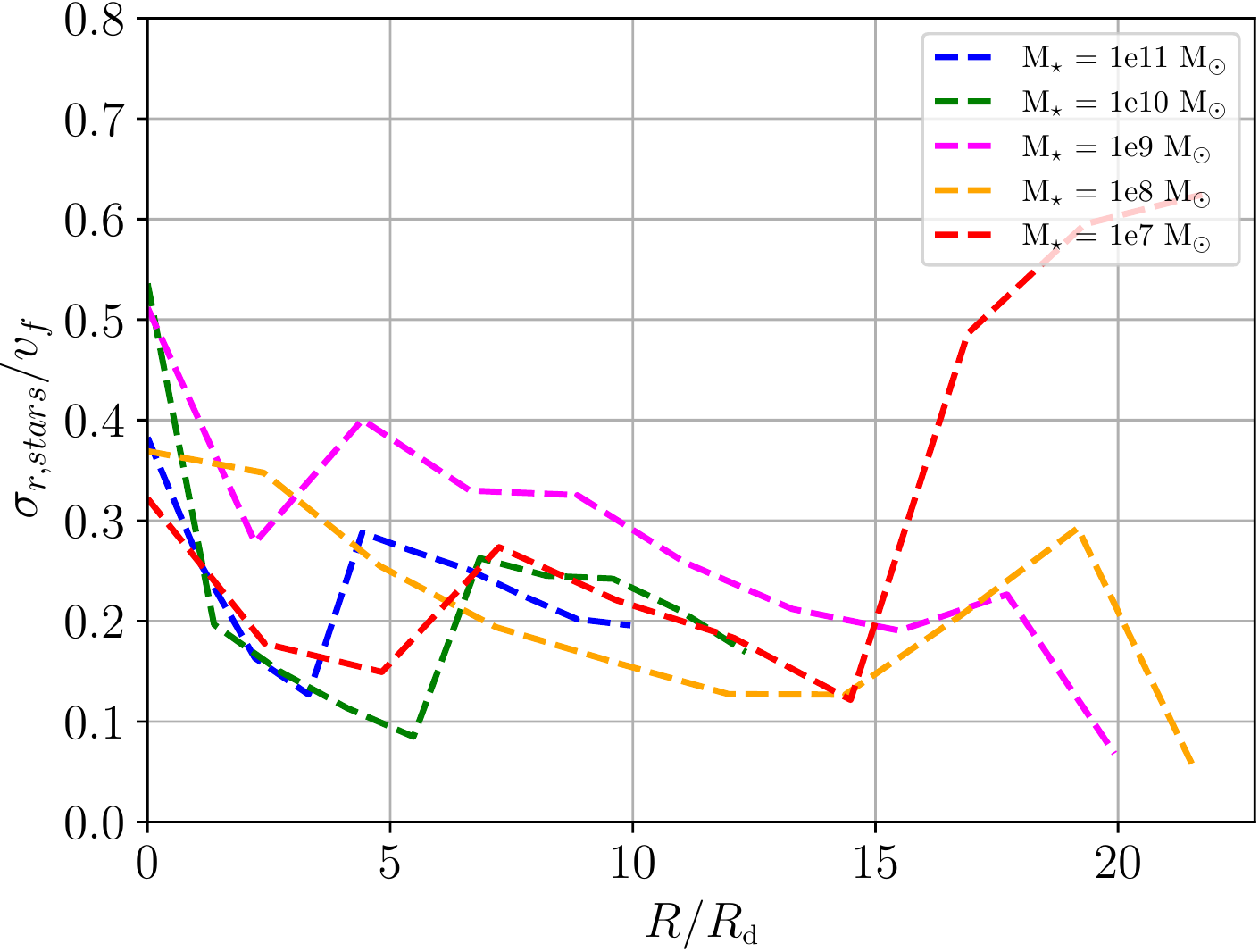}
    \caption{$\sigma_r/v_{_f}$ of all stellar particles as a function of galactocentric distance after 1~Gyr (left) and 5~Gyr (right), shown for all our models as indicated in the legend.}
    \label{fig:sigma_r_all}
\end{figure*}

\begin{figure*}
    \includegraphics[width=0.495\textwidth]{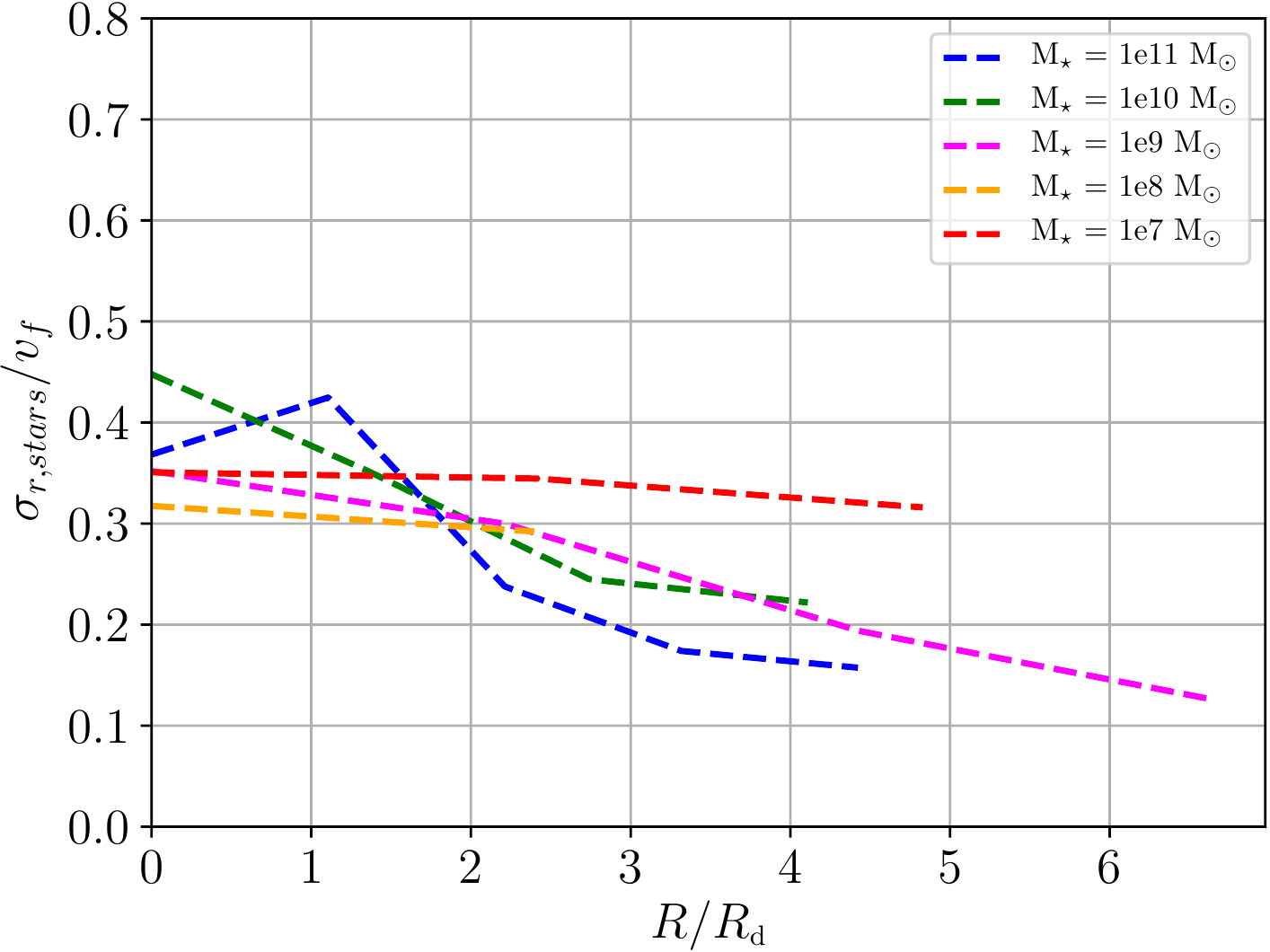}
    \hfill
    \includegraphics[width=0.495\textwidth]{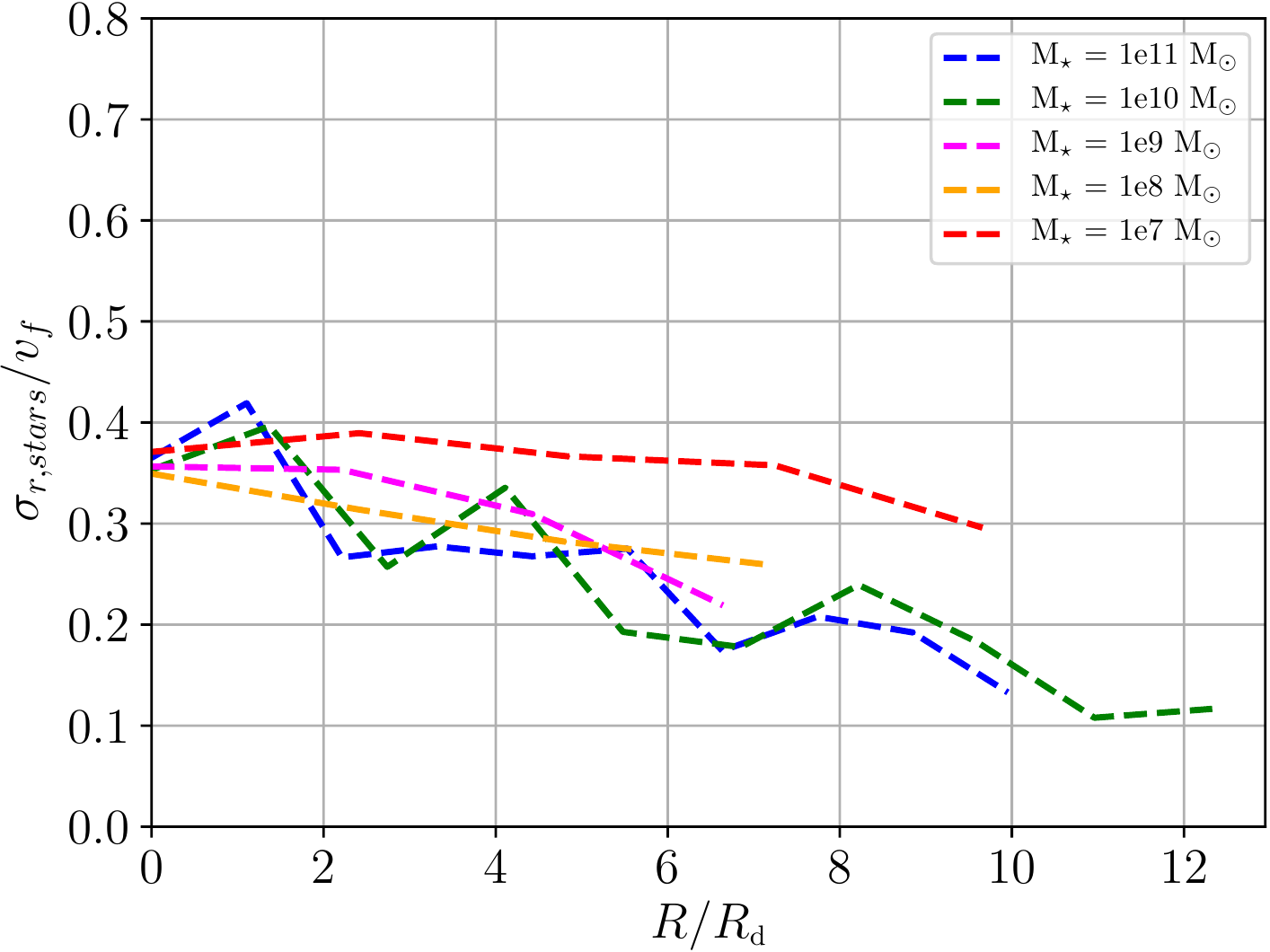}
    \caption{$\sigma_r/v_{_f}$ of newly formed stellar particles as a function of galactocentric distance after 1 Gyr (left) and 5 Gyr (right).}
    \label{fig:sigma_r_new}
\end{figure*}

In this section, we present the ratio between the radial velocity dispersion $\sigma_r$ and the asymptotic circular velocity $v_{_f}$ (Eq.~\ref{eq:asymptoticvelocity}) after 1~Gyr and 5~Gyr. To find $\sigma_r$, we divide the galaxy into annuli and find the radial velocity $v_r$ of all particles in each annulus.
\begin{eqnarray}
    v_r ~=~ \frac{x v_x + y v_y}{\sqrt{x^2 + y^2}} \, .
\end{eqnarray}
We then find the dispersion in $v_r$ for the particles in each annulus. The results are shown in Fig.~\ref{fig:sigma_r_all}, while Fig.~\ref{fig:sigma_r_new} shows a version of this for only the particles that formed during the simulation. For comparison, $\sigma_r \approx 26 - 33$~km/s in the solar neighbourhood of the MW for stars aged between $1 - 5$~Gyr \citep{Yu_2018}. This translates to $\sigma_r/v_{_f} \approx 0.2$. Model 1e10 is analogous to the MW in mass, so it is reassuring that at $5 \, R_\textrm{d}$, the simulated value of $\sigma_r/v_{_f}$ is similar to that of the MW. 

\section{Face-on and edge-on views}
\label{Face_on_view}

\begin{figure*}
    \includegraphics[width=0.495\textwidth]{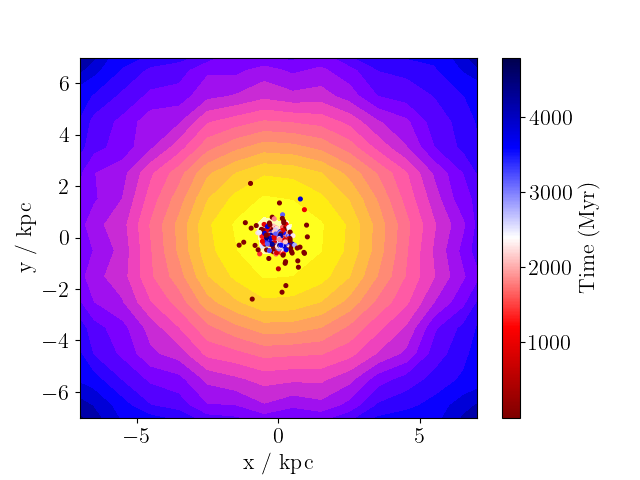}
    \hfill
    \includegraphics[width=0.495\textwidth]{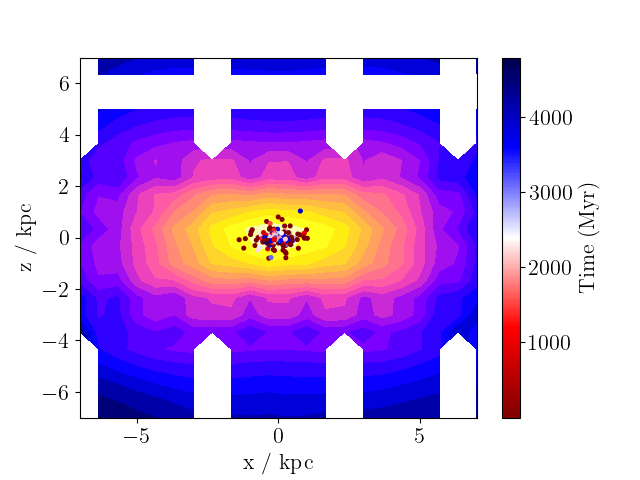}
    \caption{Model 1e7 after 5~Gyr, shown face-on (left) and edge-on (right). The colour shows the gas density. The newly formed stellar particles are shown as coloured dots, with the colour indicating the formation time as indicated on the colour bar.}
    \label{fig:Face_edge_1e7}
\end{figure*}

\begin{figure*}
    \includegraphics[width=0.495\textwidth]{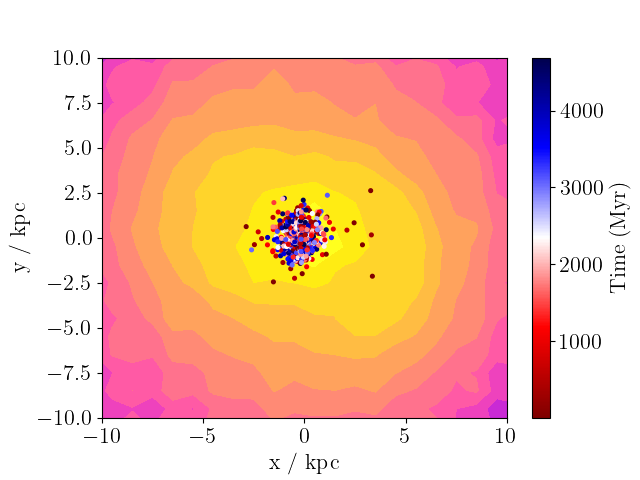}
    \hfill
    \includegraphics[width=0.495\textwidth]{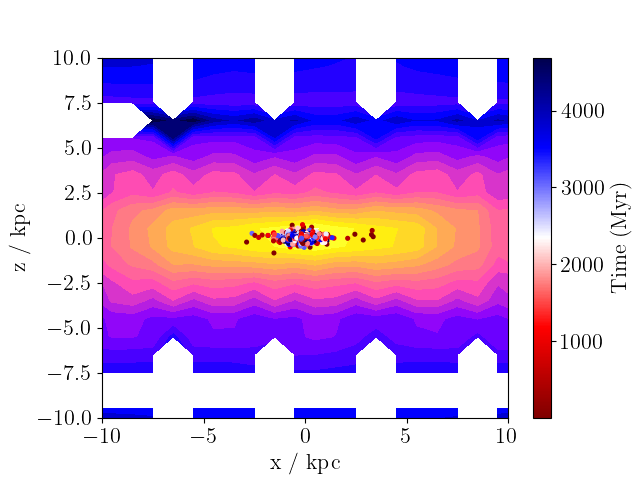}
    \caption{Similar to Fig.~\ref{fig:Face_edge_1e7}, but for model 1e8.}
    \label{fig:Face_edge_1e8}
\end{figure*}

\begin{figure*}
    \includegraphics[width=0.495\textwidth]{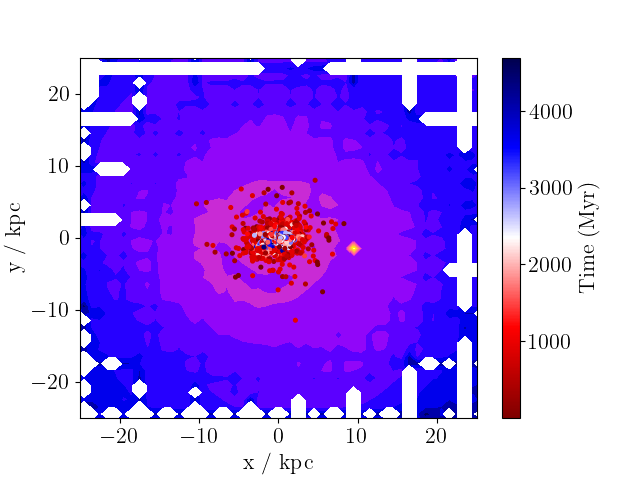}
    \hfill
    \includegraphics[width=0.495\textwidth]{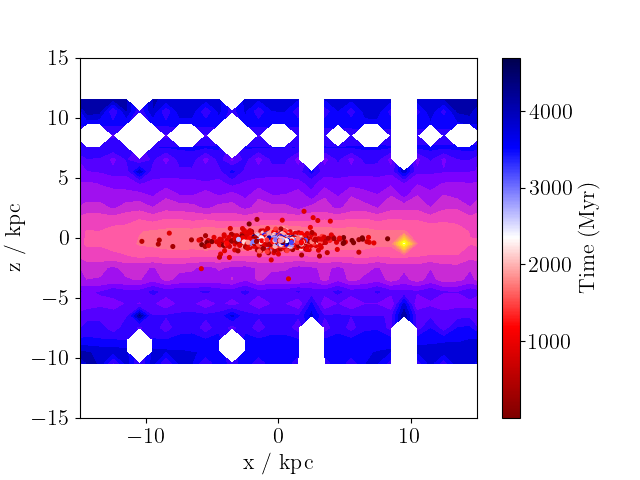}
    \caption{Similar to Fig.~\ref{fig:Face_edge_1e7}, but for model 1e9.}
    \label{fig:Face_edge_1e9}
\end{figure*}

\begin{figure*}
    \includegraphics[width=0.495\textwidth]{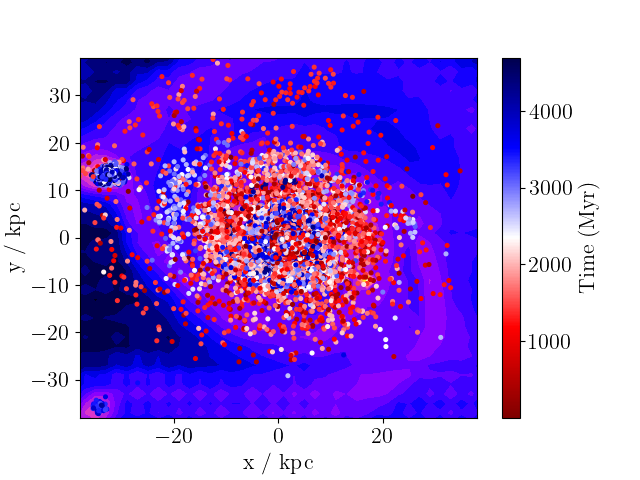}
    \hfill
    \includegraphics[width=0.495\textwidth]{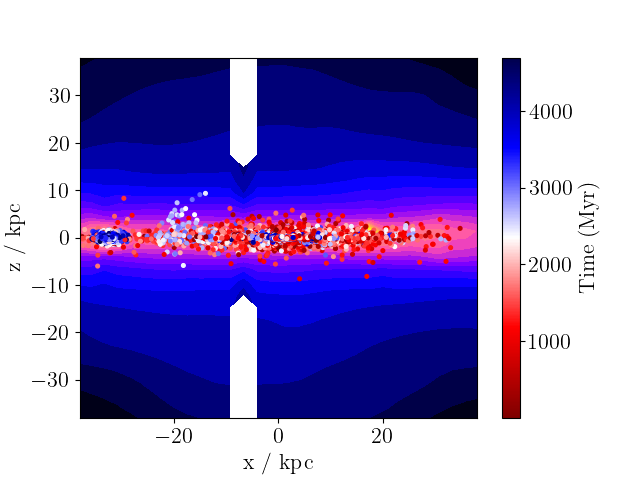}
    \caption{Similar to Fig.~\ref{fig:Face_edge_1e7}, but for model 1e10.}
    \label{fig:Face_edge_1e10}
\end{figure*}    

\begin{figure*}
    \includegraphics[width=0.495\textwidth]{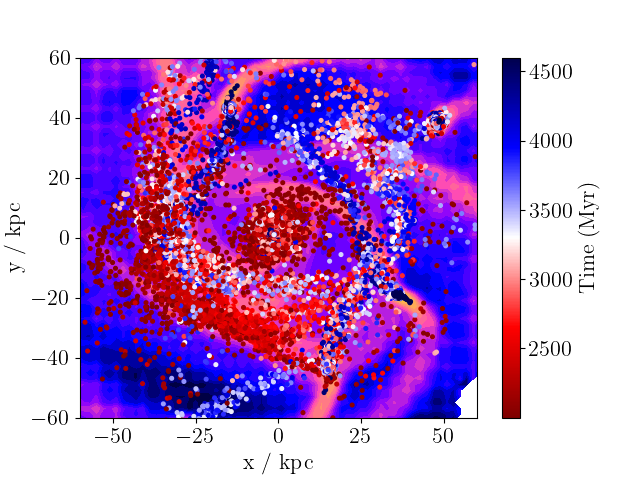}
    \hfill
    \includegraphics[width=0.495\textwidth]{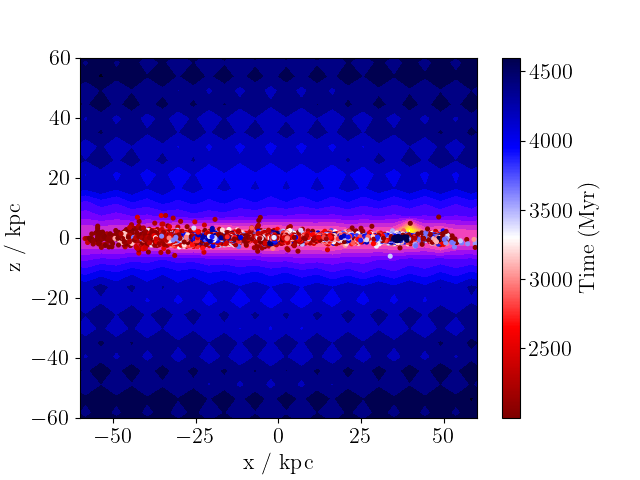}
    \caption{Similar to Fig.~\ref{fig:Face_edge_1e7}, but for model 1e11.}
    \label{fig:Face_edge_1e11}
\end{figure*}

The gas data extracted from \textsc{rdramses} can be binned in 2D to obtain $xy$ and $xz$ projections. Figs.~\ref{fig:Face_edge_1e7}, \ref{fig:Face_edge_1e8}, \ref{fig:Face_edge_1e9}, \ref{fig:Face_edge_1e10}, and  \ref{fig:Face_edge_1e11} show these projections for models 1e7, 1e8, 1e9, 1e10, and 1e11, respectively, for the final snapshot at 5~Gyr. All the new stellar particles that formed during the simulation are plotted on the gas distribution for the corresponding model. The stellar particles are colour-coded based on their birth time, which helps to provide some information on the age distribution of the stars. Note that in the edge-on views, there are white gaps in the image because the gas density is zero in those pixels and we use a logarithmic colour scheme.

\section{Effect of resolution}
\label{Effect_of_resolution}

\begin{figure}
   \includegraphics[width=0.495\textwidth]{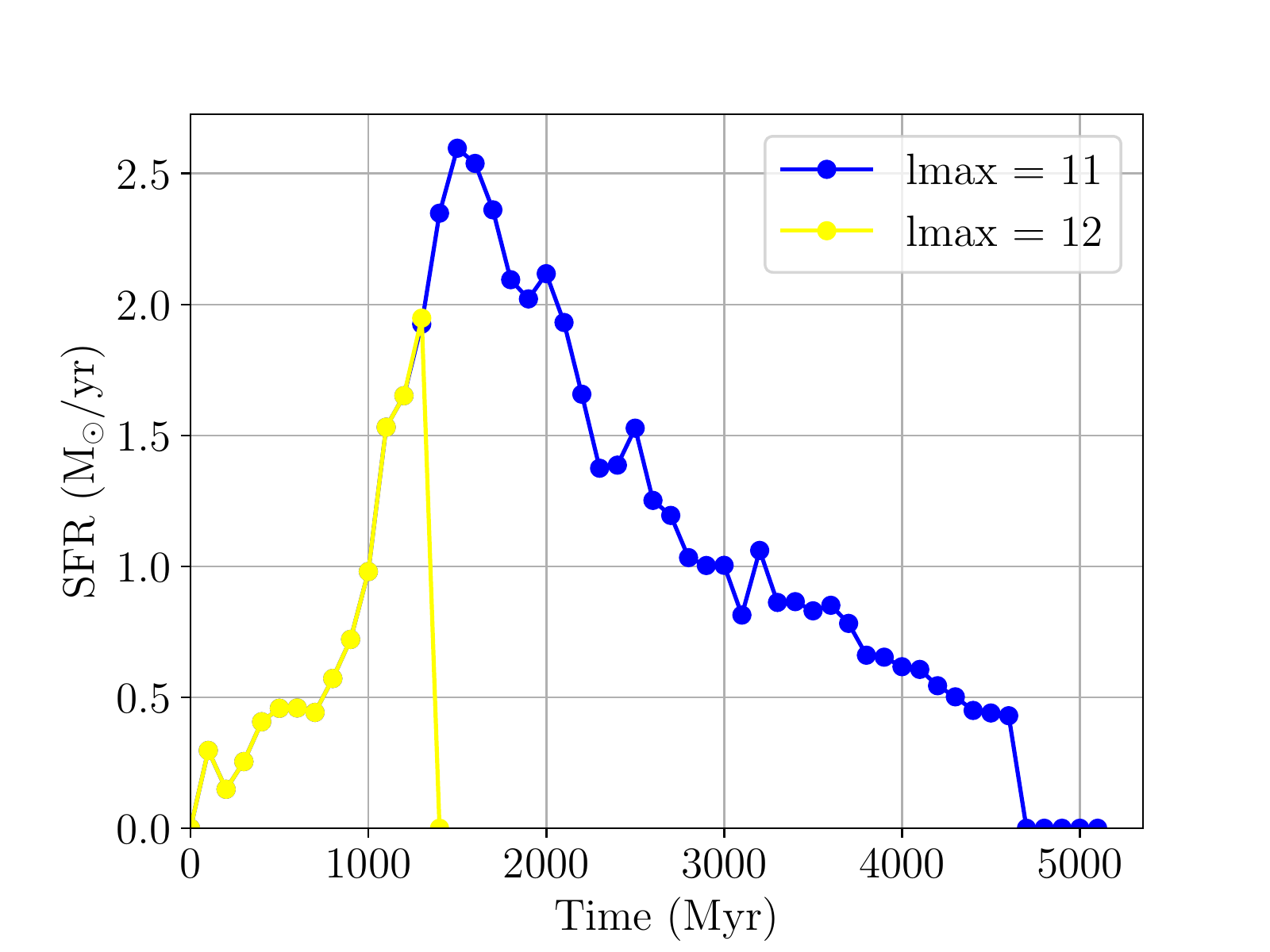}
   \caption{The SFH of model 1e11 with \textit{levelmax = 11} and \textit{levelmax = 12}}
   \label{fig:SFH_lmax11}
\end{figure}    

In Section~\ref{sec:Numerical-methods}, we discussed the effect of resolution on the SFR. Fig \ref{fig:SFH_lmax11} shows the SFH of model 1e11 with $levelmax = 11$ and 12. The SFR of model 1e11 with $levelmax = 11$ gradually rises and reaches a peak at 1.5 Gyr. The SFR for the models with $levelmax = 11$ and 12 are identical to within 1\% up to 1.3 Gyr. The $levelmax = 12$ model was able to run only up to this point due to computational limitations.

\section{Annular binning for KS analysis}
\label{KS_analysis}
\begin{figure}
    \includegraphics[width=0.495\textwidth]{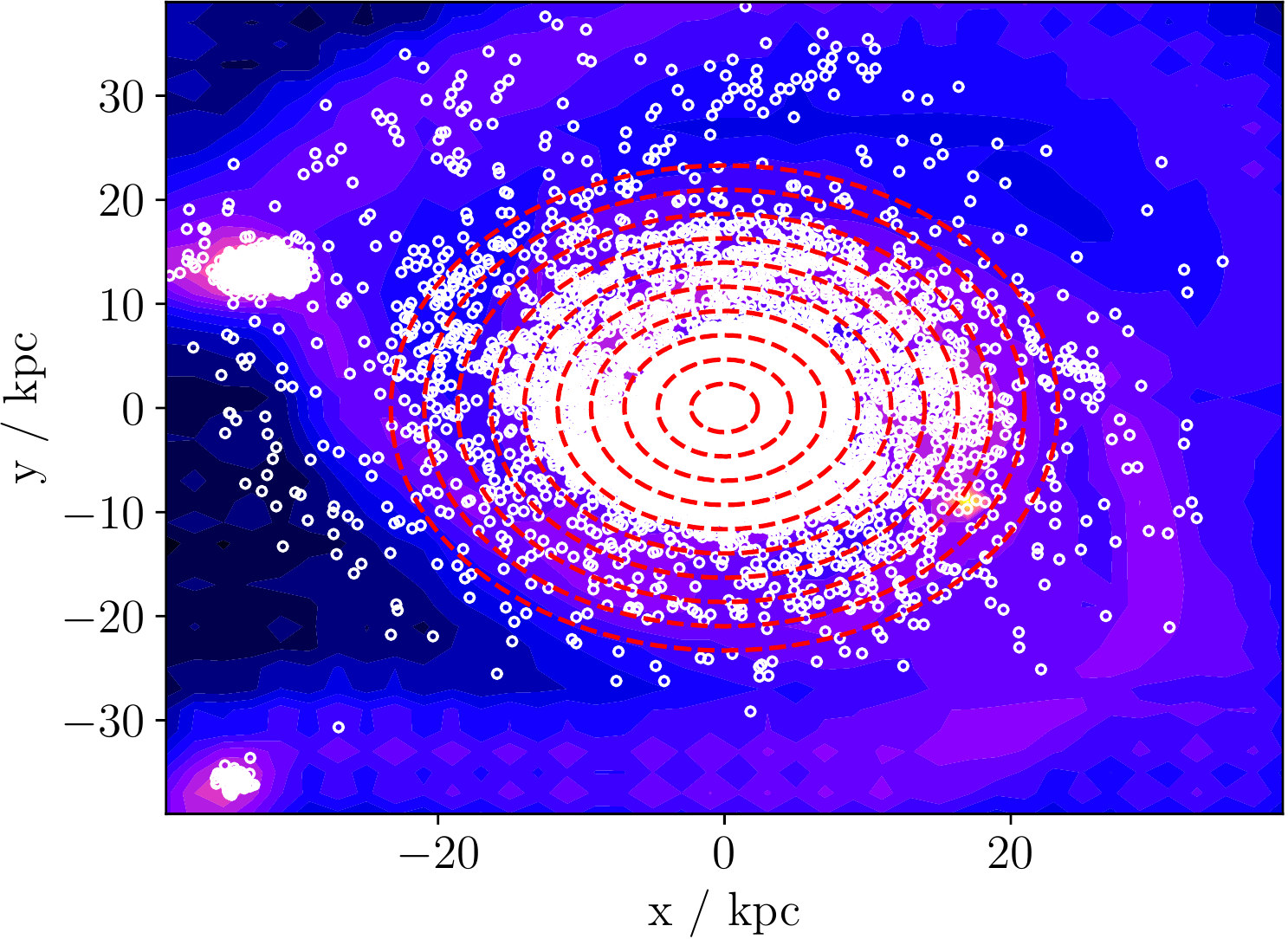}
    \includegraphics[width=0.495\textwidth]{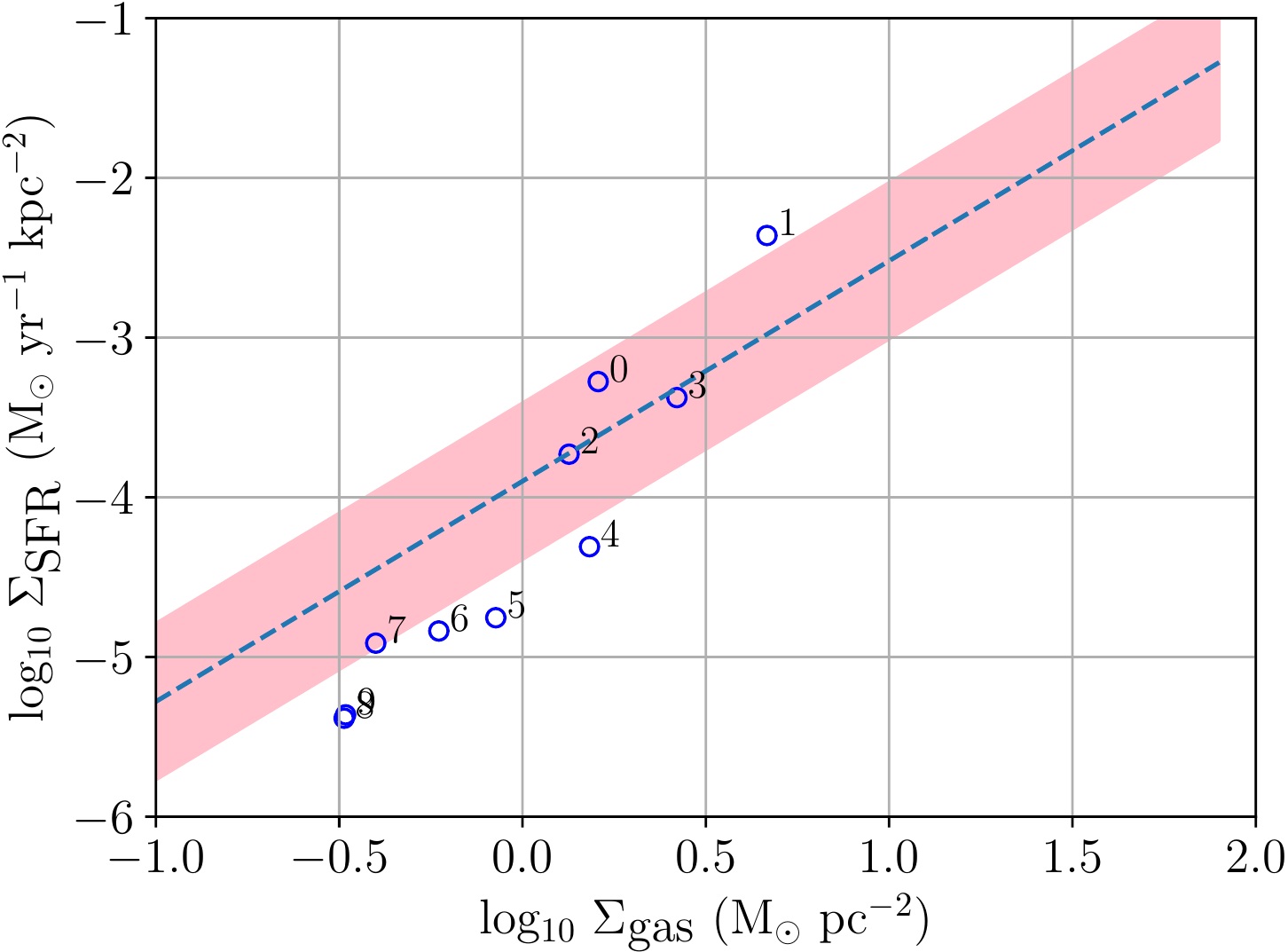}
    \caption{{\emph{Top}: The annular binning procedure used in much of this work for model 1e10 at 5 Gyr, showing circular bins out to $5 \, \widetilde{R}_\textrm{eff}$. \emph{Bottom}: How each of these bins contributes to the KS plot. The numbers indicate which bin is shown, starting from the centre and working outwards. For comparison, observations are well fit by eq.~\ref{eq:KS-law}, shown here as the dotted line surrounded by a shaded red band showing the uncertainty.}}
    \label{fig:KS_circular_bins}
\end{figure}
As mentioned in Section~\ref{sec:KS_law}, annular bins with a constant width of $\approx 10\times$ the highest resolution were constructed for each model out to $5 \, \widetilde{R}_\textrm{eff}$. Each point on the KS plot (Fig.~\ref{fig:KS_Obs}) corresponds to a particular bin. Fig.~\ref{fig:KS_circular_bins} shows an example of the annular binning for model 1e10 (top panel), along with the corresponding values from each bin on the KS plot (bottom panel). The technique applied is the same for all models. We used a maximum radius of $5 \,\widetilde{R}_\textrm{eff}$ as little star formation occurs further out.

\end{appendix}

\bsp
\label{lastpage}
\end{document}